\documentclass{bmcart}

\usepackage[utf8]{inputenc} 
\usepackage{amsmath}
\usepackage{amssymb}
\usepackage{amsthm}
\usepackage{adjustbox}
\usepackage{url}
\usepackage{multirow}
\usepackage{graphicx}
\usepackage{multirow}
\usepackage{tabularx}
\usepackage{booktabs}
\usepackage[ruled,linesnumbered,boxed]{algorithm2e}
\SetKwComment{Comment}{$\triangleright$\ }{}
\SetKwInOut{Parameter}{Parameters}
\usepackage{color}
\usepackage{soul}
\usepackage{subcaption}

\usepackage{mathptmx}       
\usepackage{helvet}         
\usepackage{courier}        
\usepackage{type1cm}        
\usepackage{makeidx}         

\newcommand{\REVISE}[1]{\textcolor{black}{#1}}

\startlocaldefs
\endlocaldefs

\begin{document}

\begin{frontmatter}

\begin{fmbox}
\dochead{Research}


\title{Layer entanglement in multiplex, temporal multiplex, and coupled multilayer networks}


\author[
   addressref={aff1},                   
   corref={aff1},                       
   email={blaz.skrlj@ijs.si}   
]{\inits{B\v{S}}\fnm{Bla\v{z}} \snm{\v{S}krlj}}
\author[
   addressref={aff2},
   email={renoust@ids.osaka-u.ac.jp}
]{\inits{BR}\fnm{Benjamin} \snm{Renoust}}


\address[id=aff1]{
  \orgname{Jo\v{z}ef Stefan International Postgraduate School},
  \city{Ljubljana},                              
  \cny{Slovenia}                                    
}

\address[id=aff2]{%
  \orgname{Median Technologies, INRIA, and Osaka University Institute for Datability Science},
  \city{Osaka},
  \cny{Japan},
}


\begin{artnotes}
\end{artnotes}

\end{fmbox}


\begin{abstractbox}

\begin{abstract}
Complex networks, such as transportation networks, social networks, or biological networks, capture the complex system they model often by representing only one type of interactions. In real world systems, there may be many different aspects that connect entities together. These can be captured using multilayer networks, which combine different modalities of interactions in a single model.
Coupling in multilayer networks may exhibit different properties which can be related to the very nature of the data they model (or to events in time-dependant data). We hypothesise that such properties may be reflected in the way layers are intertwined.
In this paper, we investigated these through the prism of layer entanglement in coupled multilayer networks. We test  over 30 real-life networks in 6 different disciplines (social, genetic, transport, co-authorship, trade, and neuronal networks). We further propose a random generator, displaying comparable patterns of elementary layer entanglement and \REVISE{transition coupling} entanglement across 1{,}329{,}696
 synthetic coupled multilayer networks.
Our experiments demonstrate difference of layer entanglement across disciplines, and even suggest a link between entanglement intensity and homophily. We additionally study entanglement in 3 real world temporal datasets displaying a potential rise in entanglement activity prior to other network activity.
\end{abstract}




\end{abstractbox}
%

\end{frontmatter}



\section{Introduction}
\label{sec:introduction}

A real world complex system often counts multiple interactions between multiple different entities. When these interactions are regrouped under multiple families of entities, multilayer network modelling becomes a tool of choice to capture the key components of the system. The use of this model emerges in all fields of science from social sciences to finances, through logistics, biology, and many more~\cite{kivela2014multilayer}.

With multilayer networks, the study of \textit{multiple viewpoints} (or aspects~\cite{kivela2019visual}) on the same network data becomes possible.
This is critical for example in social network analysis, to study the role of users in different networks, and compare them (for example the same individual may behave differently on LinkedIn, Twitter, or Facebook). These different networks form different types of links that may be overlaid.

Motivated by their practical interest, \textit{multilayer networks} also show interesting structures~\cite{battiston2014structural} that could be exploited to mine \textit{community structures} or study the roles of nodes and edges through centrality, for example. These are also possible in a traditional network analysis standpoint but often requires some kind of simplification (such as one-mode projection) but recent advances show that interesting structures can be obtained \textit{directly} from the multilayer networks~\cite{gomez2013diffusion,chen2018suppressing,vskrlj2019cbssd}.

The key concept in multilayer networks are the layers themselves. Since the structure of such networks is driven by the layers and their aspect~\cite{kivela2014multilayer}, understanding how the layers organise can reveal properties unique to a given multilayer network model~\cite{renoust2015detangler, skrlj2019patterns}. Particularly, the intertwining of edges, or \textit{layer entanglement}~\cite{renoust2014entanglement, renoust2013measuring}, shows how layers overlap to form coherent structures and substructures.

Although recent works have focused on multilayer network analysis and description~\cite{wang2018social,omodei2015characterizing}, not many have focused on a large scale analysis grouping multilayer networks of different nature -- and produced in different disciplines, while comparing them to synthetic models. One comparative study of flow analysis~\cite{de2015identifying} has particularly influenced this paper where emerging structures are described, albeit not comparing them to synthetic models.

In their seminal work, McPherson \textit{et al.}~\cite{mcpherson2001birds} discuss how ties emerge in social systems. They investigate how people similarity, \textit{i.e.} homophily, is a strong driver to the formation of ties, with the addition to make them more durable in a dynamic system. They investigate social ties in a multilayer manner, and argue for further research: \textit{``in the impact of multiplex ties on the patterns of homophily; [and] the dynamic of network change over time [...]''}.
Our original work~\cite{skrlj2019patterns} -- that we extend in this paper -- particularly resonates with the first point of McPherson \textit{et al.}, in that we displayed a link between homophily~\cite{mcpherson2001birds,borgatti2009network} in social networks and high entanglement intensity networks. 

This paper extends~\cite{skrlj2019patterns}, which originally contributed with an open source implementation of entanglement homogeneity and intensity for multiplex networks, while evaluating them over 30 real world networks. We proposed also a synthetic multiplex network generator. A generation of over 10k synthetic networks, and their comparison with the real world networks, displayed common patterns of entanglement homogeneity and intensity that could be specific to the families of applications that \textit{generated} the networks. In this extended work, we contribute with:
\begin{itemize}
\item the theoretical extension of the entanglement computation to a fully multiplex model that takes into account coupling edges;
\item the extension of our synthetic generator accordingly;
\item the computations on a wider range of real and synthetic networks (1{,}329{,}696 synthetic networks were considered);
\item \REVISE{the study of entanglement in large, temporal multiplex networks;}
\item \REVISE{an open-source implementation of all conducted experiments.}
\end{itemize}

\section{Coupled multilayer and multiplex networks}

A multilayer network can be defined as a sequence $M = \{G_l\}_{l \in L} = \{(V_{l},E_{l})\}_{l \in L}$ where \REVISE{$E_{l} \subseteq V_l \times V_l$} is a set of edges in one network $l \in L$ of the sequence~\cite{kivela2014multilayer}. 
Multilayer networks are commonly understood as layers comprised of interactions, where each layer corresponds to a specific aspect of the system. \REVISE{Coupling accounts for transitions between layers. Kivel\"{a} et al.~\cite{kivela2014multilayer} consider a multiplex network as a \textit{``diagonally coupled multilayer networks in which each layer shares at least one node with some other layer in the network''}. They consider also \textit{node-aligned multiplex networks}, which do not specifically address coupling of nodes, but assume that nodes are shared (and coupled) across all layers. In our context, we refer to \textit{coupled multilayer networks} when we specifically consider networks with coupling between nodes across layers, and simply to \textit{multiplex networks} when considering node-aligned multiplex networks. The difference between these two types of multiplex networks is only whether we consider or not the coupling between layers. In multiplex networks, nodes represent \textit{the same} entity across all layers.}

We represent a multiplex network as a structure $M'=(V_M, E_M)$, where $V_M$ is the set of nodes and $E_M$ the set of all edges (in \textit{all} layers). \REVISE{$\mathbb{V}$ denotes the super set of all nodes, and $\mathbb{E} = \mathbb{V} \times \mathbb{V}$ the super set of all edges, regardless of the layers.} 
There may exist coupling edges connecting nodes through layers, forming \textit{transition coupling}. This may concern, for example, coupled multilayer networks which are modelling transportation systems~\cite{cozzo2015structure}.
In that case, we can differentiate the \textit{elementary} layers (holding \textit{inner-layer} edges) from the \textit{transition} coupling (holding \textit{coupling} edges). Each \REVISE{transition coupling} $t = (l, l')$ between layer $l$ and $l'$ \REVISE{can be modelled similarly to a} layer, with a set of nodes and edges. If $S \subset L$ represents the subset of all \textit{elementary} layers, and $T \subset L$ the subset of all \textit{transition} coupling, we may define our coupled multilayer network $M$ as the union. It combines a multilayer network with elementary layers only, and another multilayer network with \REVISE{transition coupling} only $M = \{G_l\}_{l \in L} = M_S \cup M_T = \{G_s\}_{s \in S} \cup \{G_t\}_{t \in T}$. 
The coupling can heavily influence the structural behaviour of multilayer networks~\cite{cozzo2016characterization}. It can also influence the resilience of the network against failures~\cite{de2014navigability} and naturally the diffusion phenomena~\cite{tejedor2018diffusion} too. 

Among other examples of coupled multilayer networks, a biological system can be studied at the protein, RNA, or gene level~\cite{valdeolivas2018random}. Similarly, social networks can be studied by taking into account a person's presence on multiple platforms~\cite{mittal2019analysis}.
For computational purposes, such networks are commonly represented in the form of supra-adjacency matrices, where \textit{block-diagonal structures} connect the same node across individual layers emerges~\cite{cozzo2015structure}. Algorithms can operate on such matrices directly, and thus exploit additional information representing multiple aspects. 

Algorithms for analysis of multilayer networks can also operate on sparse adjacency data structure of the multilayer network directly. Yet, they need to take into account that a given node is present in multiple layers. Such representation is suitable for this work, as we are focused primarily on how edges co-occur across \textit{layers}. Hence, this work focuses primarily on the relations \textit{between the layers} of a given multilayer network.
We next discuss the two measures we consider throughout this work.

\section{Entanglement in multiplex networks}
\label{sec:entanglement}

We briefly recall the entanglement measures definitions from previous work~\cite{renoust2014entanglement}.

\begin{figure}[ht!]
\centering
\captionsetup{width=.90\linewidth}
\includegraphics[width=.95\linewidth]{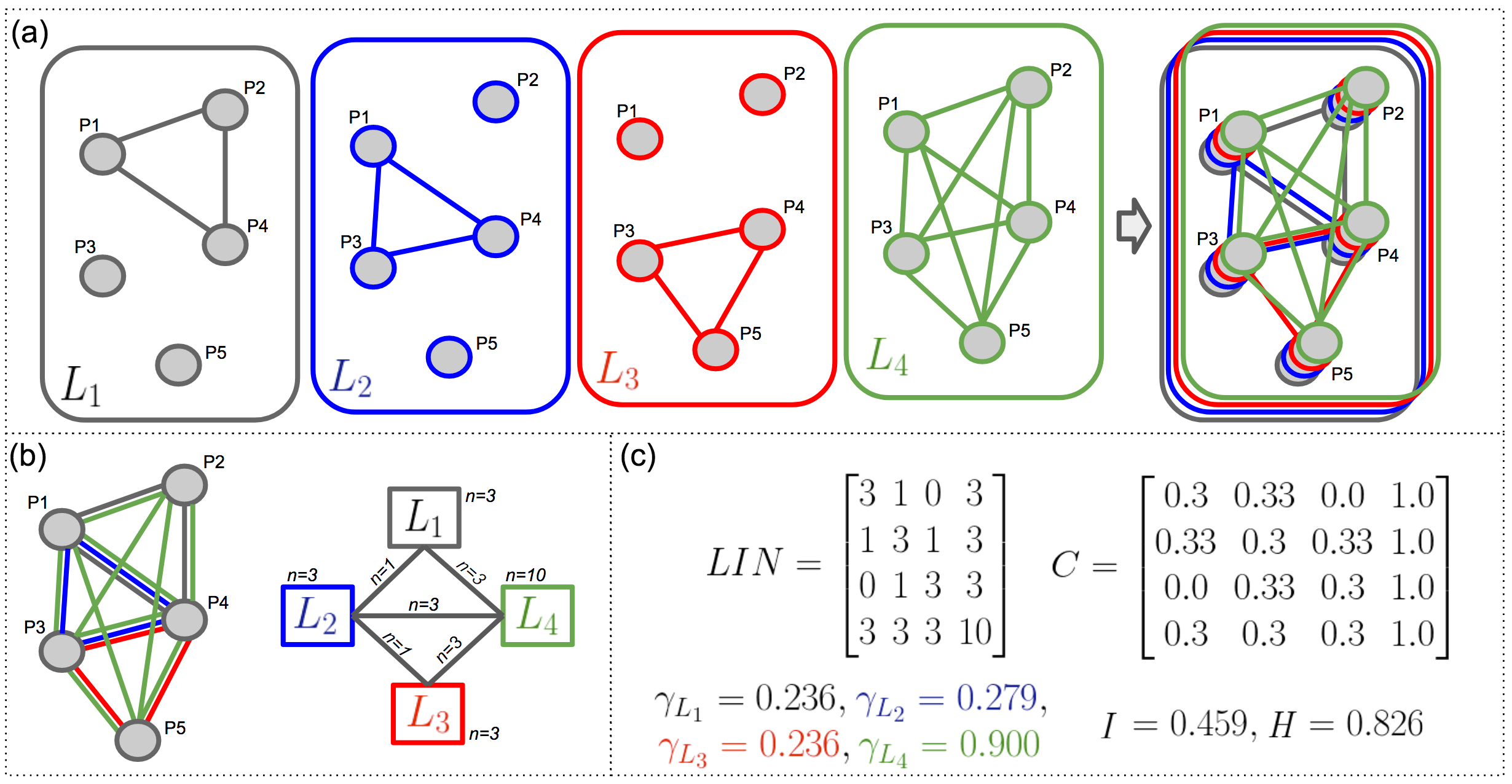}
\caption{A toy example of layer entanglement computation: a) separated layers considered in a multiplex network; b) constructing the layer interaction network from the example; c) measuring entanglement from the example.}
\label{fig:toyexample}
\end{figure}

\subsection{Layer interaction network}
\label{subsec:LIN}

Recall our multiplex network $M=(V_M, E_M)=\{G_l \}_{l \in L}$. As mentioned earlier, such a network really distinguishes itself from classical graphs through the use of different layers to connect nodes. These layers may have different patterns and may overlap together. There may even exist latent dependencies among these layers.
To investigate this matter, each layer could be abstracted to one single node and form a new graph, the \textit{Layer Interaction Network} (hereafter $LIN$)~\cite{renoust2014entanglement}. Visualizing the \REVISE{$LIN$} is a key component for multiplex network visualization such as in Detangler~\cite{renoust2015detangler}.
In the \REVISE{$LIN$}, $LIN=(L,F)$, each node $u_{l}, u_{l'}, u_{l''} \ldots $ corresponds to a layer $l, l', l'', \ldots \in L$ of the multiplex network $M$, and each edge $f \in F$ captures when two layers overlap through edges. 
More formally, there exists an edge $f=(u_l, u_{l'})$ whenever there exists at least two nodes $v, v' \in V_M$ with the condition that there exists at least one edge connecting these two nodes on each layer $e_M = (v,v') \in l$ and $e'_M = (v,v') \in l'$. 
The \REVISE{$LIN$} can be interpreted as an edge-layer co-occurrence graph, and the weight of an edge $f = (u_l, u_{l'})$, denoted as $n_{l, l'}$ equals the number of times layers $l$ and $l'$ co-occur. 
By extension, $n_{l, l}$ is the number of edges on layer $l$. This process is illustrated in Figure~\ref{fig:toyexample}b.

\subsection{Layer entanglement}
\label{subsec:entanglement}

The analysis of layer entanglement is inspired by the analysis of \textit{relation content} in social networks~\cite{burt1985relation}. The idea is to study the redundancy between relation content, each forming in our formalism a different layer. The layer entanglement measures the ``influence'' of a layer in its neighbourhood.

This measure is recursively defined: the entanglement $\gamma_l$ of a layer $l$ is defined upon the entanglement of the layers it is entangled with. Similarly to the eigen centrality~\cite{Wasserman1994}, this translates into the recursive equation:
\REVISE{$$\gamma_l.\lambda=\sum_{l'\in T}{\frac{n_{l,l'}}{n_{l,l}}\gamma_{l'}}.$$} The entanglement of a layer $\gamma_l$ can be retrieved from a vector $\vec{\gamma}$ which corresponds to the right eigenvector (associated to the maximum eigenvalue $\lambda$) of the layer overlap frequency matrix with corresponding overlap, defined as:
$$
\REVISE{
C = (c_{l,l'}), \quad \textrm{where} \quad c_{l,l'} = \frac{n_{l,l'}}{n_{l,l}} \quad \textrm{and} \quad c_{l,l} = \frac{n_{l,l}}{|\mathbb{E}|}}
$$
this metric was initially introduced in~\cite{burt1985relation}, then later constructed using the weights in the $LIN$~\cite{renoust2014entanglement} (see Figures~\ref{fig:toyexample} and~\ref{fig:maxHI}).

\subsection{Entanglement intensity and homogeneity}
\label{subsec:inthom}

\begin{figure}[ht!]
\centering
\captionsetup{width=.90\linewidth}
\includegraphics[width=.95\linewidth]{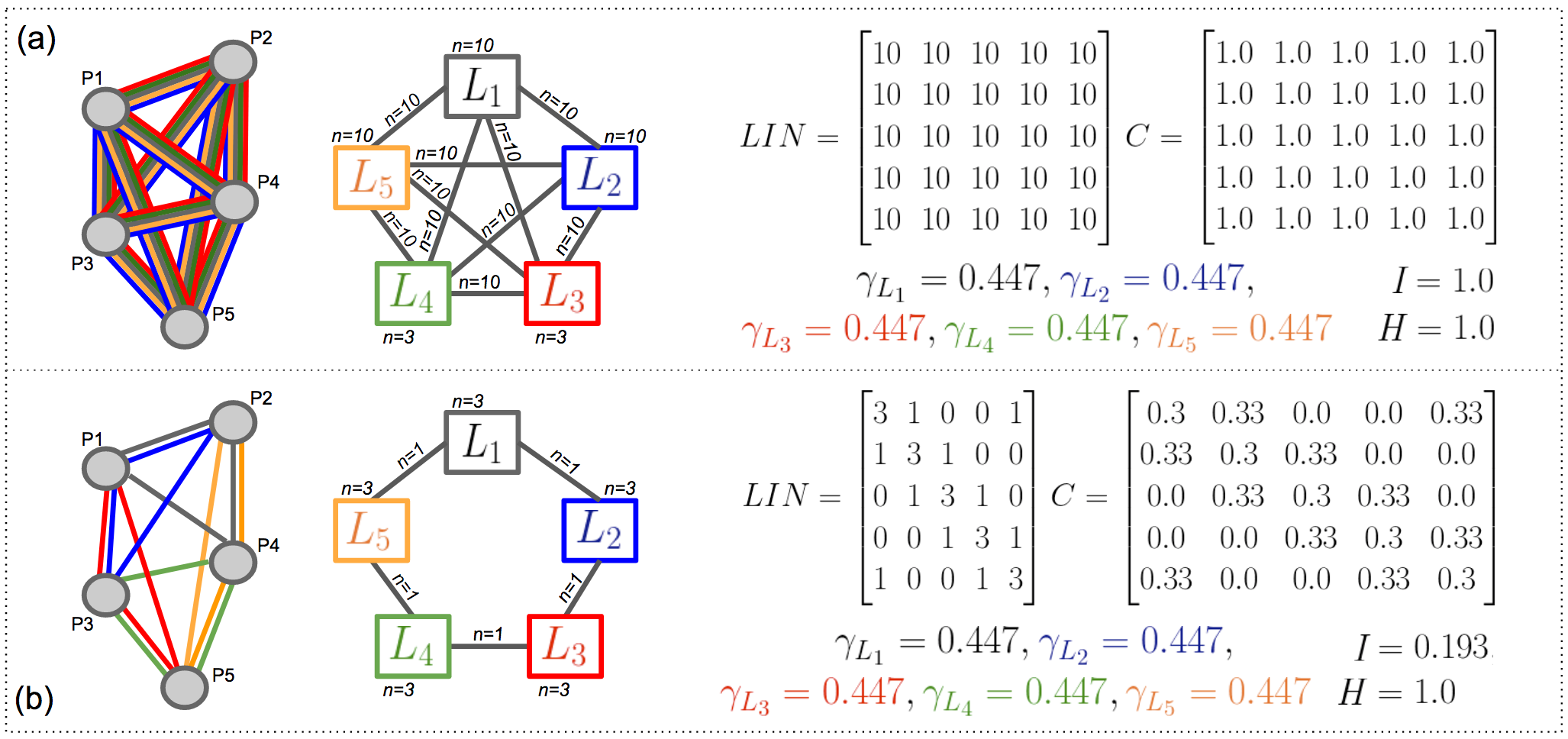}
\caption{Two very different cases of maximum homogeneity $\mathit{H}=1$, the multiplex network and the $LIN$ are shown, with matrices and entanglement measures. a) all layers are saturating all edges, so we have maximum intensity $\mathit{I}=1$; b) layers are well balanced, but we may have a lot more interactions possible.}
\label{fig:maxHI}
\end{figure}

The layer entanglement $\gamma_l$ measures the share of layer $l$ overlapping with other layers. The more a group of layers interacts together, the more the nodes they connect will be cohesive in view of \textit{these} layers, hence the more $\gamma_l\; \forall l \in L $ values will be similar (their share of entanglement will be similar). This is captured by the \textit{entanglement homogeneity}~\cite{renoust2014entanglement} which is then defined as the following cosine similarity:
\REVISE{$$\mathit{H}=\frac{<\vec{\mathbf{e}_L},\vec{\gamma}>}{\lVert\vec{\mathbf{e}_L} \rVert \lVert\vec{\gamma} \rVert} \in [0,1].$$ With $\vec{\mathbf{e}_L}=[1,1, \ldots, 1]_L$ the vector of size $L$ all filled with 1's.} Optimal homogenity is not necessarily reached only when all nodes are connected through all layers, but also when all nodes are connected in a very balanced manner between all layers (see Figure~\ref{fig:maxHI}). Homogeneity thus permits various \textit{symmetries} in a given \REVISE{$LIN$}.

When a maximum overlap is reached through all layers in the network, the frequencies in the matrix $C$ (of size $|L| \times |L|)$ are saturated with $C_{i,j}=1$. This gives us a theoretical limit to measure the amount of layer overlap through the \textit{entanglement intensity}~\cite{renoust2014entanglement}, defined as: $$\mathit{I}=\lambda/|L|.$$ In practice, both entanglement intensity and homogeneity have been used to measure the coherence of clusters of documents~\cite{renoust2013measuring}.

\subsection{\REVISE{Transition coupling} entanglement}
\label{sec:interlayerdefinition}
We have defined the layer entanglement which measures overlap between layers of a multiplex network, but many multiplex networks include another critical parameter which is coupling edges~\cite{battiston2014structural}.
The coupling often measures the transition of nodes \textit{between} layers, hence the transition of nodes are captured by edges connecting nodes \textit{across} layers.

\begin{figure}[ht!]
\centering
\captionsetup{width=.90\linewidth}
\includegraphics[width=0.95\linewidth]{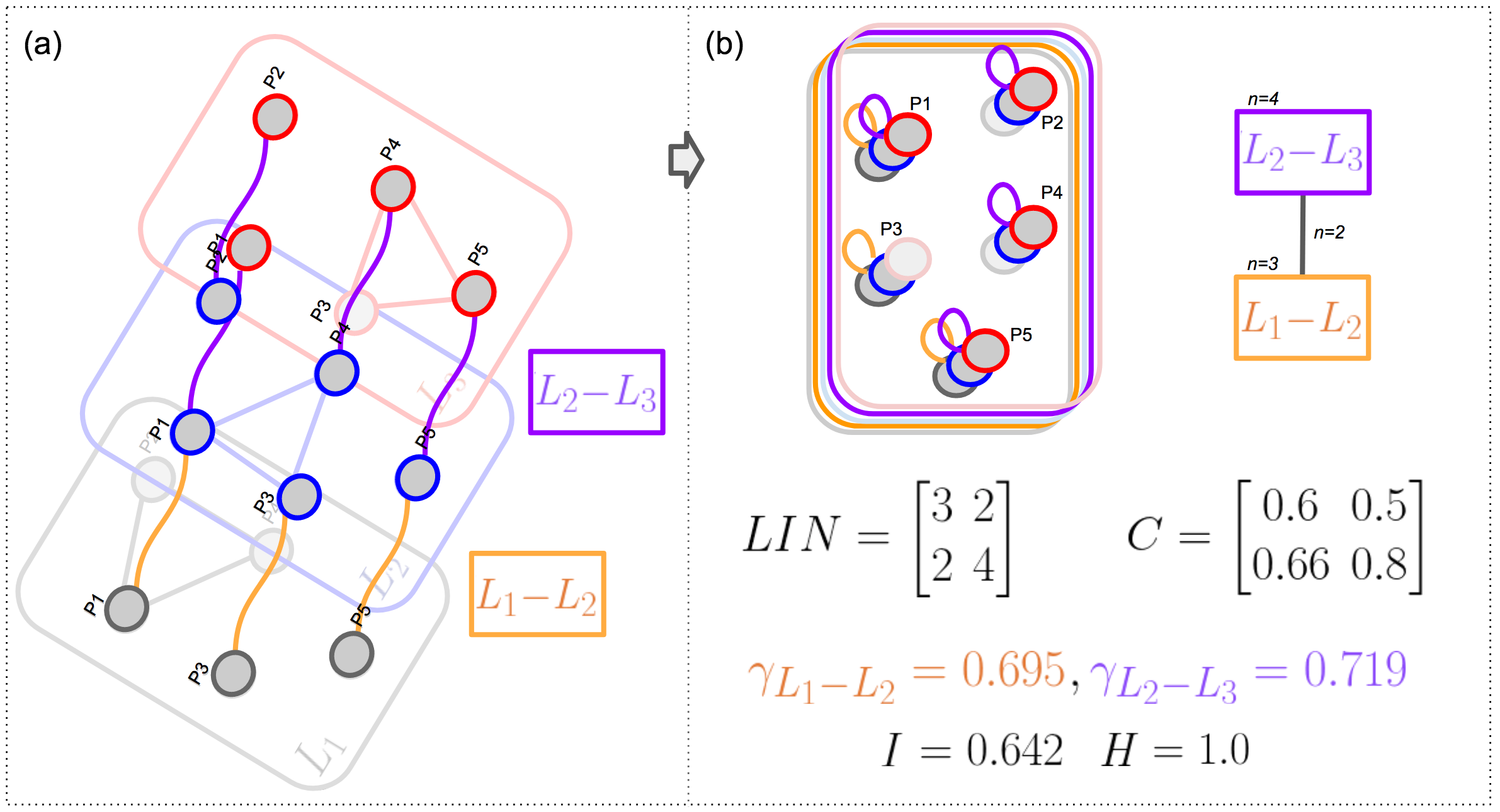}
\caption{Computing entanglement on the \REVISE{transition coupling} edges. (a) Coupling edges are illustrated in orange ($L_1-L_2$ edges) and in purple ($L_2-L_3$ edges). (b) Computing the corresponding $LIN$ and entanglement measures. Coupling edges of a same node resemble loops except they are defined across two layers. We may notice that: the \REVISE{transition coupling} $L_2-L_3$ shows a slightly higher index since there are more transitions for this coupling; the homogeneity $H$ is (almost) maximal since both layers are (almost) equally intertwined (only 2 layers, actual $H \approx 0.99986$).} 
\label{fig:transition_layers}
\end{figure}

\begin{figure}[ht!]
\centering
\captionsetup{width=.90\linewidth}
\includegraphics[width=0.95\linewidth]{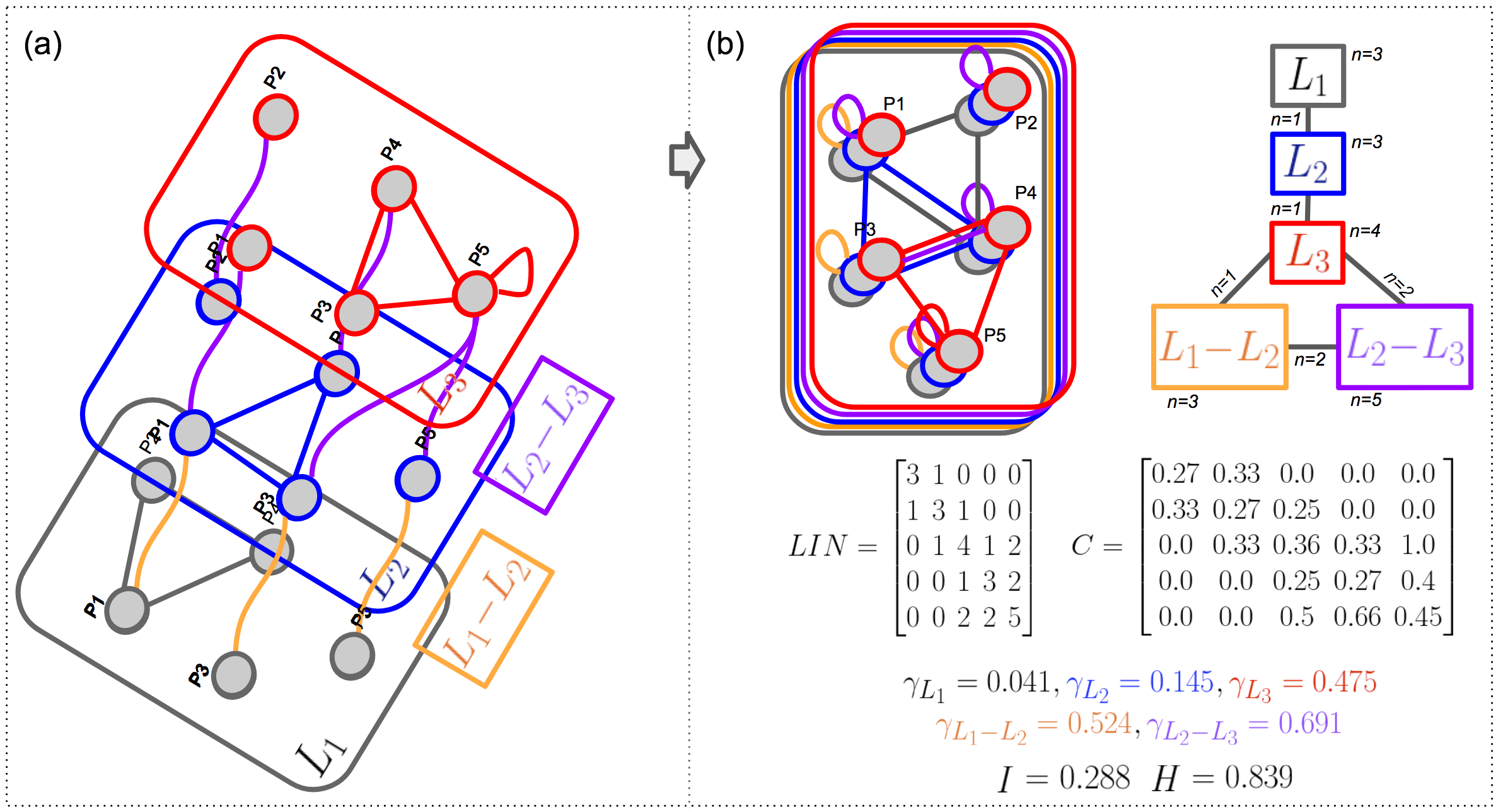}
\caption{Computing entanglement on both inner-layer and coupling edges. (a) Note that in contrast to the example in Figure~\ref{fig:transition_layers}, we have added a loop to node $p_5$ in layer $L_3$ (in red) and a coupling edge connecting nodes $p_3$ of layer $L_2$ to $p_5$ in $L_3$. (b) Computing the corresponding $LIN$ and entanglement measures. We can notice that the most intertwined \REVISE{transition coupling} displays the highest entanglement index. Because there is limited overlap between elementary layer edges and \REVISE{transition coupling} edges, entanglement intensity $I$ is rather low.} 
\label{fig:all_layers}
\end{figure}

Recall our multiplex graph $M = (V_M, E_M)$. Suppose $S$ is the set of \textit{elementary} layers, we can then have transitions between any pair of elementary layers $l \in S$ and $l' \in S$. 
Let $u_l = (u,l), u \in V_M, l \in S$, the connection of a node $u$ within a layer $l$. 
A \REVISE{transition coupling} edge $e$ can be defined as follows: $e=(u_l,v_{l'}) \in E_M$ such that $e$ connects nodes $\{u, v\} \subseteq V_M$ across layers $l \neq l', \{l, l'\} \subseteq S$. 
Coupling edges often connect a same node across two layers and may be used to model a physical transition, such as a change from subway to train in a station of a transportation network. 
As a consequence, a pair of layers $(l, l')=t$ forms a \REVISE{transition coupling} $t \in T$ when there exists at least one such edge $e=(u_l,v_{l'}) \in E_M$. 
Note that taken together, these elementary and \REVISE{transition coupling} subsets form the set of all layers $S \cup T = L$, and that the size of $T$ is bounded by the size of $S$ such that $|T| \leq \frac{1}{2}|S|(|S|-1)$.

Now, given this definition, nothing limits the computation of entanglement (introduced in previous Sections~\ref{subsec:LIN} to~\ref{subsec:inthom}) only to the elementary layers part of $M_S$, as illustrated in Figure~\ref{fig:transition_layers}.
\REVISE{Entanglement can also be used to characterise the coupling between these \textit{elementary} layers if applied only to the edges of the \textit{transition} coupling $M_T$.}

\REVISE{The nature of coupling often captures a very distinct characteristic of the network in comparison to its elementary layers. A \REVISE{transition coupling} edge mostly connects the same node across layers, while elementary layers do not always display loops. These cases may happen on rare occasions, one example being an underground path connecting subway stations being modelled as a \REVISE{transition coupling}, but the literature is very poor of such examples. It is however technically possible to consider both elementary layers and \REVISE{transition coupling} in one multiplex network $M$ to compute entanglement (as shown in Figure~\ref{fig:all_layers}), but we keep this discussion for the Appendix. In practice, the intensity and homogeneity greatly differ between them, and often results in clearly separated components of the $LIN$.}

\section{A coupled multilayer network generator}
\label{sec:multiplex-generator}

In this section, we describe an algorithm which generates synthetic coupled multilayer networks, \textit{i.e.} multilayer networks which share some nodes across some layers, but do not guarantee that all nodes are being shared between all layers. \REVISE{These kinds of networks make the link between general multilayer networks and node-aligned multiplex networks (for which the assumption is that all nodes are shared through all layers~\cite{kivela2014multilayer}).}

The algorithm is based on the following observations. Let $M = (V_M, E_M)$ represent a coupled multilayer network with layer set $L$.
Each node is associated to a random number of layers $\{l_1, l_2, \ldots, l_i\} \subseteq L$.
Now for each layer $l_i \in L$ there is a set of nodes $V_{l_i} \subseteq V_M$ which forms a potential set of edges of size $|E_{l_i}| = \frac{1}{2}|V_{l_i}|(|V_{l_i}|-1)$. We introduce $o$, a parameter determining the probability of a node occurring at a given layer. We then introduce the probability $p$ of an edge to be created between any pair of nodes belonging to a layer so we may avoid cliques to form on each layer. 
We referred in our previous work to the edge dropout \cite{skrlj2019patterns}, which is $d = 1-p$ as the share of links we drop from the clique model.
Intuitively, the more similar a given random multiplex is to a clique over each layer, the higher its elementary layer intensity should be.  
\REVISE{Hence, high intensity implies larger probability that two given nodes will have an edge between them on more than one layer.}
The generator also accounts for coupling by adding \REVISE{transition coupling} edges. These coupling edges are connecting nodes across two layers. We introduce $q$, the probability for a same node to be connected across two layers. The higher $q$, the more nodes will be connected through layers. Note that in our initial work \cite{skrlj2019patterns}, neither $o$ nor $q$ were considered ($o$ was in fact picked uniformly).

\begin{algorithm}[t]
\Parameter{Number of nodes $n$, number of layers $m$, inner-layer edge probability $p$, coupling edge probability $q$,
}
\KwResult{A coupled multilayer network $M$}
$M$ $\leftarrow$ emptyMultilayerObject\;
\For{node in $[1 \dots n]$}{
    layerNodes $\leftarrow$ assignNodeToLayers(node, $o$, $m$)\\\Comment*[r]{Nodes are assigned to layers among $m$ with probability $o$.}
    
    update($M$, layerNodes)\Comment*[r]{Update global network.}
}
\For{layer $l_i$ with corresponding node set $V_{l_i}$}{
    nodeClique $\leftarrow$ generator of node pairs from $V_{l_i}$\Comment*[r]{With or without possible loops.}
    innerLayerEdges $\leftarrow$ sampleWithProbability(nodeClique, $p$)\Comment*[r]{Sample via $p$.}
    update($M$,innerLayerEdges)\Comment*[r]{Update global network.}
}
\For{layers $l_i, l_j$ with shared node set $V_{l_i, l_j}$}{
    sameNodeTransitionCouplingEdges $\leftarrow$ sampleWithProbability($V_{l_i, l_j}$, $q$)\Comment*[r]{Sample via $q$.}
    update($M$,sameNodeTransitionCouplingEdges)\Comment*[r]{Update global network.}
}
\Return{$M$}\;
\caption{A coupled multilayer network generator.}
 \label{algo:rep}
\end{algorithm}

The purpose of this generator is to offer a simple \textit{testbed} for further exploration, as well as additional evidence of the relation between homogeneity and intensity on many random, synthetic networks. The Algorithm~\ref{algo:rep} represents the proposed procedure.

The generator first randomly assigns the same node index to the many layers (lines 2-5). Once assigned, the layers are processed by applying sampling on $|V_{l_i}| \choose 2$ possible edges in layer $l_i$. 
Note that in line 7, this whole clique is virtually generated. 
The global multiplex is updated during this process (lines 6-10). 
These steps are then repeated for \REVISE{each \REVISE{transition coupling}} \textit{i.e.} pairs of elementary layers (lines 11-14).
\REVISE{The implementation thus uses a generator, for which \textit{lazy evaluation} avoids potential combinatorial explosion when considering a large number of nodes and low edge probability.}

\subsection{Some theoretical properties of the generator}
\REVISE{In this section we show two properties of the proposed generator. We denote $n = |V_M|$ the parameter setting the number of nodes of the network, $m = |L|$ the parameter setting the number of edge layers in the network, and $p$ the inner-layer edge probability. 
Let $\phi \in \mathbb{N}^{+}$ represent the number of possible edges. Then $\phi \leq m \cdot \binom{n}{2}.$
Let $o=1$. Each layer can have at most $n$ nodes. Assuming they form a clique, each layer is thus comprised of $n \choose 2$ edges. As there are $m$ layers, there can be at most $m \cdot \binom{n}{2}$ edges --- a clique of $n$ nodes in each layer (assuming $p = 1$). We refer to this bound as $\phi \leq m \cdot \binom{n}{2}$.}

\REVISE{In the limit, as $p \rightarrow 1$, a full clique needs to be constructed, assuming each node is projected across all layers. The complexity \textit{w.r.t.} the number of layers and edges is: $\mathcal{O}(m \cdot \binom{n}{2}) = \mathcal{O}(|E_M|).$
Note that, even though theoretically, the proposed generator creates a clique and then samples from it, current, lazy implementation only \textit{generates} the edges needed to satisfy a given $p$ percentage. In practice, only when $p \approx 1$, the generator needs larger portions of space (and time). As such, fully connected networks do not represent real systems, we were able to generate a multitude of very diverse networks. This generator-based implementation does not imply that large spatial overheads are not possible: such situations occur when very dense networks are considered.}

\REVISE{
We next discuss the impacts of $q$ parameter.
The number of coupling edges has a worst case complexity of $\mathcal{O}(\binom{m}{2}\cdot n)$ since $q$ directly depends on the number of layers available.
Let $l_a$ and $l_b$ represent a given pair of layers, where each layer consists of all $n$ possible nodes. As each node couples only to itself, there are at most $n$ edges between $l_a$ and $l_b$. As there are $\binom{m}{2}$ possible layer pairs, if nodes are in each pair fully coupled, the network can have at most $\binom{m}{2} \cdot n$ coupling edges.}

However, is that also the case when considering only \REVISE{transition coupling}? Consider the following example of a multiplex network without the coupling edges. No matter what $p$ is employed, if $q \approx 0$, coupling intensity will be low -- very few coupling edges are introduced, the observed \REVISE{$LIN$} will be very sparse. Hence, we posit that the distribution of intensity shall be \textit{constant} with respect to a given $p$. The proof of this claim is by contradiction. We assume that $p$ would indeed influence coupling entanglement intensity. Since transition coupling intensity is defined solely based on the coupling edges, this claim would imply a dependency between $p$ and $q$, which is by the definition (and design) not the case. Even if the nodes are \textit{isolated} in each layer, transition coupling intensity can be high. Note also that the node positioning, governed by $o$, directly impacts both elementary and transition coupling entanglement, since there is higher possibilities for edges to overlap when nodes belong to many layers. These points are illustrated in our empirical evaluation Section~\ref{sec:evaluation} and further in the Appendix materials.

\section{Layer entanglement in temporal multiplex networks}
\label{sec:ent-description-model}

Analysis of temporal multiplex networks has shown promising results in multiple fields of science, such as for example healthcare and transportation \cite{sannino2017visibility}.

 Since patterns of layer interaction networks result in typical entanglement values,
considering temporal entanglement means textitasizing particular topologies of a temporal multiplex network. For example, a high intensity among members in a multiplex social network communicating through different social media corresponds to a synchronization of communications between them. When such a synchronization corresponds to the preparation of a particular event, understanding such synchronization could help forecast the event.
 
In this section, we first discuss how we define temporal multiplex networks and entanglement time series. We limit the following discussion to the consideration of entanglement between elementary layers only, \textit{i.e.} only inner-layer edges.

\subsection{Temporal multiplex networks and entanglement}

\begin{figure}[ht!]
    \centering
\captionsetup{width=.90\linewidth}
    \includegraphics[width = .95\linewidth]{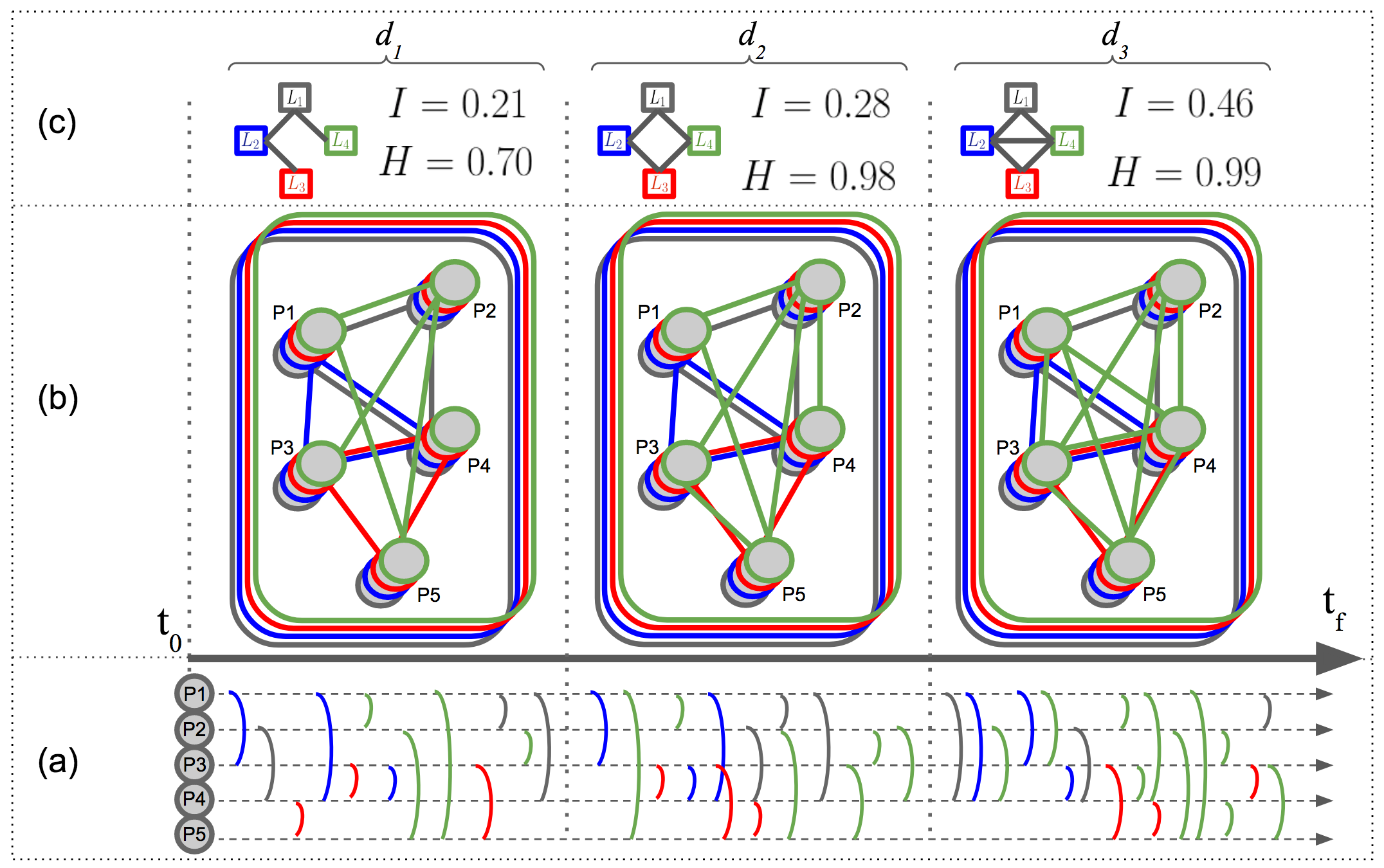}
    \caption{Converting temporal edges of multiple types into temporal entanglement series. a) Edges of different types are defined over time between $t_0$ and $t_f$. b) Time frames $d_1$, $d_2$, and $d_3$ are defined so we may construct the three corresponding multiplex network slices. c) For each slice, we can compute a $LIN$ and the corresponding entanglement intensity $I$ and homogeneity $H$, which compose the series once taken together among all slices.}
    \label{schematic:multiplex}
\end{figure}

Real-life networks often evolve over time, making them behave differently at different points. In our current setting, we define the temporal aspect of our network such as each edge $e_t$ is defined at a specific time point $t$. A multiplex network $M_{d}$ can then be defined for a given time window ${d}$. A time window ${d} = [t_0, t_f]$ covers a time frame (beginning at $t_0$ and ending $t_f$), and the multiplex network $M_d$ is defined such as each edge exists within the time window: 
\begin{equation*}
    M_d=(V_M,\{e_t \in E_M\}_{t\in d}).
\end{equation*}
\noindent The second scenario we considered is that of \textit{moving time windows}. Here, edges from the $f$ \textbf{past} windows are considered when constructing a given network $M$, \textit{i.e.},
\begin{equation*}
    M_f=(V_M,\{e_t \in E_M\}_{t\in \{d-f,\dots,d-1\}}).
\end{equation*}
Our intuition is to compare the shape of a network at different moving time windows. For example, we could compare political social networks under different rulers of a country~\cite{renoust2016face, renoust2016visual}. To do so, we can simply compute entanglement homogeneity and intensity for each time window and compare them. Since our computation only focuses on edge, we consider the network as multiplex, the nodes are shared across all time frames. 

Slicing the time windows is a very different topic and many options are open~\cite{gomez2013diffusion, beck2014state}. For example, it could be achieved manually, with equal time slices, moving window, or with \textit{volume of changes}. In our context, we consider the identification of time window through slices of equal duration in time, but the principle can be extended. We refer to the duration $r$ in time of the slices as \textit{time resolution}.

We may now investigate entanglement homogeneity and intensity properties with respect to time resolution ($r$), and verify if patterns of intensity/homogeneity variation can be predicted. 
Note that one challenge of slice-based modelling of temporal multiplex networks is the problem of selecting the correct resolution $r$, \textit{i.e.} how coarse (or fine)-grained the intervals must be in order to capture desired dynamics. 

In a system covering a global period of $D$, once a slicing resolution is chosen, we can observe values of homogeneity and intensity at the time series level, \textit{i.e.} for each slice $d \in D$, and define the intensity time series ${\mathbb S}_{I} = \{I_{M_d}\}, \forall d \in D$ and the homogeneity time series as ${\mathbb S}_{H}= \{H_{M_d}\}, \forall d \in D$. These intensity and homogeneity time series can now feed further processing.
Note that \REVISE{$\mathbb{S}_{I_f}$ and $\mathbb{S}_{H_f}$} are defined analogously (entanglement for the past $f$ slices, moving in the increments of one slice).
The whole processing from temporal edges to time series is illustrated in Figure~\ref{schematic:multiplex}.

In our following evaluation (Section~\ref{sec:temporal}), we explore $\mathbb{S}_I$ and $\mathbb{S}_H$ when also considering a moving window of previous $f$ time slices. The rationale for considering past $f$ slices up to the considered time point is that such information only includes past data, and could indicate whether entanglement can be also used for \textit{forecasting purposes}. The second option considered, where only the current time slice was plotted, can shed insight on whether online monitoring based on $I$ or $H$ is a sensible option. 

\section{Empirical evaluation}
\label{sec:evaluation}


We now study entanglement intensity and homogeneity across different series of networks.
We first investigate entanglement measures across different parameters of synthetic settings. We follow with investigations on a large panel of real world networks. We finish our study with the study of entanglement in temporal multiplex networks.

\begin{figure}[ht!]
    \centering
    \captionsetup{width=.90\linewidth}
    \begin{tabular}{cc}
\subcaptionbox{Lower elementary $I$}{\includegraphics[width = 0.45\linewidth]{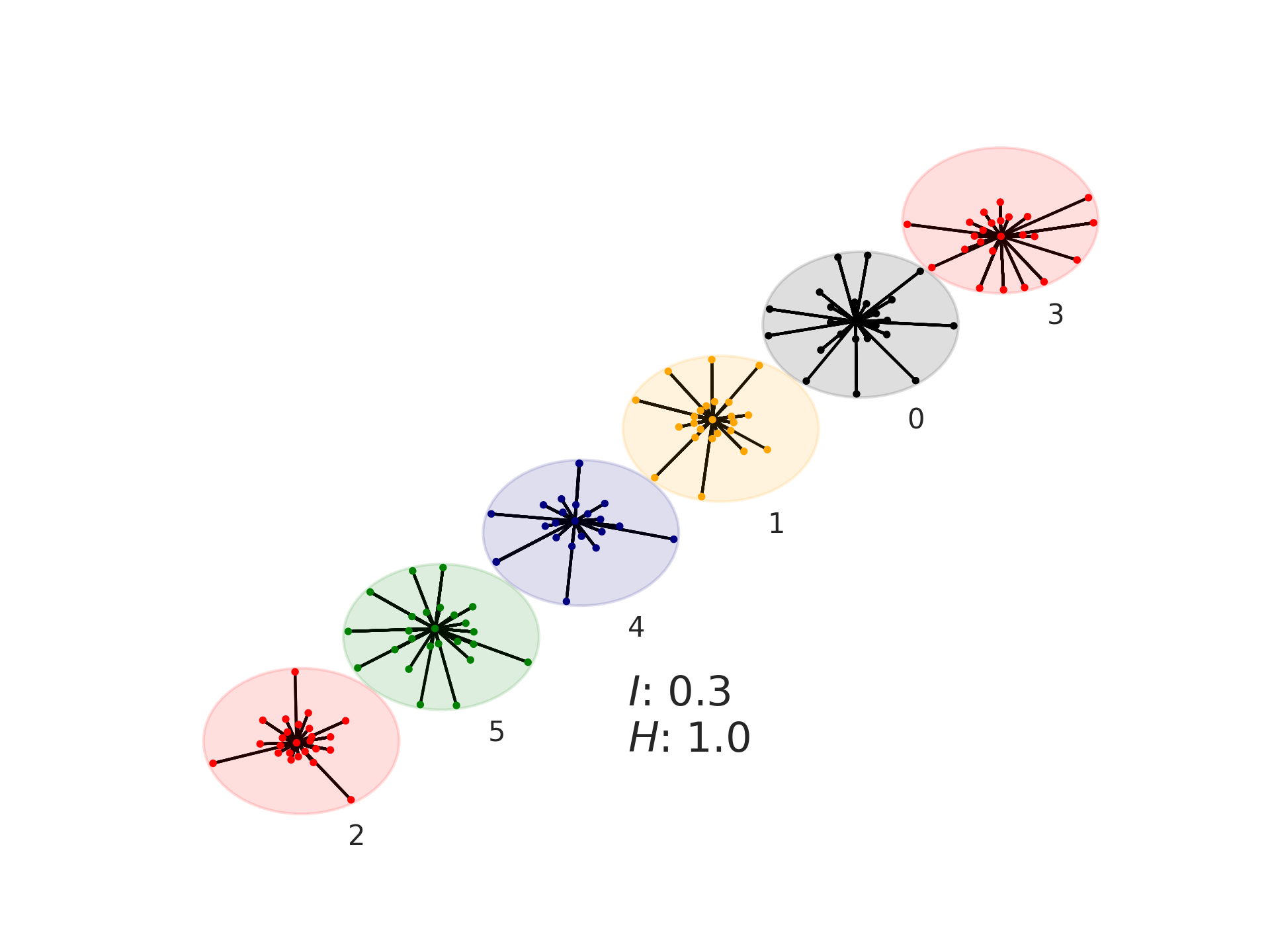}} &
\subcaptionbox{Higher elementary $I$.}{\includegraphics[width = 0.45\linewidth]{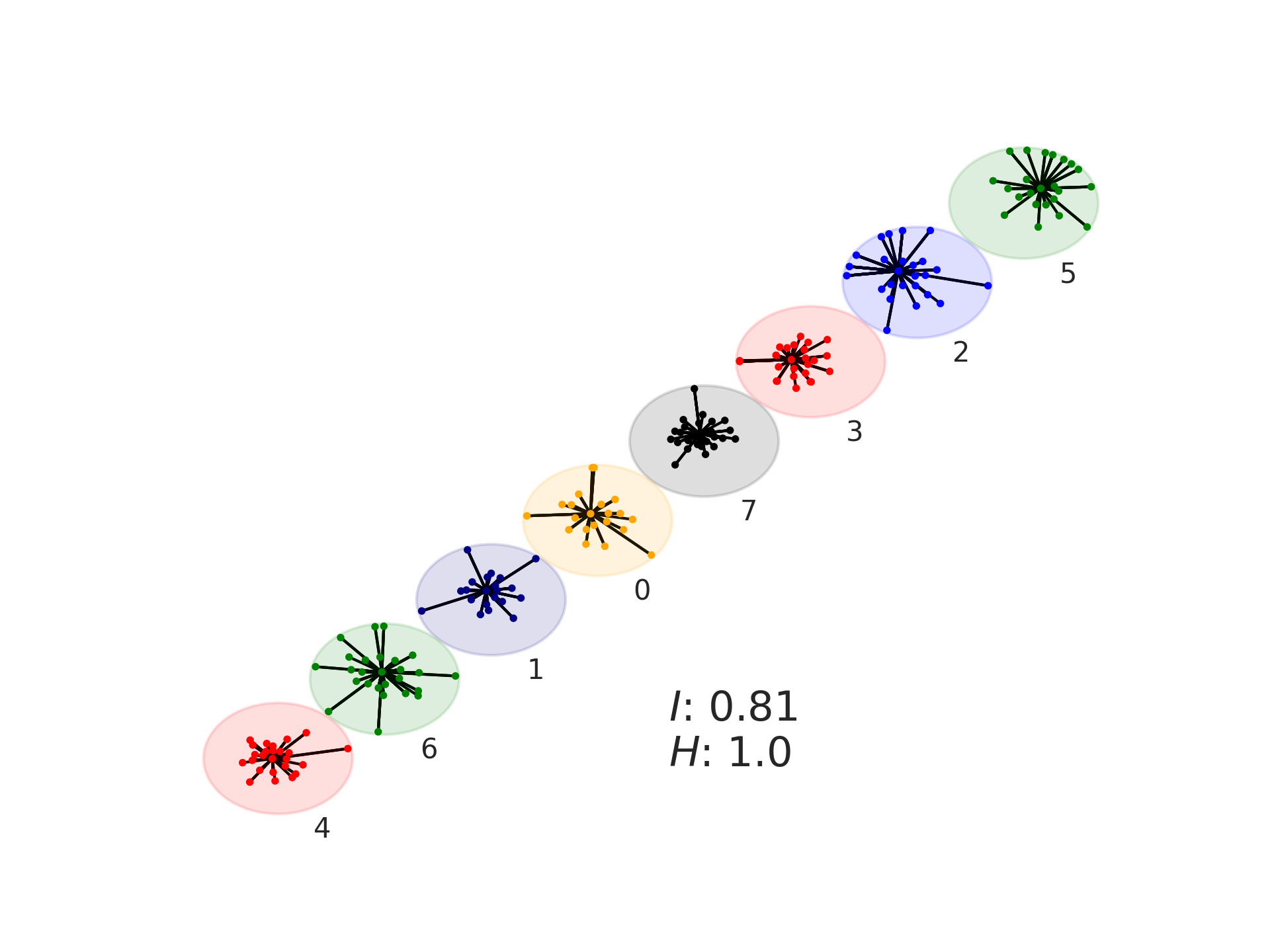}} \\
\end{tabular}
    \caption{\label{fig:viz-inn}Visualization of inner-layer edges in synthetic coupled multilayer networks.}
\end{figure}

\subsection{Entanglement in synthetic networks}
\label{sec:multilayer}

In this first study, we compare entanglement measures over a series of synthetic multiplex networks, using our proposed generator.

\REVISE{
We consider for all our generations, the following key parameters:
\begin{itemize}
    \item Number of nodes ($n$) from 10 to 200 in increments of 10.
    \item Number of layers ($m$) in 1,2,3,4,6,7,9,10.
    \item Layer assignment probability ($o$), from 0 to 1 in increments of 0.05
\item Edge probability ($p$) from 0 to 1 in increments of 0.05.
    \item \REVISE{Transition coupling} edge probability ($q$) from 0 to 1 in increments of 0.05.
\end{itemize}
}

\subsubsection{Multiplex networks without \REVISE{transition coupling}}

 A first generation concerns multiplex networks settings in which \REVISE{transition coupling} is not specified (for example, friendship over different social platforms), \REVISE{so we do not consider parameter $q$ here.}
 
 We have generated in total 1{,}329{,}696 synthetic networks (a couple are illustrated in Figure~\ref{fig:viz-inn}).

\begin{figure}[ht!]
\centering
\captionsetup{width=.90\linewidth}
\includegraphics[width = 0.75\linewidth]{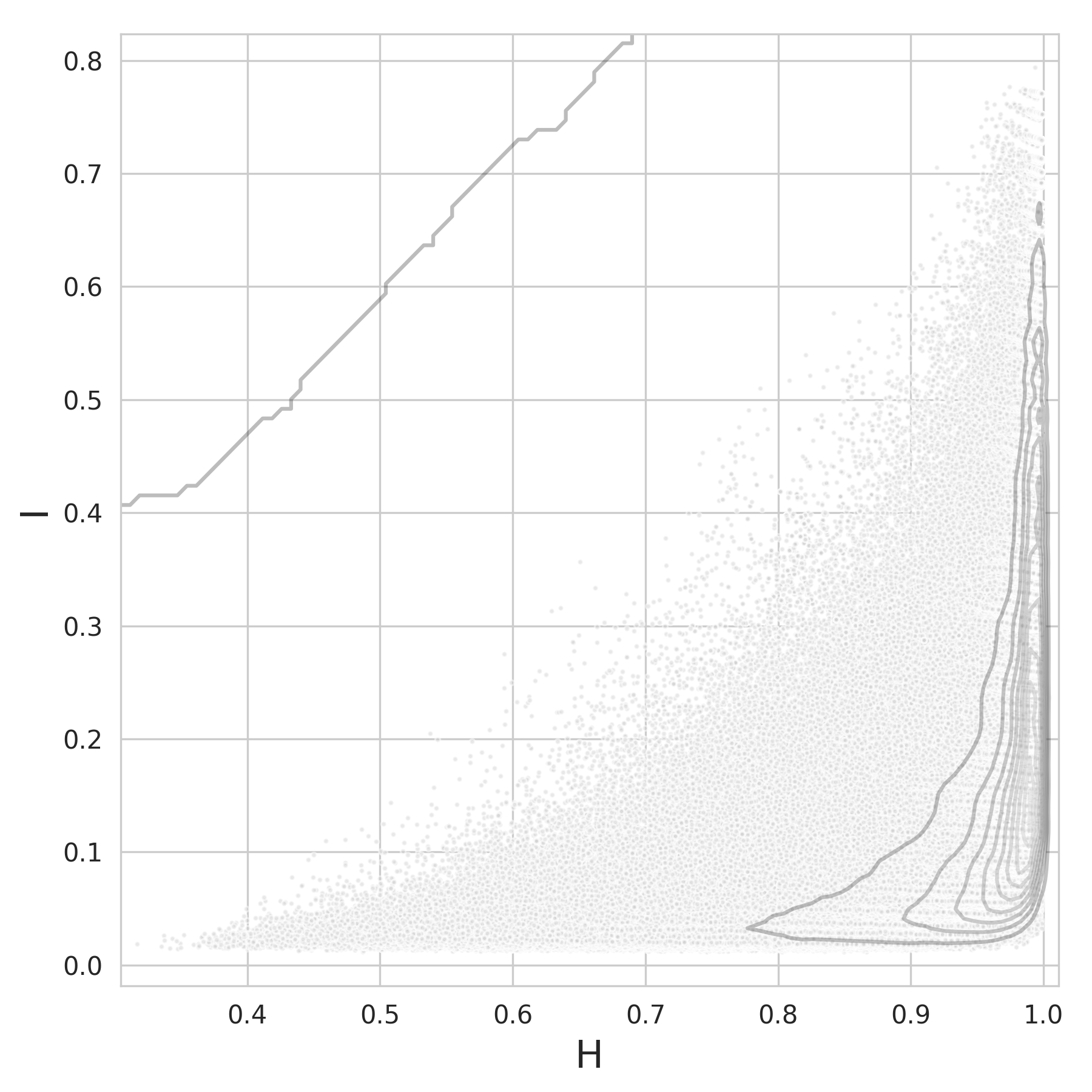}
\caption{Homogeneity and intensity $H \times I$ on 1{,}329{,}696 synthetic multiplex networks without \REVISE{transition coupling} with density lines (Gaussian kernel density estimation).}
\label{fig:syn_parameters}
\end{figure}

\begin{figure}[ht!]
\centering
\captionsetup{width=.90\linewidth}
\begin{tabular}{cc}
\subcaptionbox{Elementary $H \times n$}{\includegraphics[width = 0.45\linewidth]{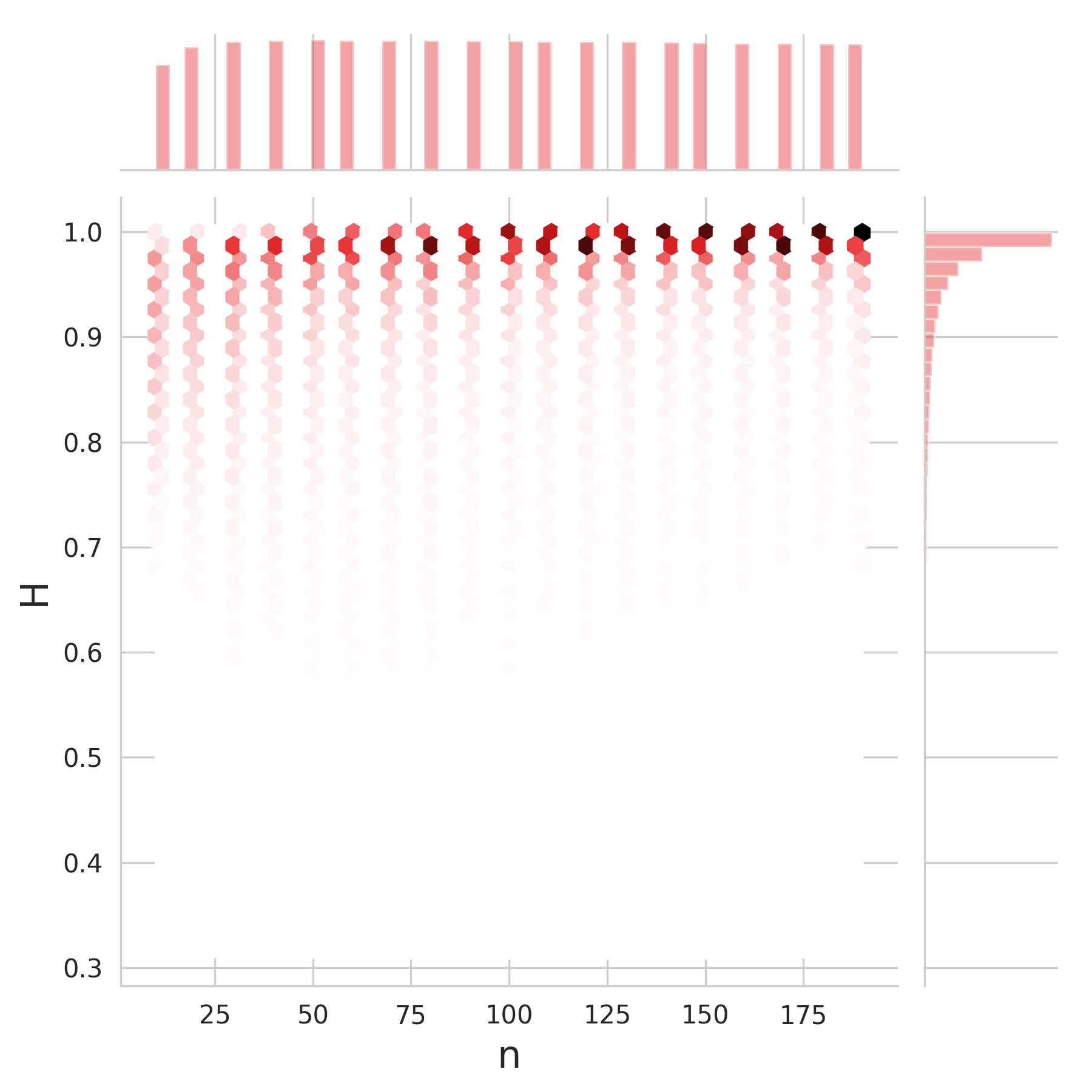}} &
\subcaptionbox{Elementary $I \times n$}{\includegraphics[width = 0.45\linewidth]{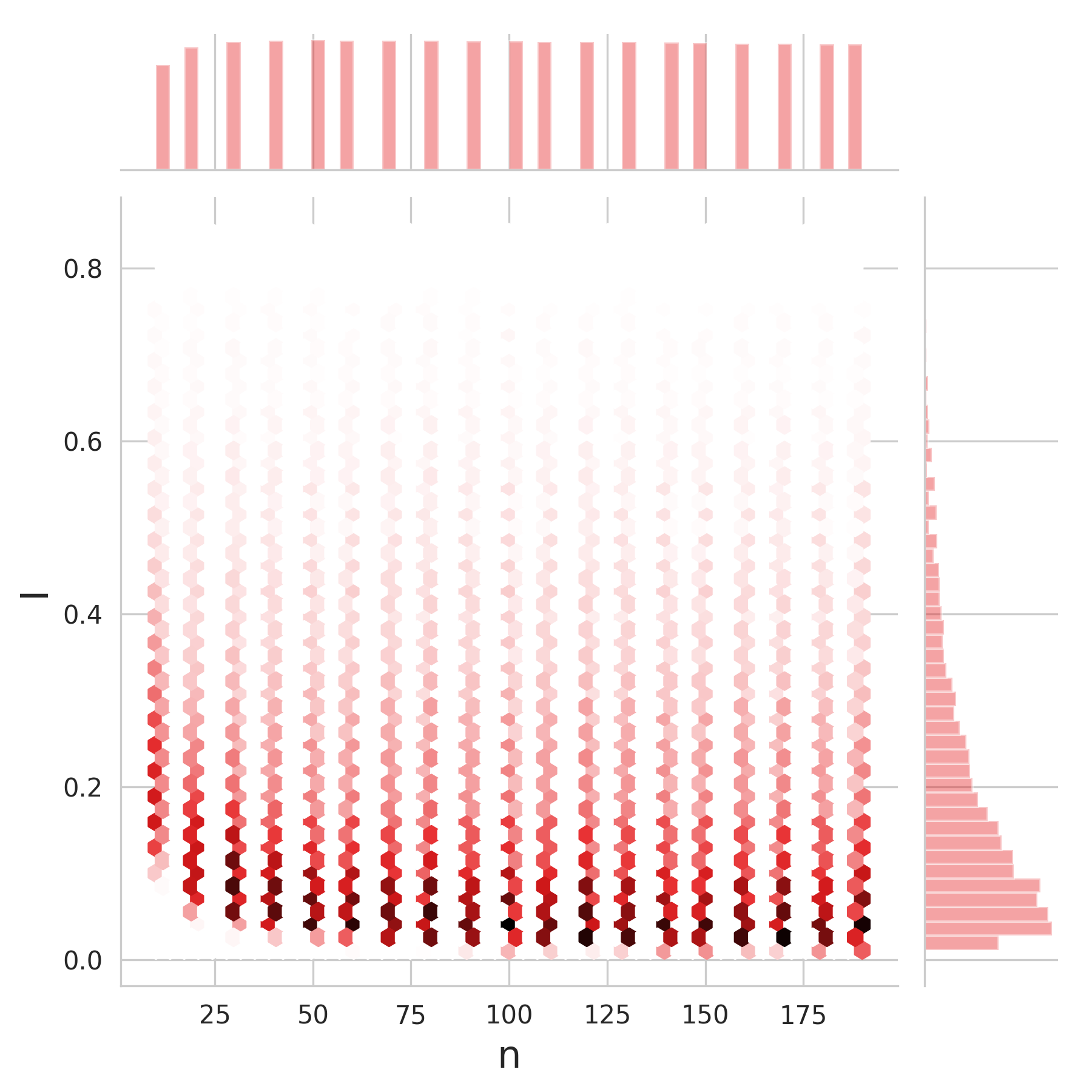}} \\
\subcaptionbox{$H \times m$}{\includegraphics[width = 0.45\linewidth]{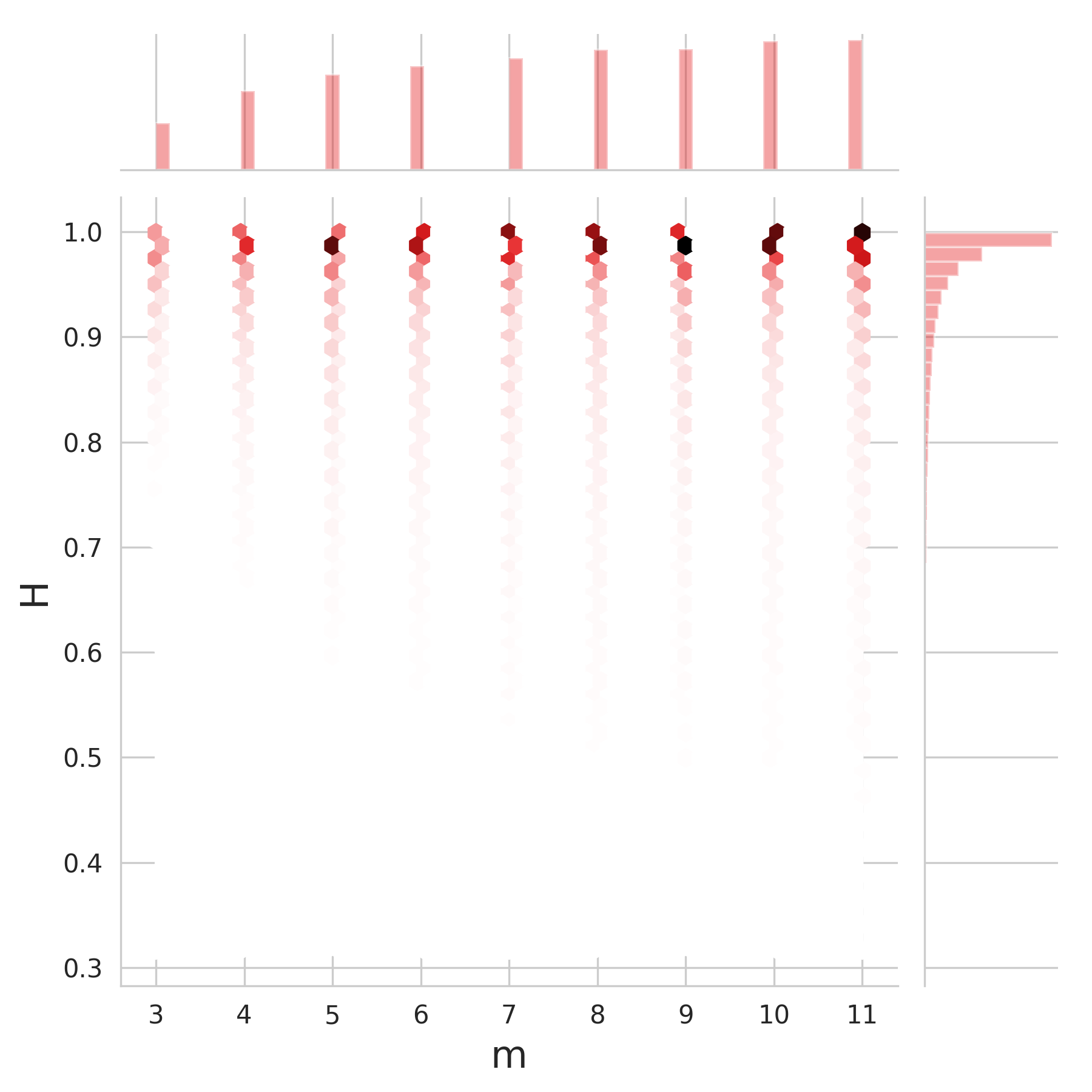}} &
\subcaptionbox{Elementary $I \times m$}{\includegraphics[width = 0.45\linewidth]{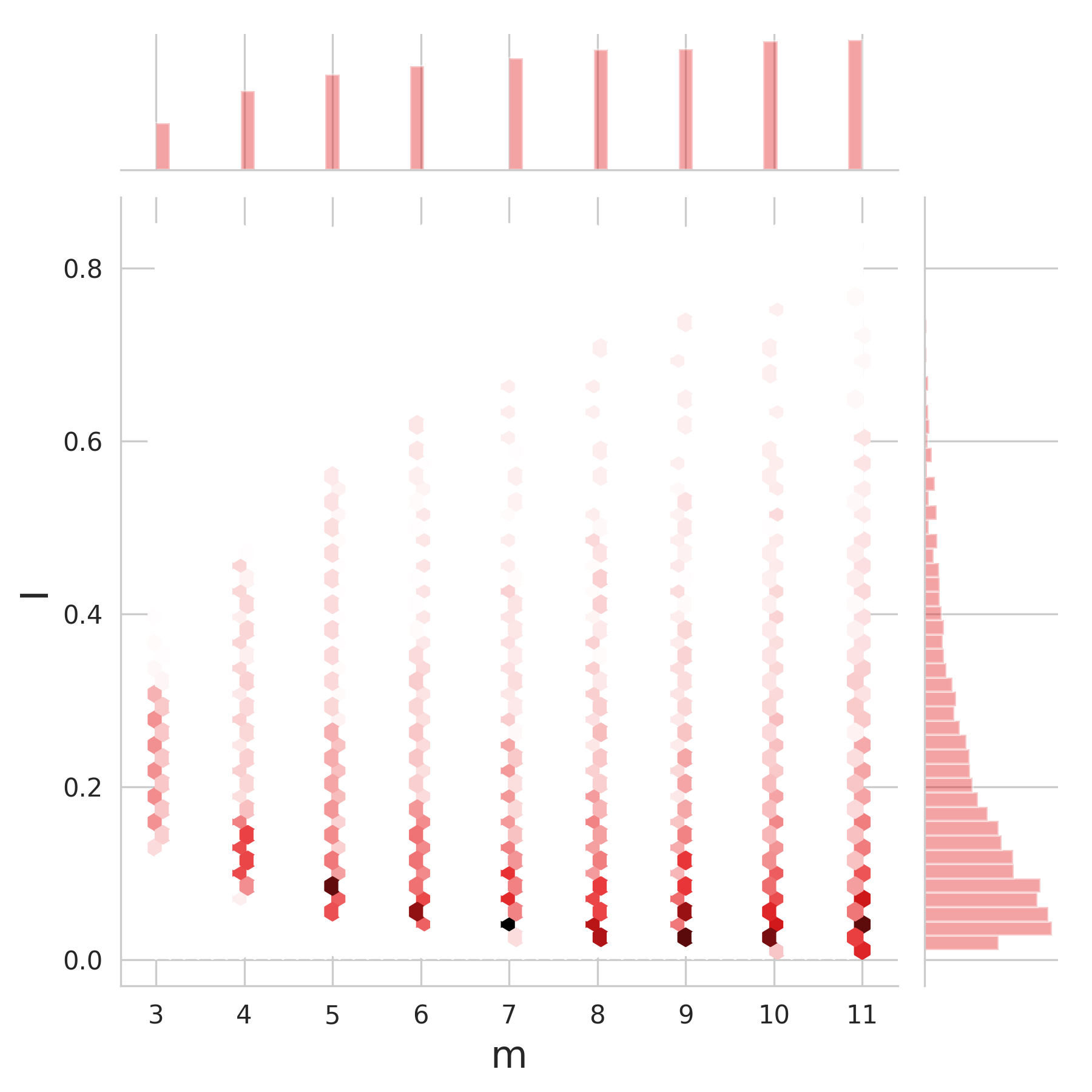}} \\
\end{tabular}
\caption{Results on synthetic multiplex networks without considering \REVISE{transition coupling}. Dependency on the number of nodes $n$ (a, b) and layers $m$ (c, d) on the elementary layer entanglement. The intensity (b, d) shows some influence on each parameter.}
\label{fig:inner_nm}
\end{figure}

\begin{figure}[ht!]
\centering
\captionsetup{width=.90\linewidth}
\begin{tabular}{cc}
\subcaptionbox{Elementary $H \times o$}{\includegraphics[width = 0.45\linewidth]{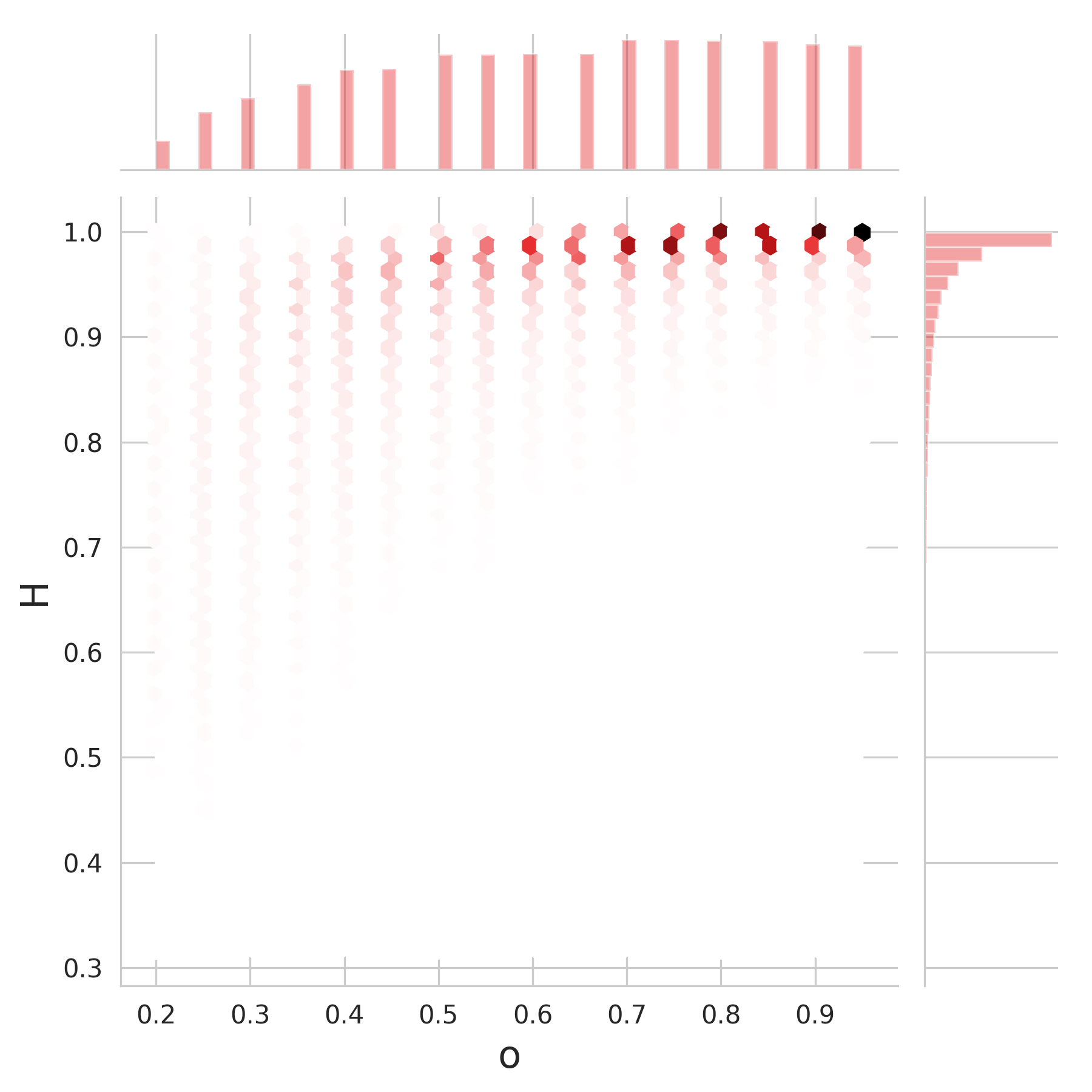}} &
\subcaptionbox{Elementary $I \times o$}{\includegraphics[width = 0.45\linewidth]{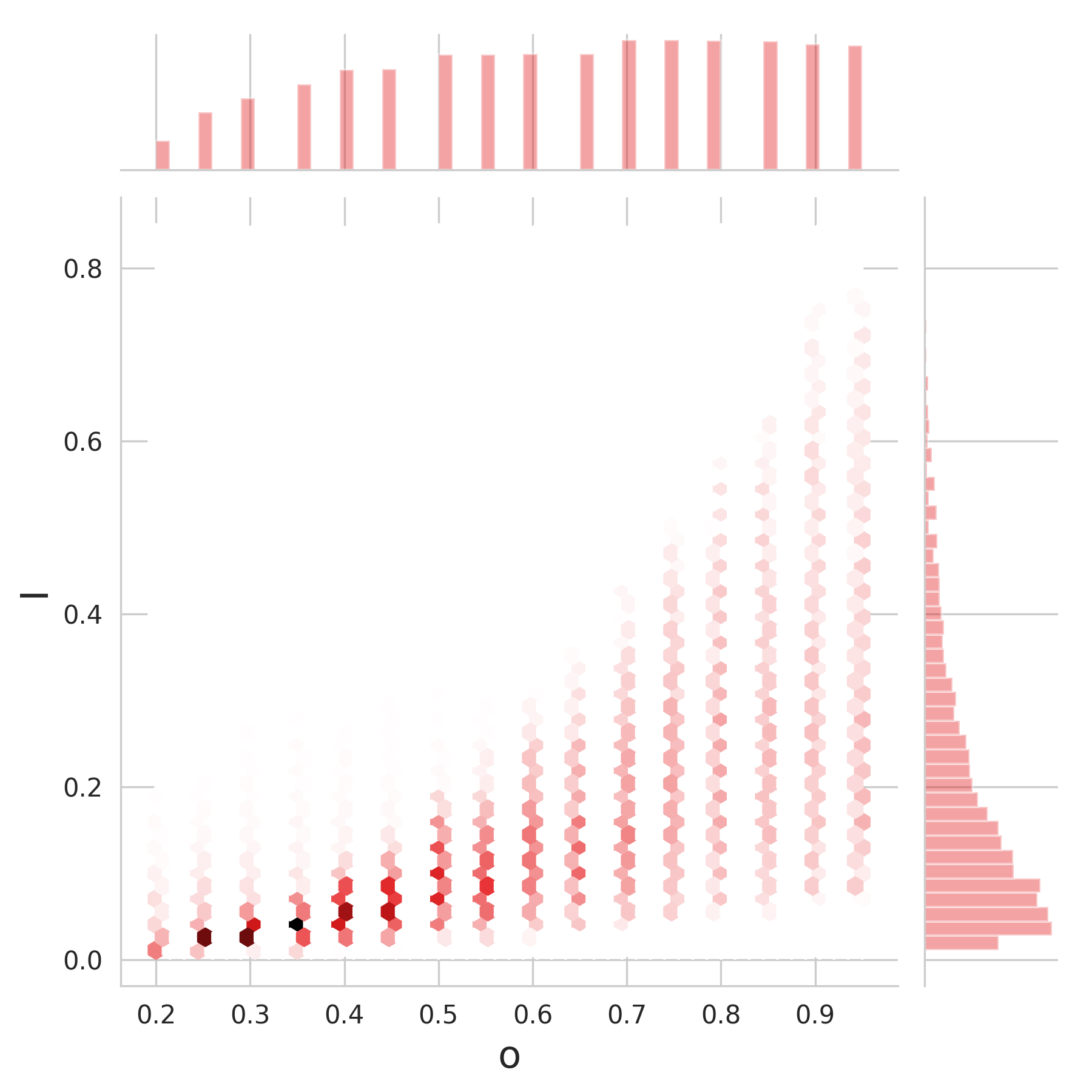}} \\
\subcaptionbox{Elementary $H \times p$}{\includegraphics[width = 0.45\linewidth]{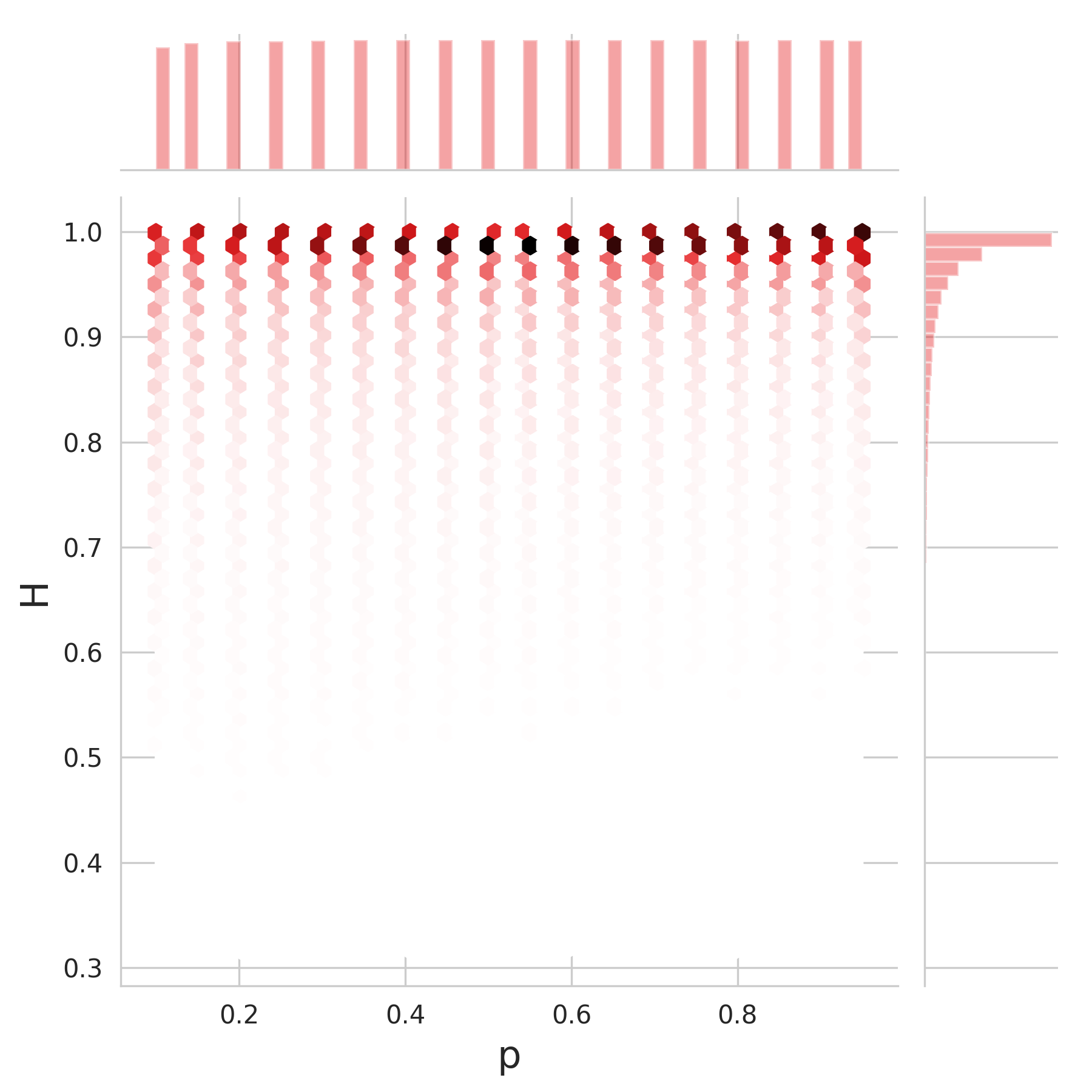}} &
\subcaptionbox{Elementary $I \times p$}{\includegraphics[width = 0.45\linewidth]{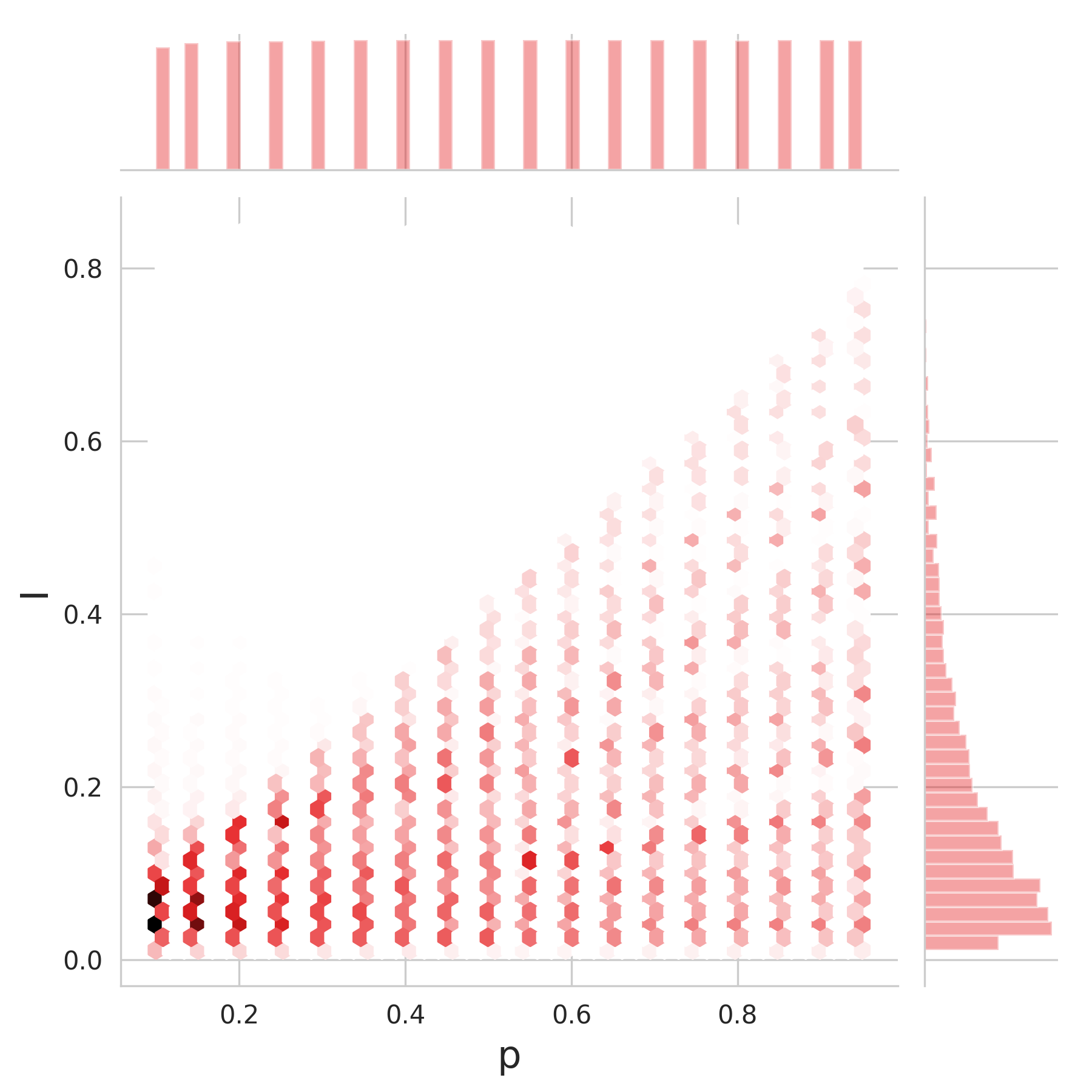}} \\
\end{tabular}

\caption{Results on synthetic multiplex networks without considering \REVISE{transition coupling}. There is small dependency on the layer assignment probability $o$ to nodes, since the higher it is, the more overlap may occur. The homogeneity (c) shows less dependency to the inner-layer edge probability $p$ than intensity (d), which also increases the likelihood of layer overlap.}
\label{fig:syn_dropout}
\end{figure}

We measure entanglement intensity $I$ and homogeneity $H$ on each generated network (averaged over all connected components \REVISE{of the layer overlap frequency matrix}). We investigate the role of the different parameters over the entanglement measures, as illustrated in Figures ~\ref{fig:syn_parameters},~\ref{fig:inner_nm} and~\ref{fig:syn_dropout}.

There is an obvious dependency between entanglement intensity and homogeneity since we cannot obtain low homogeneity with high intensity values (Figure~\ref{fig:syn_parameters}). This is due to the nature of both measures. With a high intensity, most of the layers are overlapping over most of the network. As a consequence, there is little space for permutations in the way layers overlap, this means the entanglement of all individual layers $\gamma_l$ tends to align, hence resulting in high values of homogeneity.
This leads to a denser production of high homogeneity networks as illustrated by the density lines in Figure~\ref{fig:syn_parameters}.

The number of nodes $n$ and edges $m$ do not show a strong dependency with homogeneity, but a slight one on intensity. Higher values of $n$ and $m$ make it easier to obtain sparser networks, with the consequence of resulting lower values of intensity. We further illustrate these in Figure~\ref{fig:inner_nm}. This effect mitigates quickly with higher numbers of nodes and layers.

We further explore the \REVISE{layer} assignment probability of a node $o$, and the inner-layer edge probability $p$ in Figures~\ref{fig:syn_dropout}.
 There is a first dependency appearing on the layer assignment probability $o$, for which higher values tend to produce higher homogeneity (Figure~\ref{fig:syn_dropout}b). Higher homogeneity is reached when all layers contribute equally, meaning that a higher $o$ shows more chances for each layer to contain most of the nodes. 
We may also observe apparent linear trend between the edge probability $p$ (sparseness) and entanglement intensity (Figure~\ref{fig:syn_dropout}d). This trend confirms that sparser networks (\textit{i.e.} lower $p$) are less ``intensely'' overlapping over edges. As intensity directly measures this property, this result outlines one of the \textit{desired} properties of the proposed network generator.

\subsubsection{Multiplex networks with \REVISE{transition coupling}}

\begin{figure}[ht!]
    \centering
    \captionsetup{width=.90\linewidth}
    \begin{tabular}{cc}
\subcaptionbox{Lower transition $I$.}{\includegraphics[width = 0.45\linewidth]{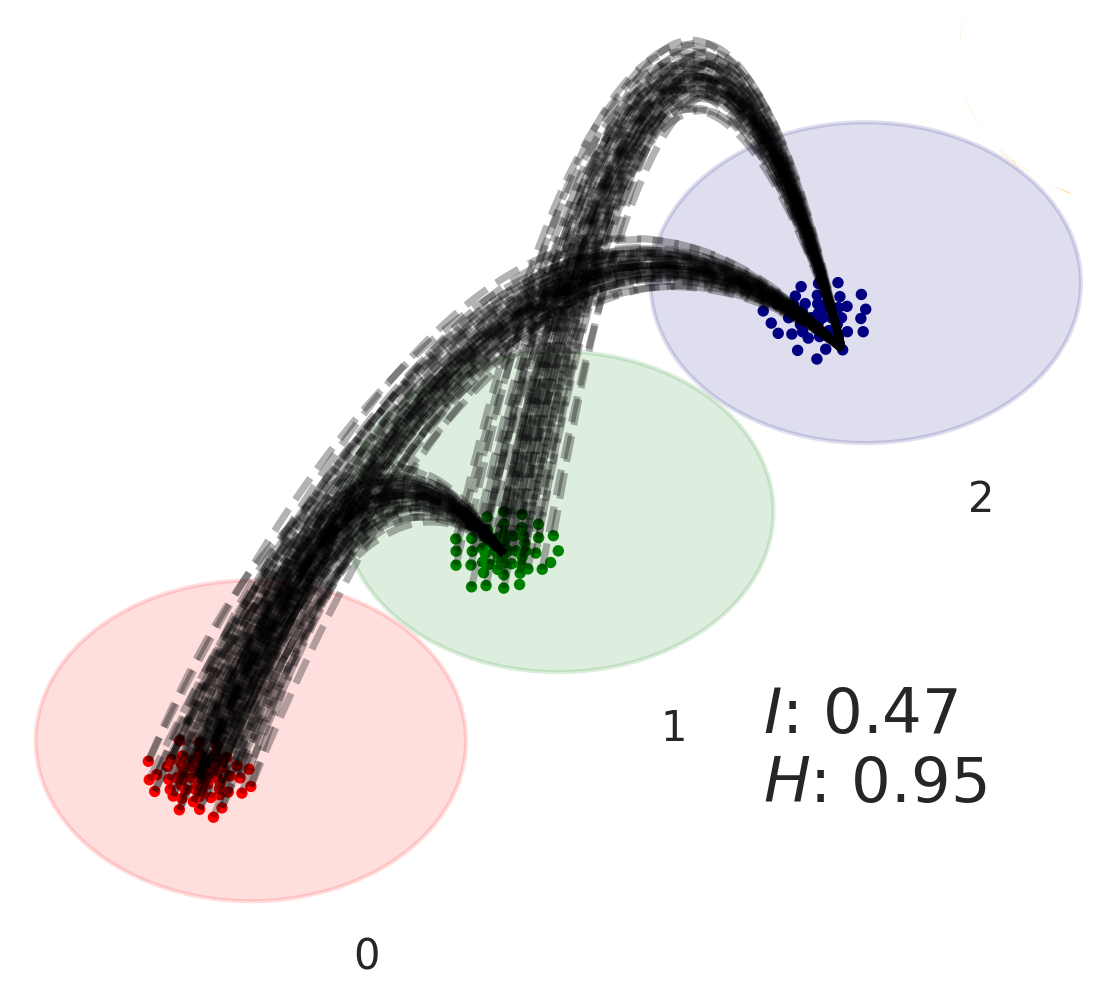}} &
\subcaptionbox{Higher transition $I$.}{\includegraphics[width = 0.45\linewidth]{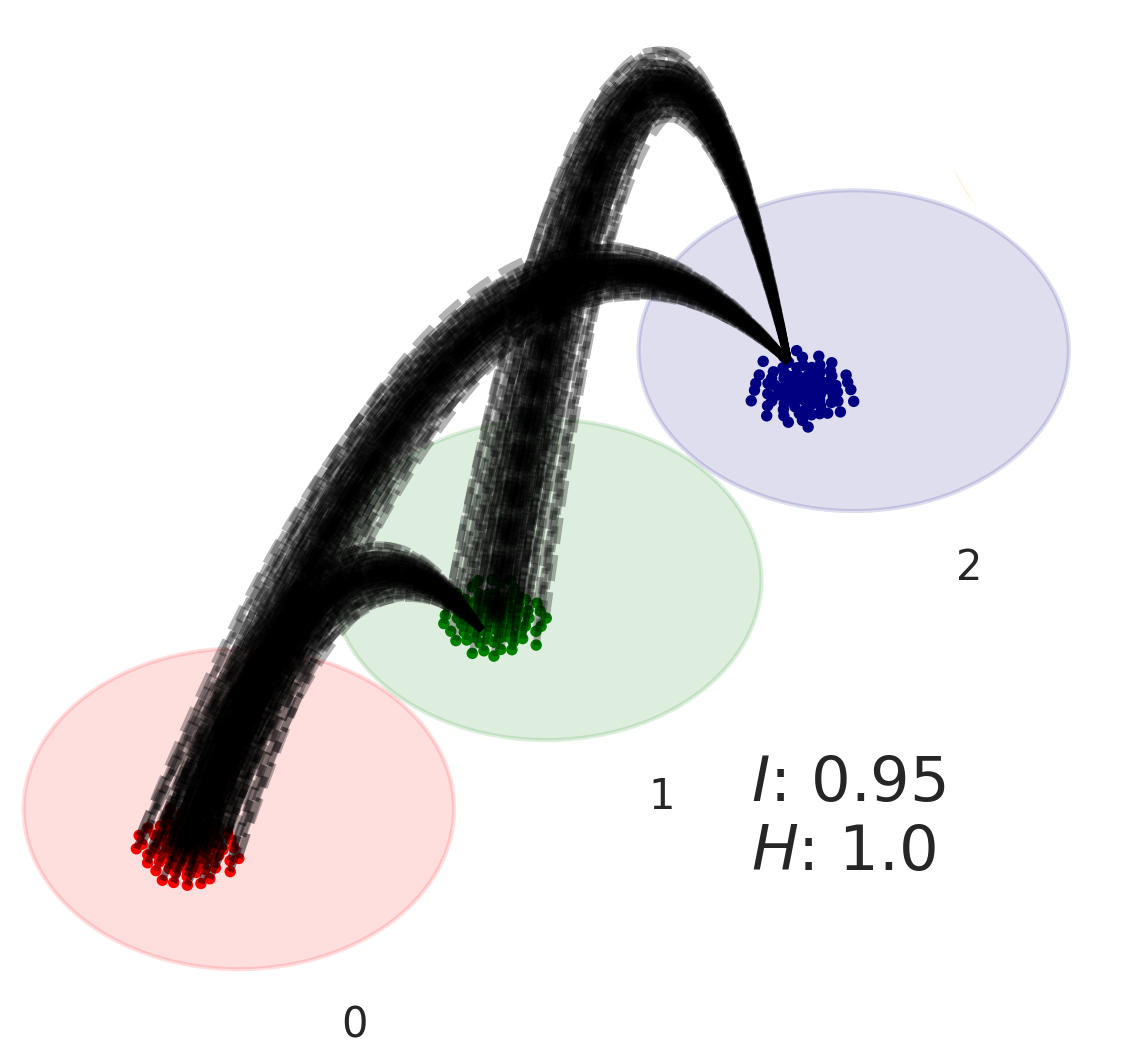}} \\
\end{tabular}
    \caption{\label{fig:viz-int}Visualization of coupling edges in synthetic coupled multilayer networks.}
\end{figure}

\begin{figure}[ht!]
\centering
\captionsetup{width=.90\linewidth}
\includegraphics[width = 0.75\linewidth]{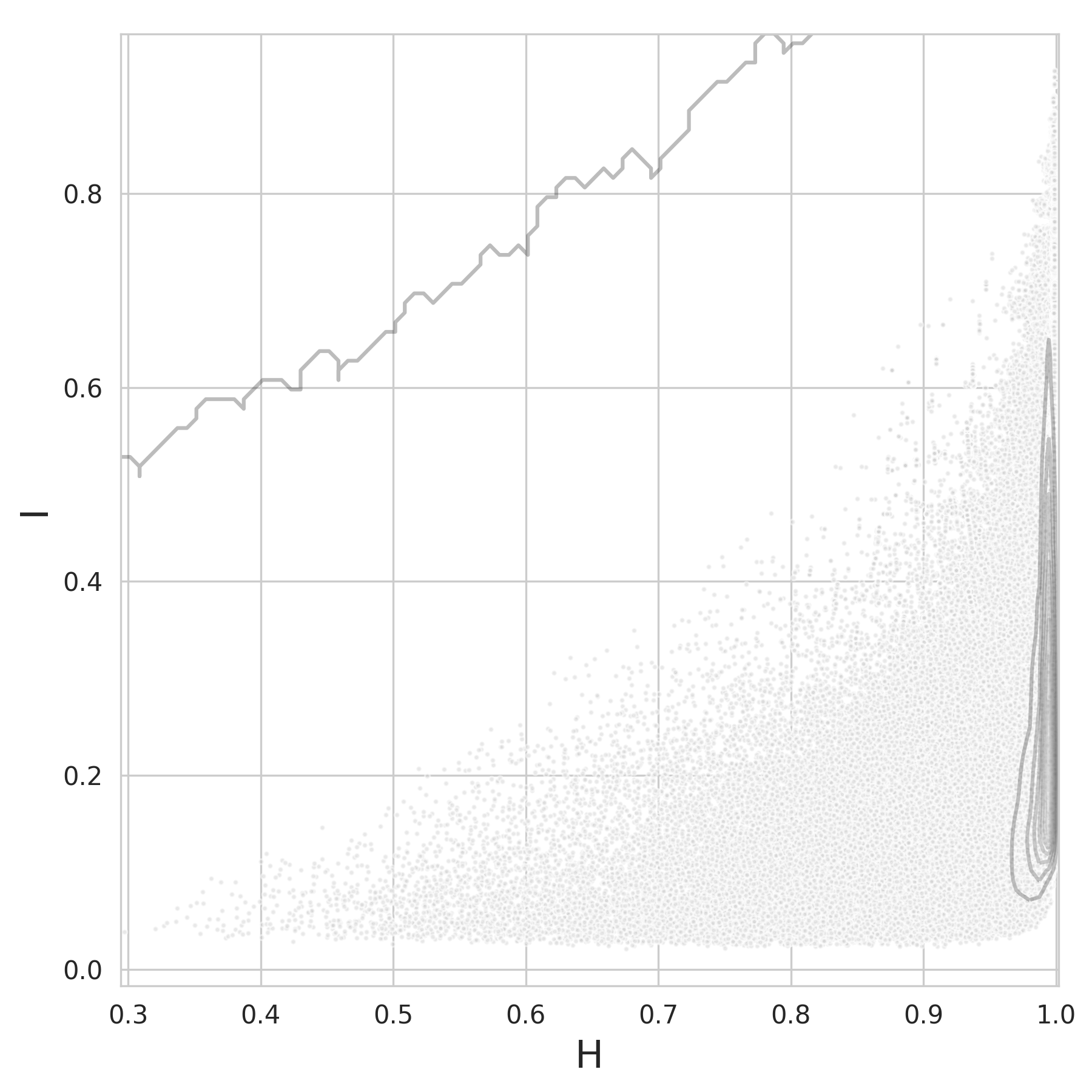}
\caption{Homogeneity and intensity $H \times I$ results on 1{,}329{,}696 synthetic multiplex networks considering their \REVISE{transition coupling} with density lines (Gaussian kernel density estimation).
}
\label{fig:multilayer-synth}
\end{figure}

A second experiment is focusing on multiplex graphs with \REVISE{transition coupling}, \textit{i.e.} considering \REVISE{only the} coupling edges in our 1{,}329{,}696 generated networks (illustrated in Figure~\ref{fig:viz-int}). This experiment reproduces the previous one, but focusing on the \REVISE{transition coupling} entanglement.
Results are shown in Figure~\ref{fig:multilayer-synth} and~\ref{fig:coupled-gen}, dependency on the number of nodes and layers is illustrated in Appendix. From Figure~\ref{fig:multilayer-synth}, the shape is globally the same, with the difference in a skewed density of high-homogeneity without a dense production of very low intensity generated networks (from the density lines).

\begin{figure}[ht!]
    \centering
    \captionsetup{width=.90\linewidth}
    \begin{tabular}{cc}
\subcaptionbox{Transition ($H \times o$).}{\includegraphics[width = 0.45\linewidth]{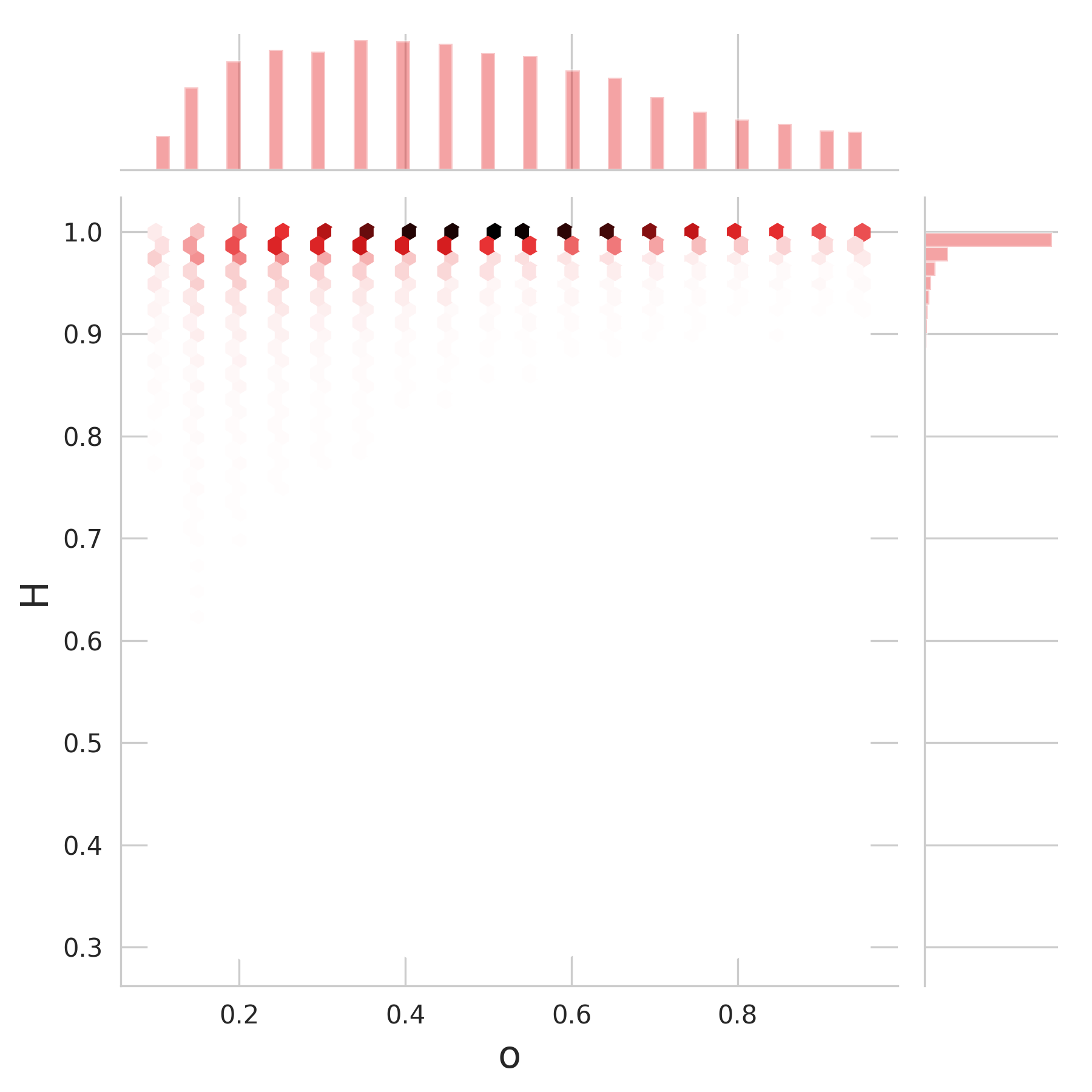}} &
\subcaptionbox{Transition ($I \times o$).}{\includegraphics[width = 0.45\linewidth]{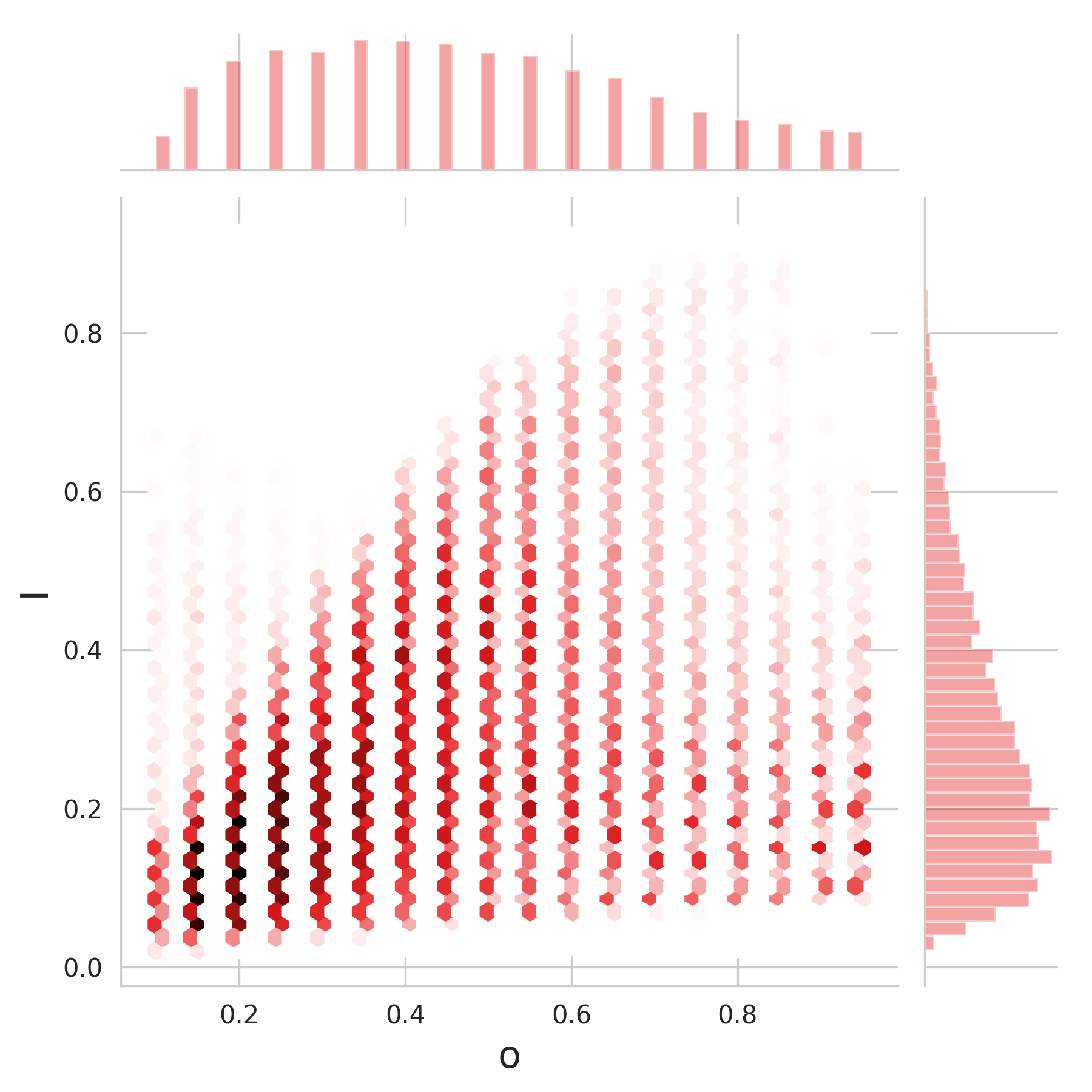}} \\
\subcaptionbox{Transition ($H \times q$).}{\includegraphics[width = 0.45\linewidth]{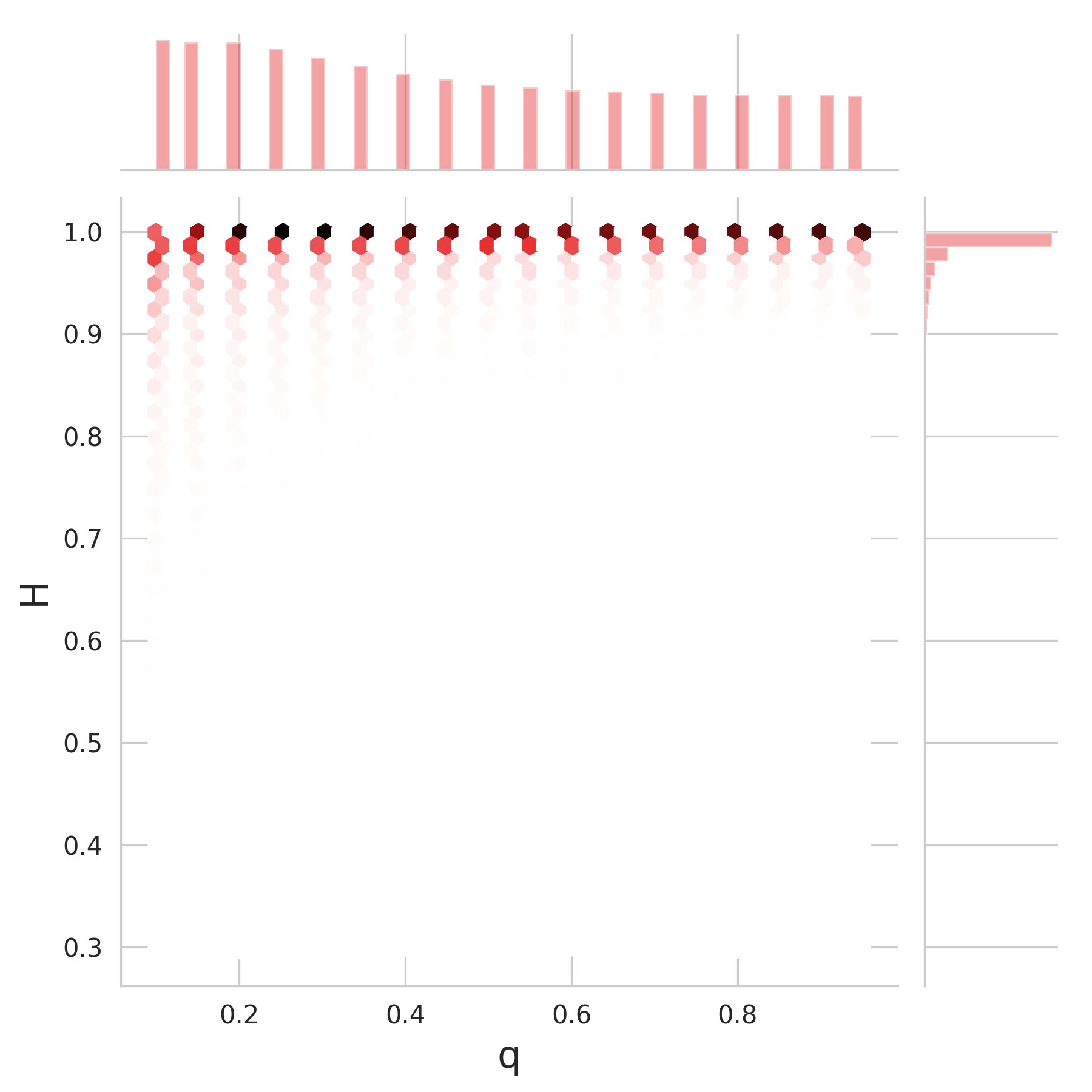}} &
\subcaptionbox{Transition ($I \times q$).}{\includegraphics[width = 0.45\linewidth]{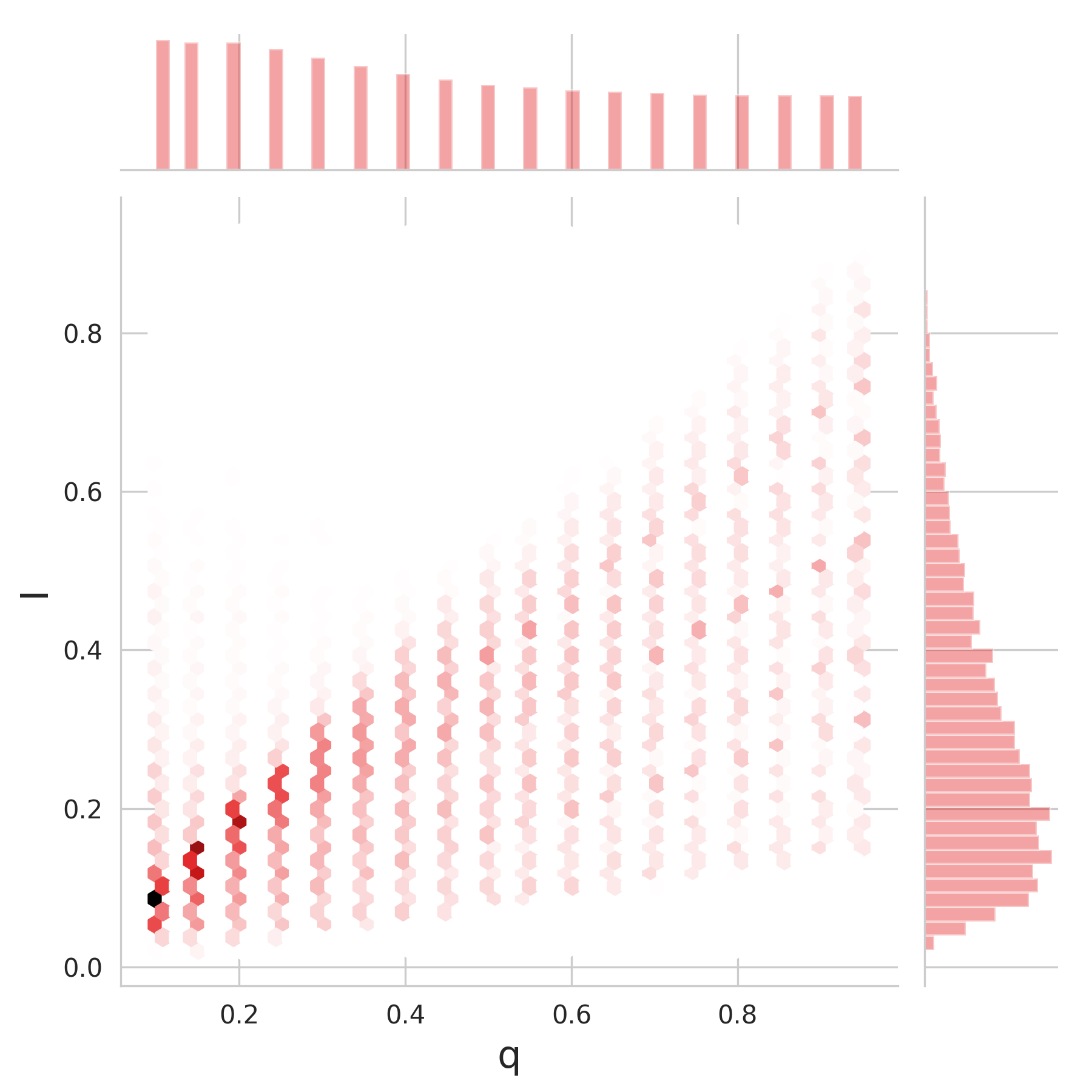}} \\
\end{tabular}
    
    \caption{\label{fig:coupled-gen}Homogeneity and intensity distributions in \REVISE{transition coupling} entanglement \textit{w.r.t.} \REVISE{$o$} and $q$.}
\end{figure}

The profile is sensibly the same than that of the previous experiment, except that the layer assignment probability $o$ appears to have a more diffuse impact, and the direct dependency is this time observed on the coupling edge probability $q$. Comparison with parameter $p$ obviously does not influence entanglement, but can be found in Appendix materials for additional inspection.

Overall, the networks with \REVISE{transition coupling} are more \textit{saturated} when compared to the ones without transition. The reason may be that we only consider here \REVISE{transition coupling} edges that only connect the \textit{same node} across layers. 

For the interested reader, we also illustrate in the Appendix material the independence of parameters $q$ over the elementary layer entanglement and $p$ over the \REVISE{transition coupling} entanglement. We also report there the computation of entanglement over the combined elementary layers and \REVISE{transition coupling}, which displays a dependency on both $p$ and $q$ parameters. \REVISE{Finally, we have computed the layer correlation coefficient, as suggested in~\cite{nicosia2015measuring}, confirming the role of the different parameters of our generator.}

\subsection{Multiplex network comparison across disciplines}
\label{sec:results-comparison}

\begin{figure}[htbp]
\centering
\captionsetup{width=.90\linewidth}
\begin{tabular}{cc}
\subcaptionbox{\REVISE{Real networks: $H$}}{\includegraphics[width = 0.40\linewidth]{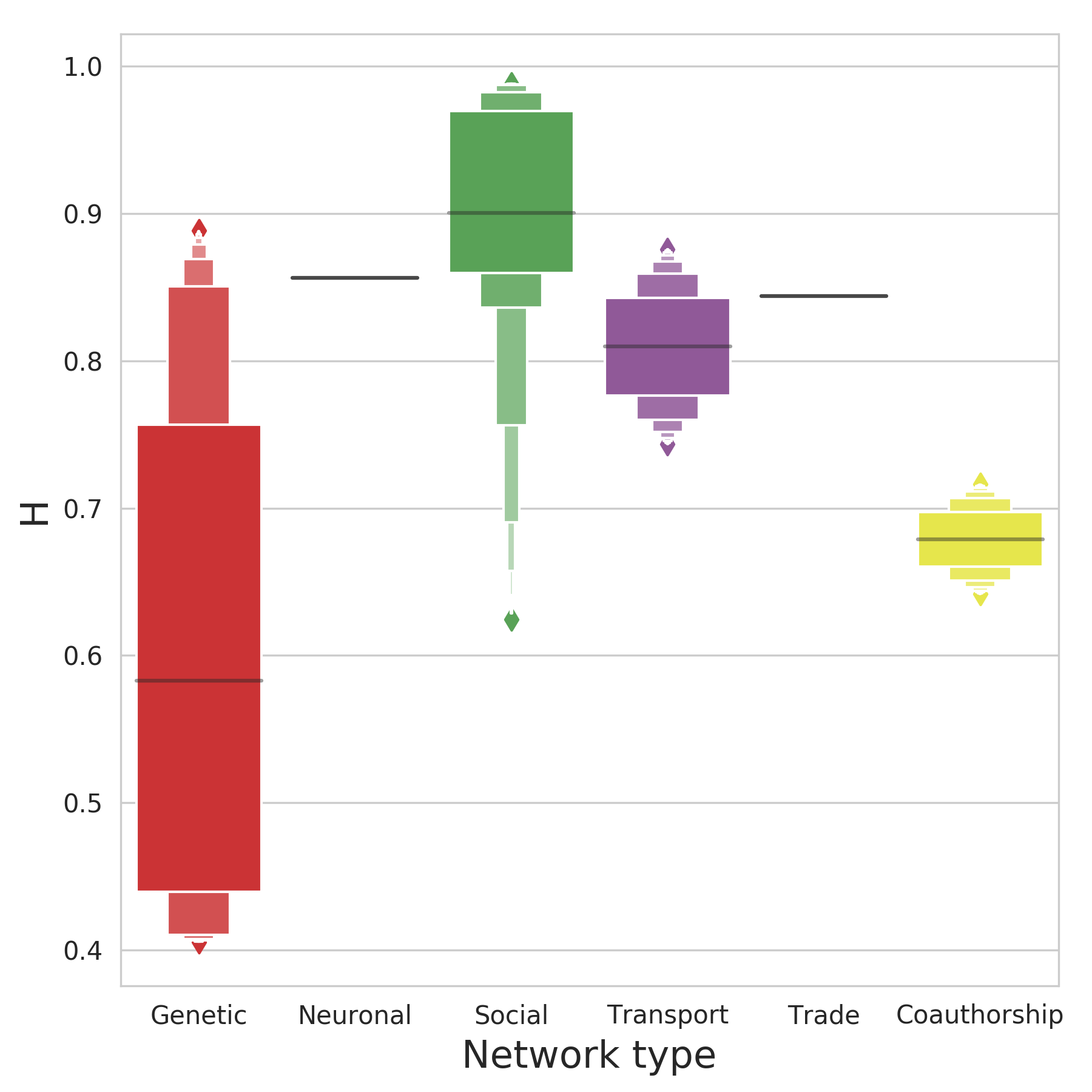}} &
\subcaptionbox{\REVISE{Real networks: $I$}}{\includegraphics[width = 0.40\linewidth]{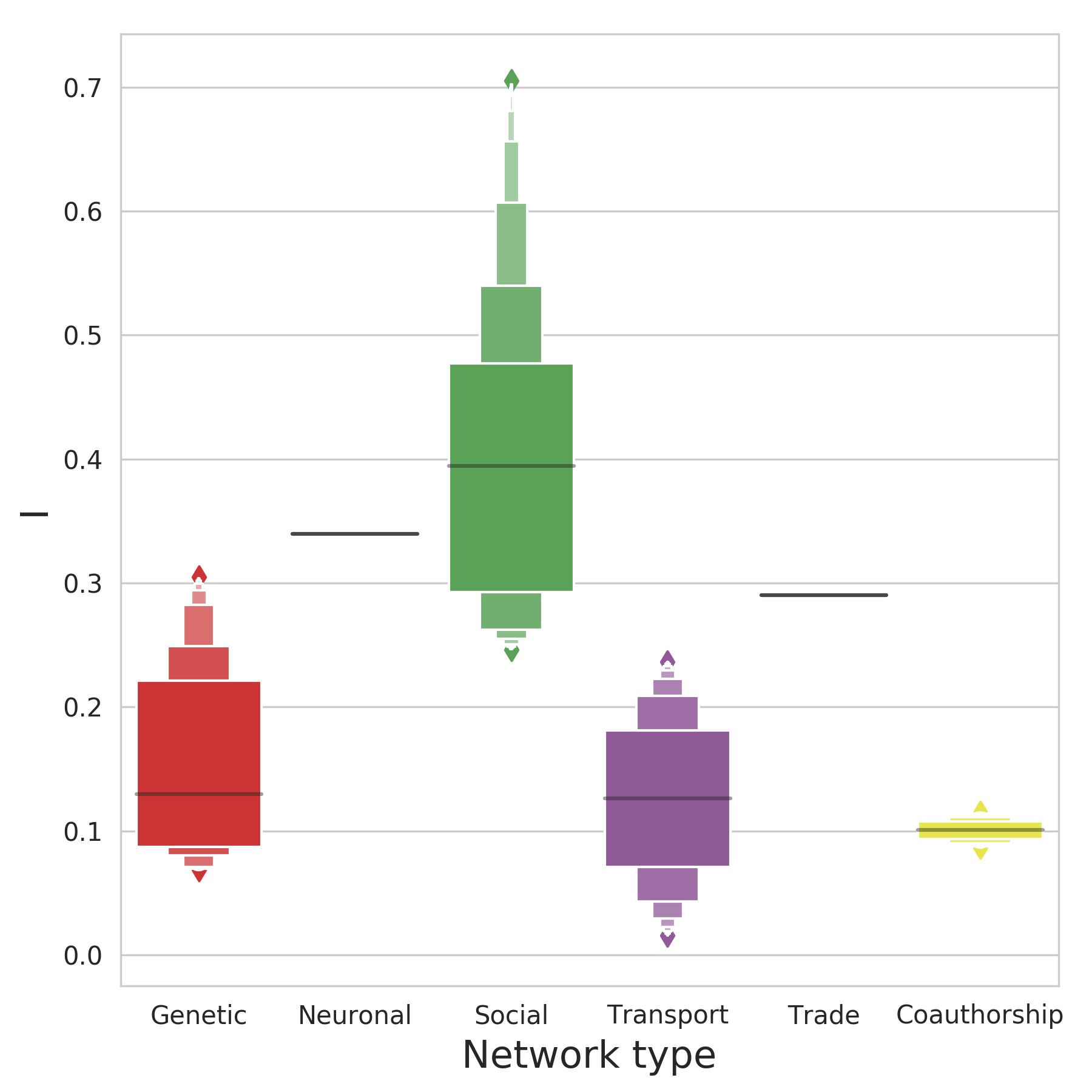}} \\
\end{tabular}
\caption{Entanglement homogeneity and intensity compare for each category of networks, showing quite diverse set of properties proper to the different families of networks.}
\label{fig:real}
\end{figure}

\begin{table}[ht!]
\centering
\caption{Real multiplex networks and their properties. The ID in the second column corresponds to Figure~\ref{fig:real_overlay}.}
\resizebox{0.9\textwidth}{!}{
\begin{tabular}{lccccccccc}\toprule
                   Dataset &    ID &      Type &    Nodes &    Edges &  Number of layers &  Mean degree &      CC  & Intensity & Homogeneity \\ \midrule
          arXiv-Netscience~\cite{de2015identifying} & 34 &Coauthorship &    26796 &    59026 &                13 &         4.41 &    3660  & 0.114786 & 0.641670 \\
               PierreAuger~\cite{de2015identifying} & 26 &  Coauthorship &      965 &     7153 &                16 &        14.82 &     131  & 0.086551  & 0.716156 \\
               Arabidopsis~\cite{stark2006biogrid} &  0  &   Genetic &     8765 &    18655 &                 7 &         4.26 &     387  & 0.111636  & 0.408940 \\
                       Bos~\cite{stark2006biogrid} &  1   &  Genetic &      369 &      322 &                 4 &         1.75 &      82 & 0.160341 & 0.582015 \\
                   Candida~\cite{stark2006biogrid} &    5   & Genetic &      418 &      398 &                 7 &         1.90 &      50 & 0.284783  & 0.888476 \\
                  Celegans~\cite{stark2006biogrid} &    7  & Genetic &     4557 &     8182 &                 6 &         3.59 &     193 &0.115718  & 0.420231 \\
                DanioRerio~\cite{stark2006biogrid} & 8   &   Genetic &      180 &      188 &                 5 &         2.09 &      45 & 0.068219 & 0.870304\\
                Drosophila~\cite{stark2006biogrid} & 9    &   Genetic &    11970 &    43367 &                 7 &         7.25 &     346 & 0.082283  & 0.405509 \\
                    Gallus~\cite{stark2006biogrid} &  12 &    Genetic &      367 &      389 &                 6 &         2.12 &      54 & 0.151845  & 0.433374 \\
           HepatitusCVirus~\cite{stark2006biogrid} &   13  &  Genetic &      129 &      137 &                 3 &         2.12 &       4 & 0.304679  & 0.777382 \\
                Homo Sapiens~\cite{stark2006biogrid} &   14  &  Genetic &    36194 &   170899 &                 7 &         9.44 &     785 & 0.101047  & 0.519648 \\
              HumanHerpes4~\cite{stark2006biogrid} &    16 &  Genetic &      261 &      259 &                 4 &         1.98 &      21 & 0.245979  & 0.595037 \\
                 HumanHIV1~\cite{stark2006biogrid} &   15   & Genetic &     1195 &     1355 &                 5 &         2.27 &      13 & 0.158347  & 0.583648  \\
               Oryctolagus~\cite{stark2006biogrid} &   24  &  Genetic &      151 &      144 &                 3 &         1.91 &      21 & 0.241322 & 0.635943 \\
                Plasmodium~\cite{stark2006biogrid} &  27  &    Genetic &     1206 &     2522 &                 3 &         4.18 &      27 & 0.249623  & 0.853694 \\
                    Rattus~\cite{stark2006biogrid} &    28 &   Genetic &     3263 &     4268 &                 6 &         2.62 &     296 & 0.126889  & 0.457888 \\
                 SacchCere~\cite{stark2006biogrid} &    29  & Genetic &    27994 &   282755 &                 7 &        20.20 &     432 & 0.070428 & 0.695150 \\
                 SacchPomb~\cite{stark2006biogrid} & 30    &   Genetic &    10178 &    63677 &                 7 &        12.51 &     286 & 0.079756  & 0.407135 \\
                   Xenopus~\cite{stark2006biogrid} &  32 &    Genetic &      582 &      620 &                 5 &         2.13 &     109 & 0.082539  & 0.829466\\
            YeastLandscape~\cite{costanzo2010genetic} &   33 &   Genetic &    17770 &  8473997 &                 4 &       953.74 &       4 & 0.132035  & 0.534030 \\
                  CElegans~\cite{chen2006wiring} &   7  &  Neuronal &      791 &     5863 &                 3 &        14.82 &       6 & 0.339461  & 0.856373 \\
                Cannes2013~\cite{omodei2015characterizing} &  6  &  Social &   659951 &   991854 &                 3 &         3.01 &   48375 & 0.269159  & 0.900587  \\
 CKM-Physicians-Innovation~\cite{coleman1957diffusion} &  3   &  Social &      674 &     1551 &                 3 &         4.60 &      12 &0.394666  & 0.988309 \\
                CS-Aarhus~\cite{magnani2013combinatorial} &   4  &   Social &      224 &      620 &                 5 &         5.54 &      13 &0.341388  & 0.894766 \\
      Kapferer-Tailor-Shop~\cite{kapferer1972strategy} &  17   &   Social &      150 &     1018 &                 4 &        13.57 &       5 & 0.438509 & 0.910168 \\
      Krackhardt-High-Tech~\cite{krackhardt1987cognitive} &  18  &    Social &       63 &      312 &                 3 &         9.90 &       3 &0.412875  & 0.838791 \\
           Lazega-Law-Firm~\cite{lazega2001collegial} &  19   &   Social &      211 &     2571 &                 3 &        24.37 &       3 & 0.516232 & 0.970364 \\
                MLKing2013~\cite{omodei2015characterizing} &   21  &   Social &   392542 &   396671 &                 3 &         2.02 &   36041 & 0.260099 & 0.624426 \\
       MoscowAthletics2013~\cite{omodei2015characterizing} &   22 &     Social &   133619 &   210250 &                 3 &         3.15 &    6323 & 0.246321  & 0.880520  \\
         ObamaInIsrael2013~\cite{omodei2015characterizing} &   23   &  Social &  3457453 &  4061960 &                 3 &         2.35 &  651141 &0.316202  &0.835469  \\
 Padgett-Florence-Families~\cite{padgett1993robust} &   25  &   Social &       26 &       35 &                 2 &         2.69 &       2 & 0.547715 & 0.986433 \\
   Vickers-Chan-7thGraders~\cite{vickers1981representing} &  31   &   Social &       87 &      740 &                 3 &        17.01 &       3 &0.705372  & 0.968908 \\
                       FAO~\cite{de2015structural} & 11  &   Trade &    41713 &   318346 &               364 &        15.26 &     571 & 0.290018 & 0.843847 \\
                     EUAir~\cite{cardillo2013emergence} &  10 & Transport &     2034 &     3588 &                37 &         3.53 &      41 &0.015499  & 0.743443 \\
                    London~\cite{de2014navigability} &  20  &  Transport &      399 &      441 &                 3 &         2.21 &       3 & 0.236502  & 0.875838 \\
                \bottomrule
\end{tabular}
}
\label{tab:summary}
\end{table}

We now consider real world static networks. 
All considered networks are summarised with their main characteristics in Table~\ref{tab:summary}\footnote{The networks are hosted at \url{https://comunelab.fbk.eu/data.php}}.  Unfortunately, we have not found a real case with a large number of \REVISE{transition coupling} edges, so we limit this evaluation to elementary layer entanglement. For each network, we computed elementary layer homogeneity and intensity, for all connected components.

\begin{figure}[ht!]
\centering
\captionsetup{width=.90\linewidth}
\includegraphics[width = 0.95\linewidth]{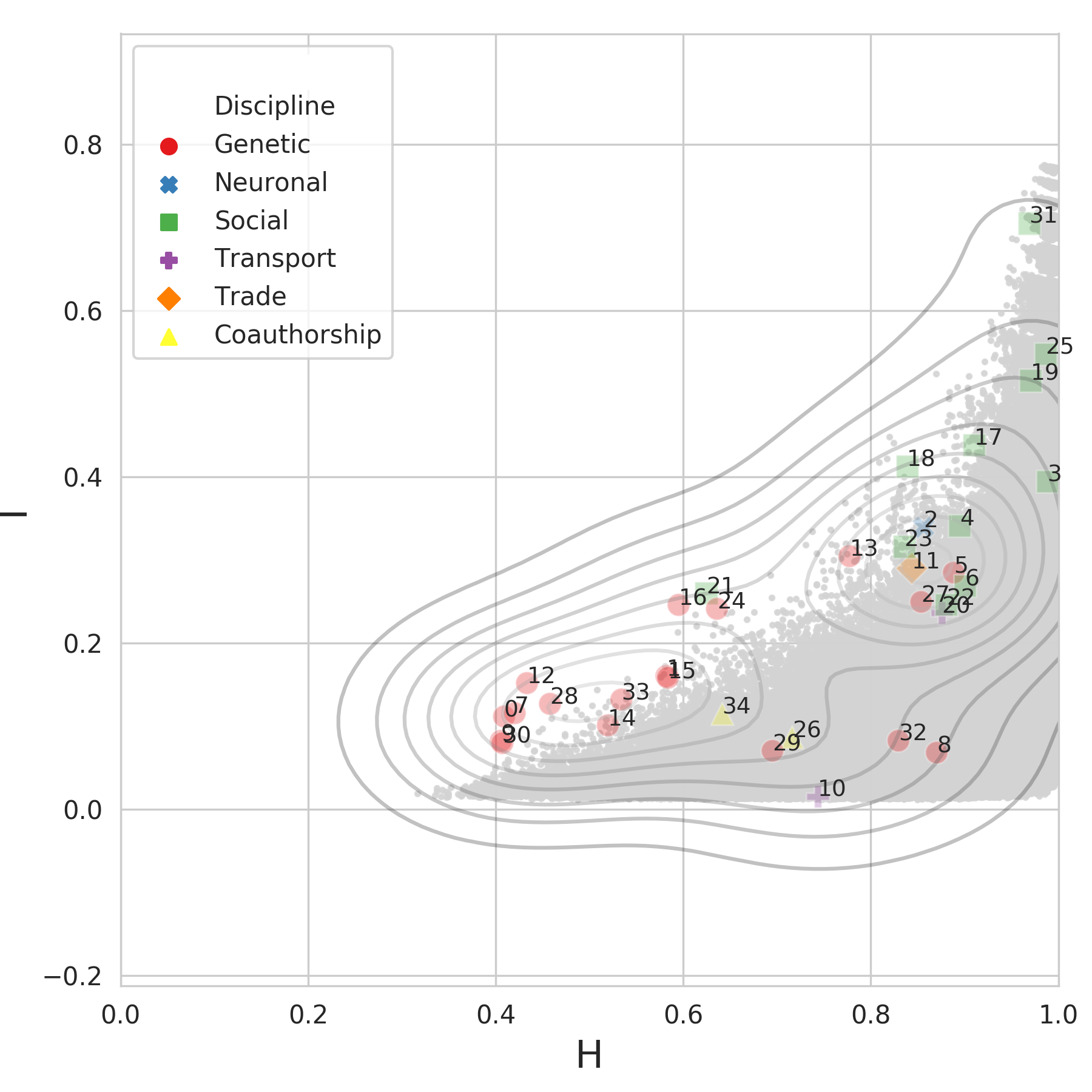}
\caption{Real networks: $H \times I$. Labels of networks map to Table~\ref{tab:summary} (ID). Grey dots represent synthetic samples of Figure~\ref{fig:syn_parameters}, with Gaussian kernel density estimation over lines over the real world samples. Social networks shows \REVISE{a tendency to fall within the high homogeneity/intensity range, coinciding with the high inner-layer edge probability parameter $p$ of synthetic networks.}}
\label{fig:real_overlay}
\end{figure}

We first investigate individual results through the distributions of each metric across network types, Figure~\ref{fig:real}. We then compare individual networks across entanglement intensity and homogeneity Figure~\ref{fig:real_overlay}.

\begin{figure}[htbp]
\centering
\captionsetup{width=.90\linewidth}
\begin{tabular}{cc}
\subcaptionbox{\REVISE{Genetic \textit{vs.} Social networks - $H$}}{\includegraphics[width = 0.40\linewidth]{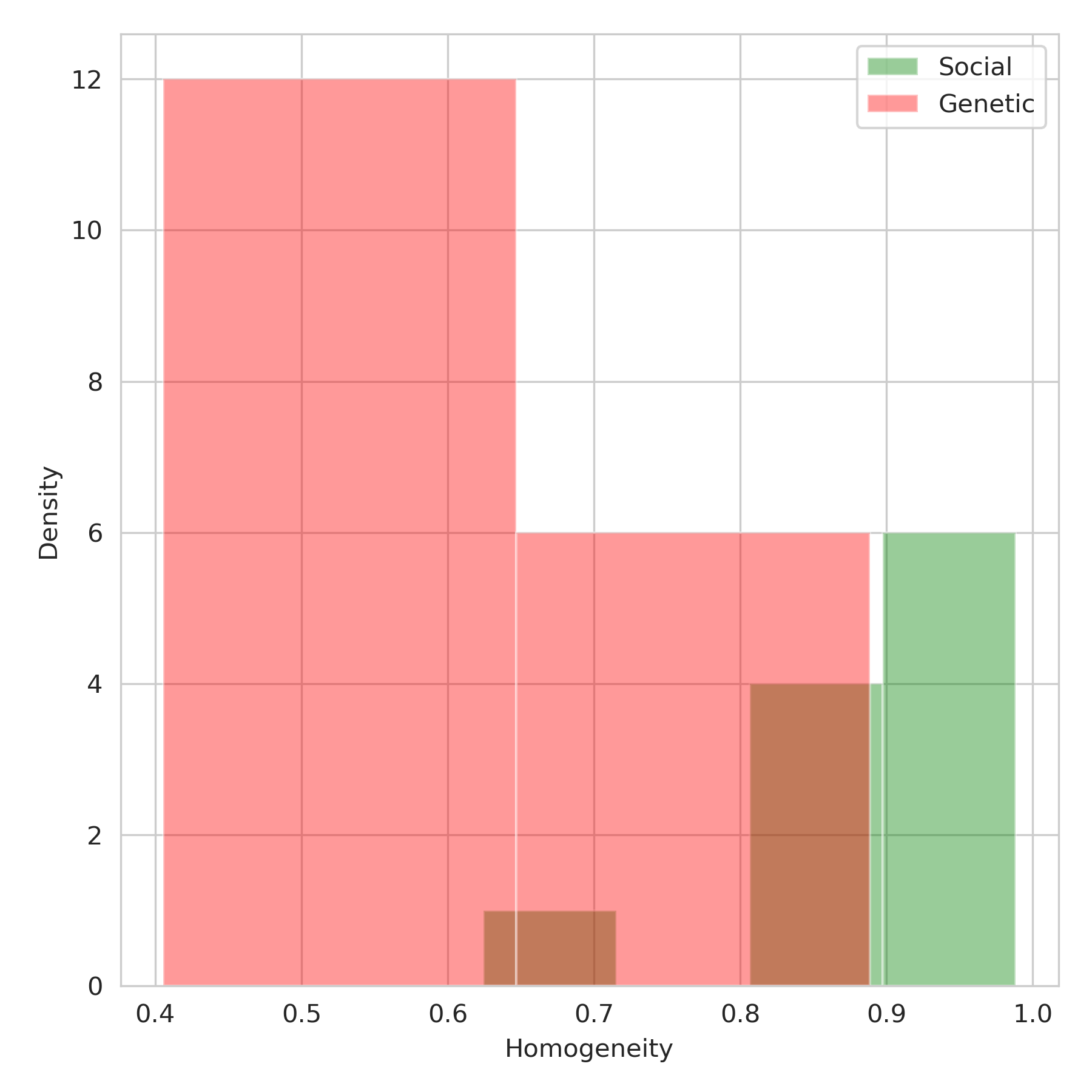}} &
\subcaptionbox{\REVISE{Genetic \textit{vs.} Social networks - $I$}}{\includegraphics[width = 0.40\linewidth]{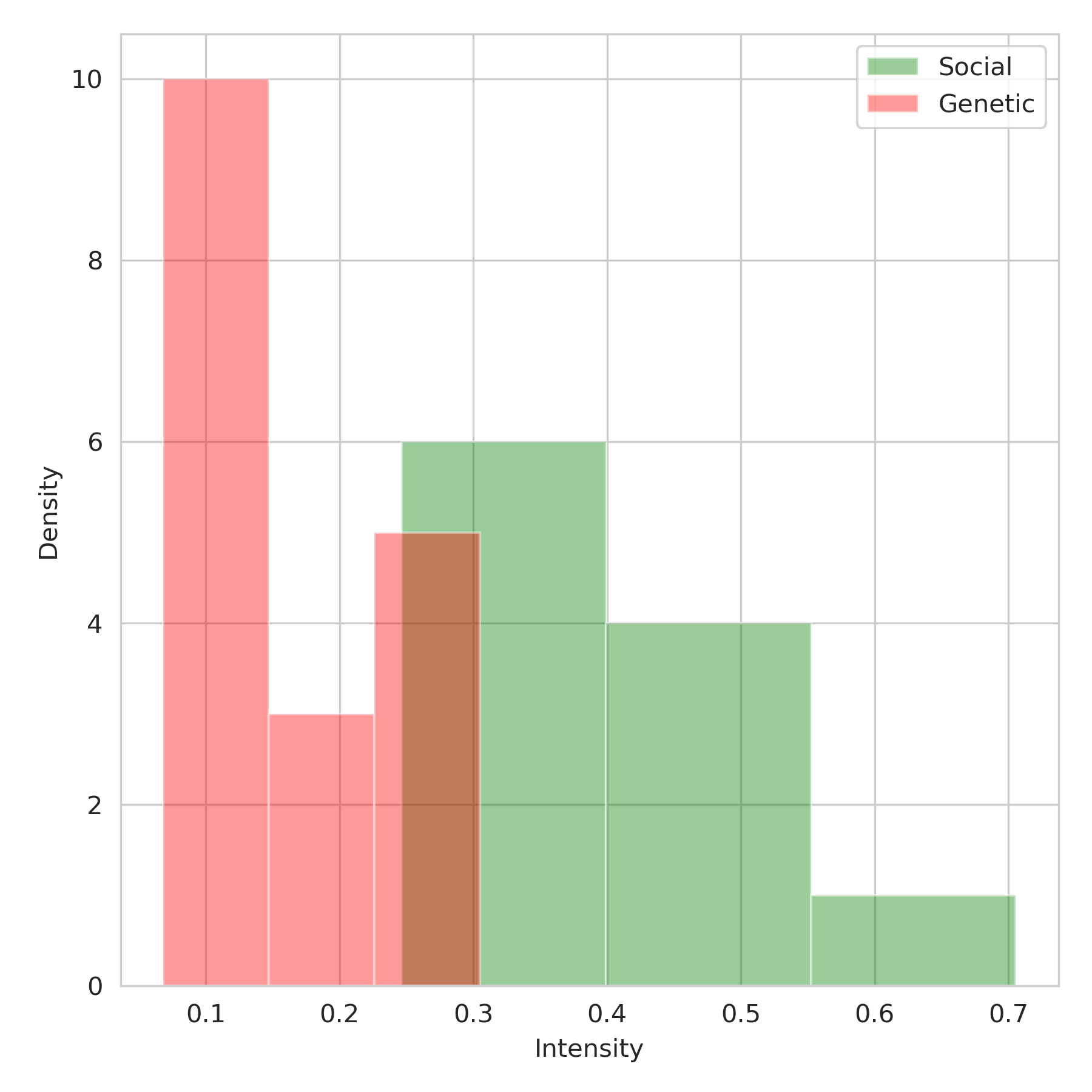}} \\
\end{tabular}
\caption{Distributions of homogeneity and intensity when genetic networks are compared to social ones.}
\label{fig:dist}
\end{figure}

Two main observations are apparent when studying the results on real networks. First, the difference between social and genetic (biological) multiplex networks becomes obvious when both entanglement intensity and homogeneity are considered (Figure~\ref{fig:real_overlay}). 
To confirm these differences, we further compare their distributions, \textit{i.e.}, the intensity and homogeneity of social \textit{vs.} genetic networks, in Figure~\ref{fig:dist}.

In addition, from Figure~\ref{fig:real_overlay}, we may observe that many genetic networks sit in relatively low intensity/homogeneity places, whereas social networks sit in the top right corner: the high entanglement homogeneity of social networks is quite noticeable. This suggests a few interpretations:
\begin{itemize}
\item genetic networks show in general very little layer overlap;
\item \REVISE{some genetic networks could be matched to synthetic networks of low inner-layer edge probability, especially when homogeneity is low, being very sparse, potentially pointing at low layer assignment probability too;} 
\item layers in social networks tend to overlap a lot;
\item social networks tend to be quite dense and may be simulated by synthetic networks with a high inner-layer edge probability;
\end{itemize}

The results on social networks indicate a high level of layer overlap and it may be due to the overall behaviour of people, which is rather similar across different networks, whatever their means of interaction. \textit{Simmelian ties, triadic closure, and homophily} (which are well studied in social sciences) are probably strong drivers of this layer overlap.

\subsection{Entanglement in temporal multiplex networks}
\label{sec:temporal}

In our last experiment, we investigate entanglement across time slices of three real-life temporal multiplex networks:
\textit{MLKing2013}, \textit{MoscowAthletics2013}, and \textit{Cannes2013} (as found in \cite{omodei2015characterizing}). Each network consists in a collection of Twitter activity related to some event. The networks are comprised of three layers of connection, namely \textit{retweets}, \textit{replies} and \textit{comments}. They can be summarised as follows. The \textit{MLKing2013} data set consists of 421{,}083 events covering a week of celebration of M.L. King's speech ``\textit{I have a dream}'' in 2013, forming 396{,}671 edges between 327{,}708 nodes. The \textit{MoscowAthletics2013} data set consists of 303{,}330 events covering two weeks of the World Championships of Athletics held in Moscow in 2013, forming 210{,}250 edges between 88{,}805 nodes.
The \textit{Cannes2013} network consists of 1{,}297{,}545 events (temporal edges) covering a month of the 2013 Cannes Film Festival, together forming a network of 930{,}419 edges and 438{,}538 nodes. Note that the networks are not trivially small, offering additional evidence of the stability of  the entanglement computation.

The networks were analysed following the methodology introduced in Section~\ref{sec:ent-description-model}. We propose two experiments with regard to time segmentation. 

\begin{figure}[ht!]
\centering
\captionsetup{width=.90\linewidth}
\subcaptionbox{\textit{MLKing2013}, static window of 1h }{\includegraphics[width = 0.45\linewidth]{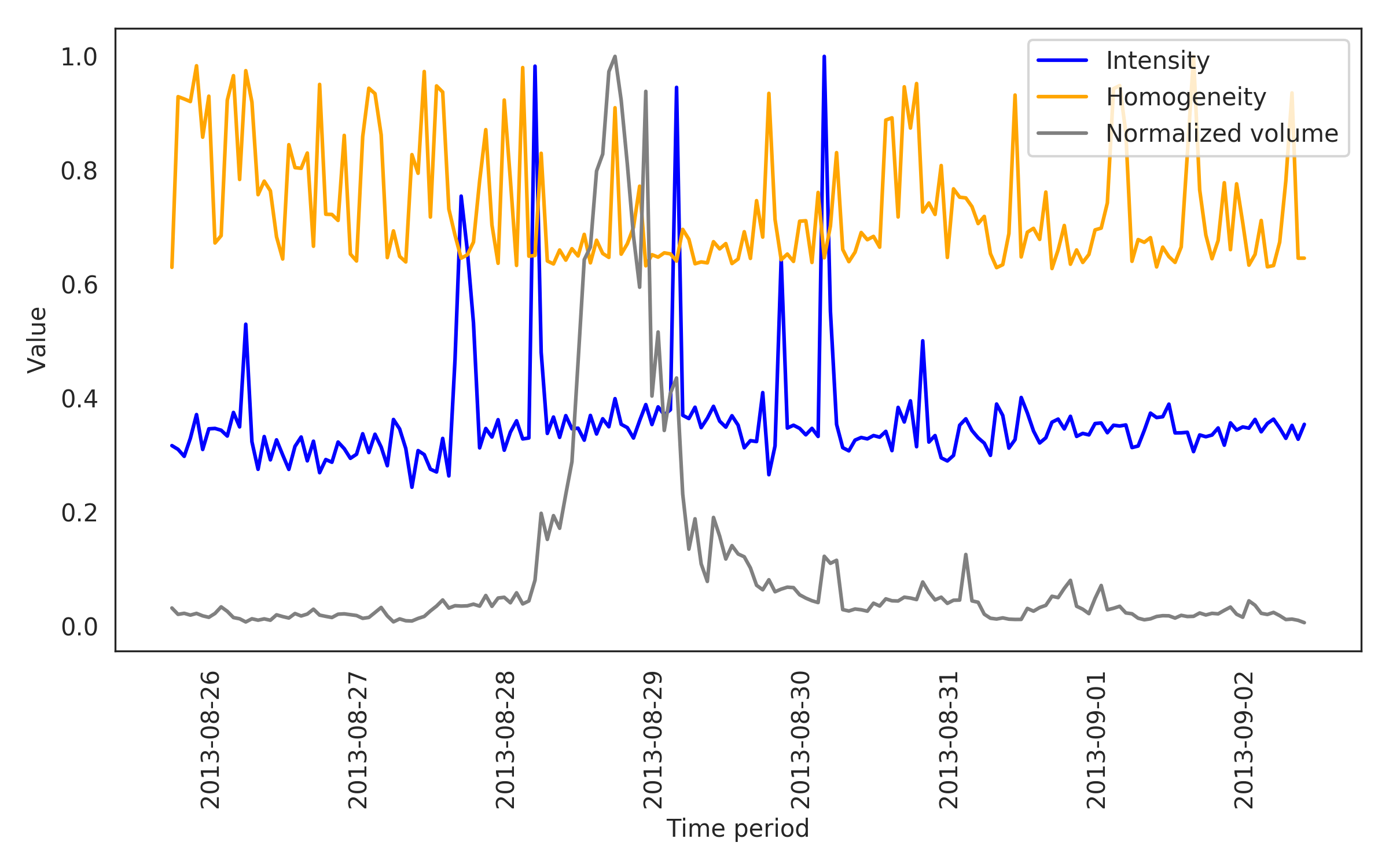}} \\
\begin{tabular}{cc}
\subcaptionbox{\textit{MLKing2013}, static window of 3h}{\includegraphics[width = 0.45\linewidth]{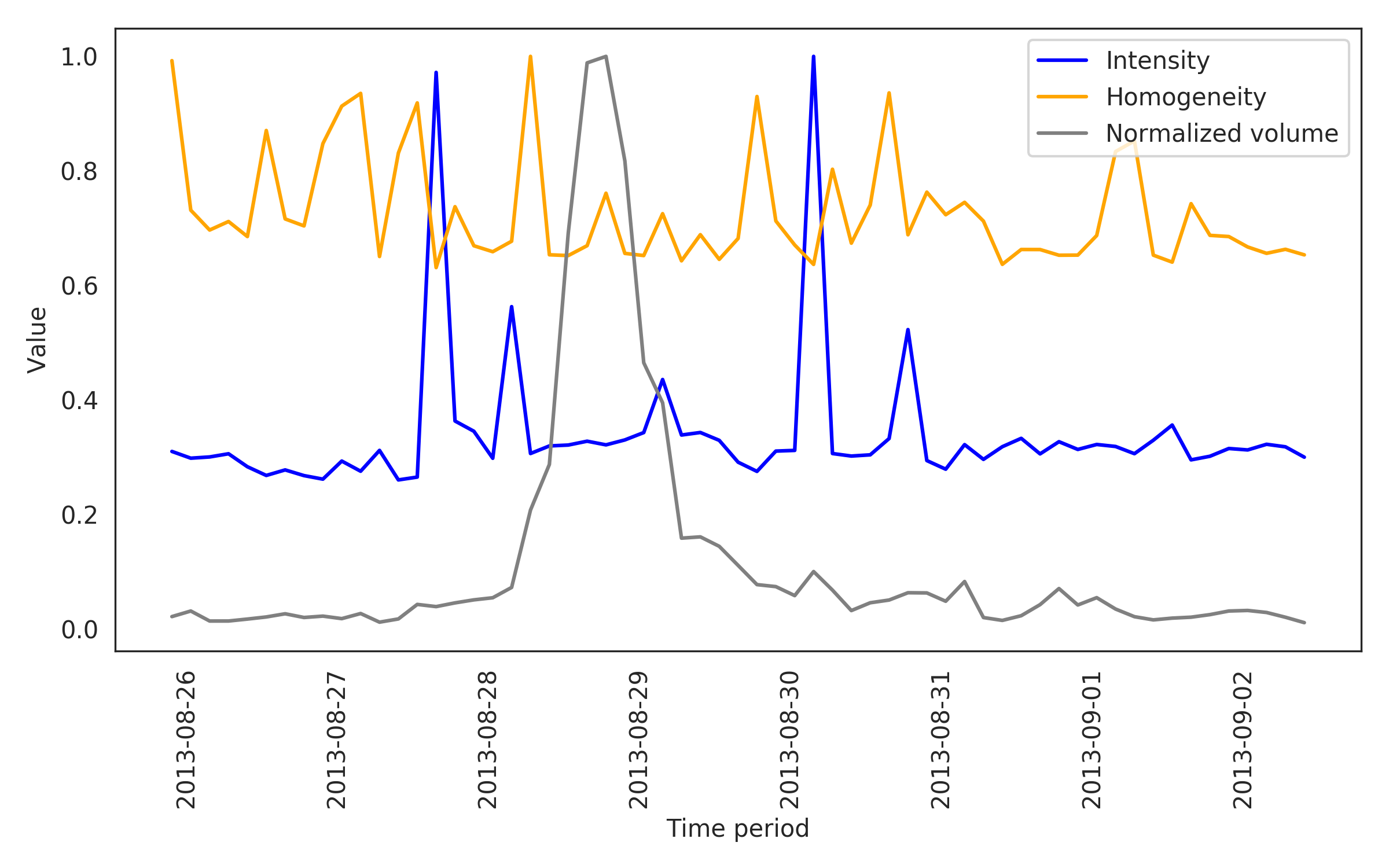}} &
\subcaptionbox{\textit{MLKing2013}, sliding window of 3h, \REVISE{with 1h steps}}{\includegraphics[width = 0.45\linewidth]{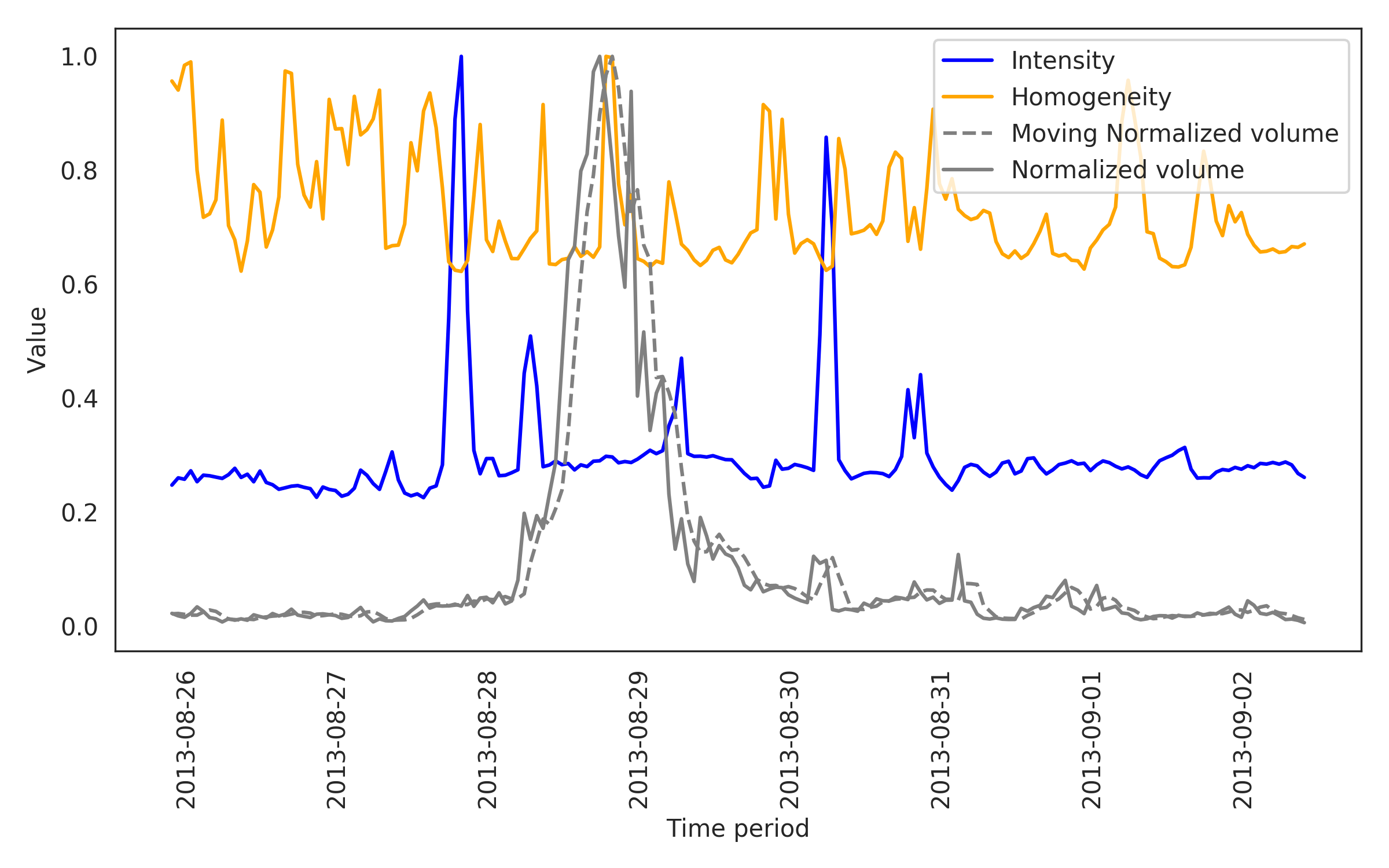}}
\end{tabular}
\caption{Visualization of temporal entanglement across \textit{MLKing2013}. In grey, volume over the period of time (dotted line for the aggregated volume over sliding window (c)). Intensity in blue and homogeneity in yellow.}
\label{fig:mlking}
\end{figure}

\begin{figure}[ht!]
\centering
\captionsetup{width=.90\linewidth}
\subcaptionbox{\textit{MoscowAthletics2013}, static window of 1h }{\includegraphics[width = 0.45\linewidth]{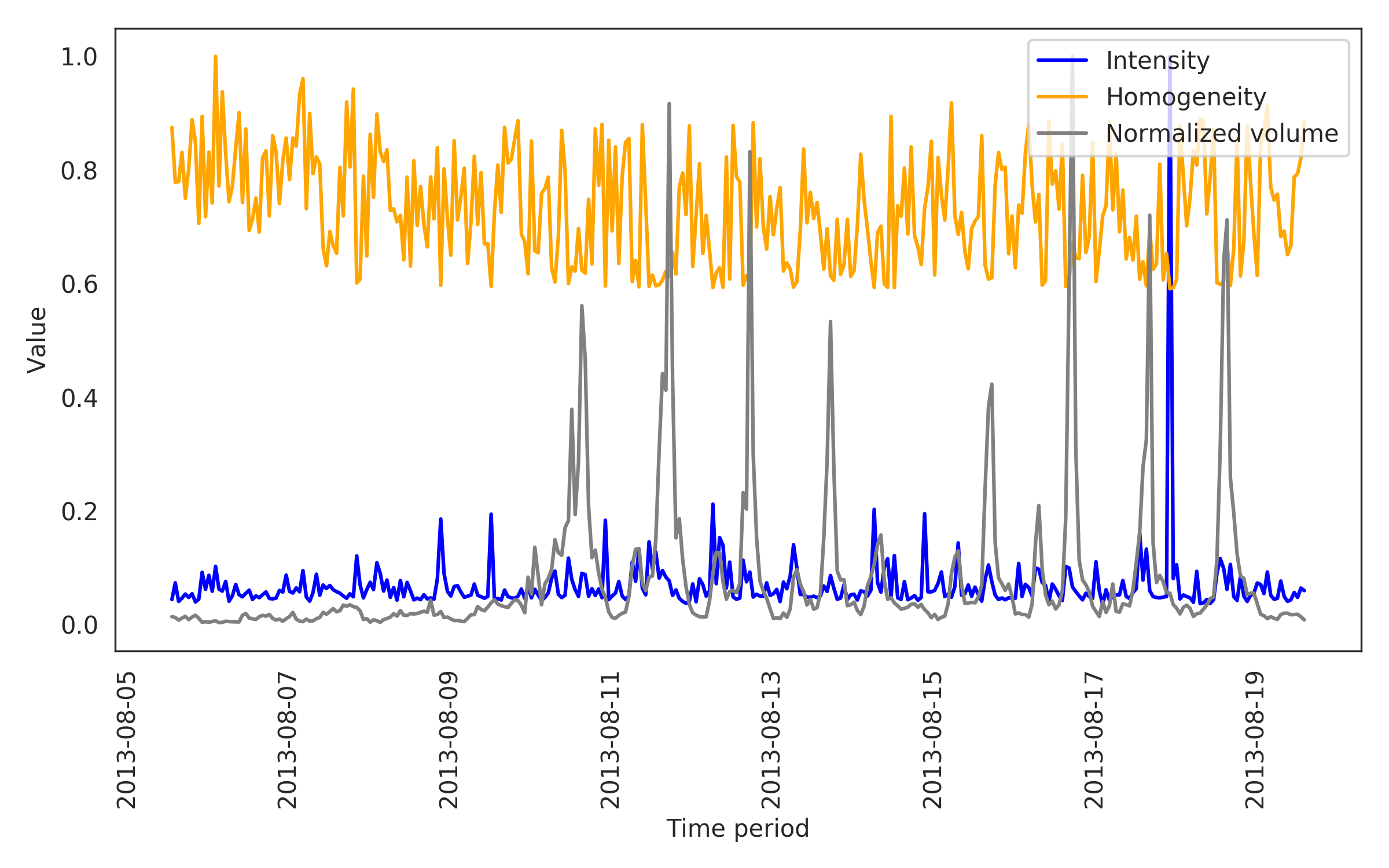}} \\
\begin{tabular}{cc}
\subcaptionbox{\textit{MoscowAthletics2013}, static window of 6h}{\includegraphics[width = 0.45\linewidth]{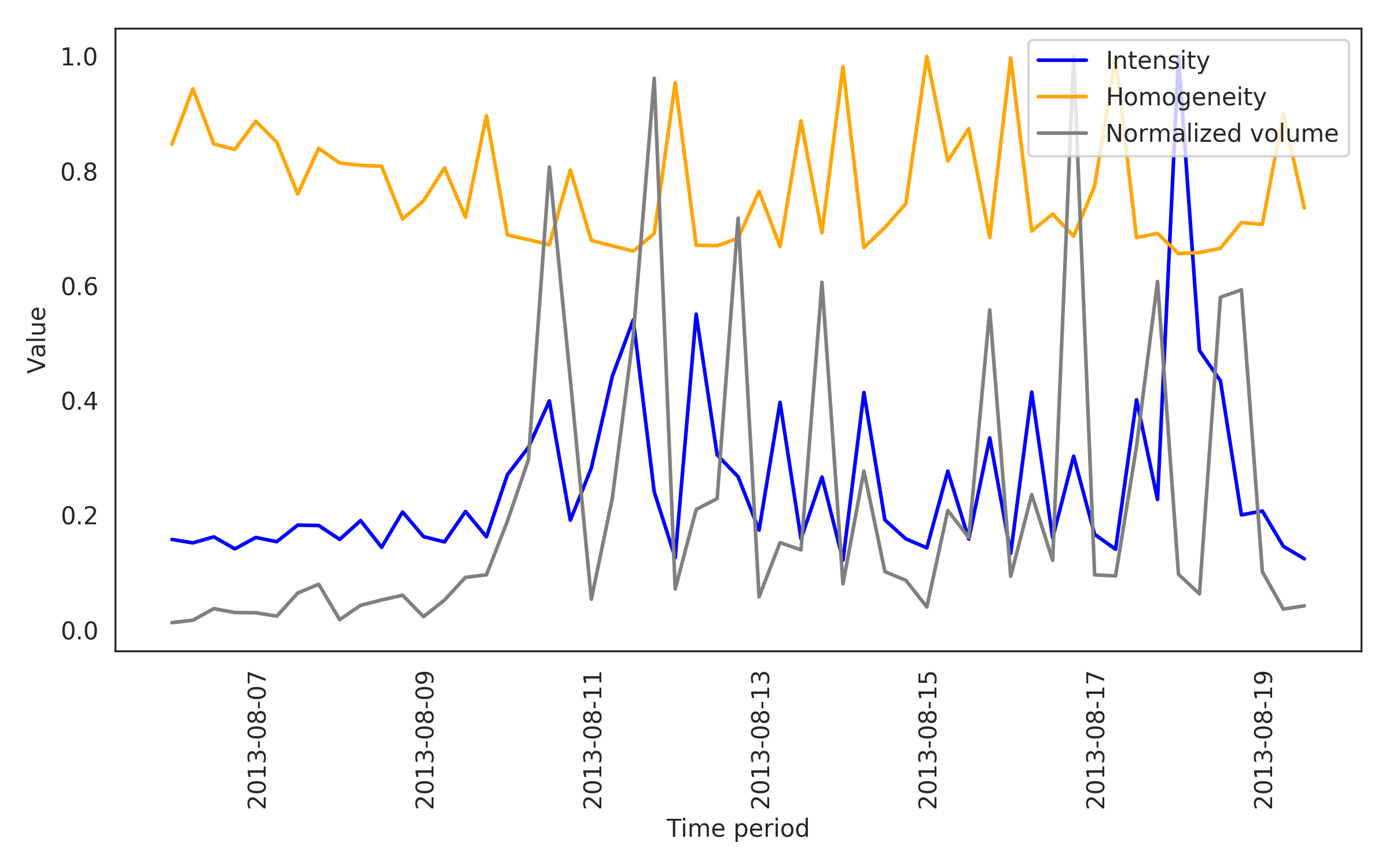}} &
\subcaptionbox{\textit{MoscowAthletics2013}, sliding window of 6h, \REVISE{with 1h steps}}{\includegraphics[width = 0.45\linewidth]{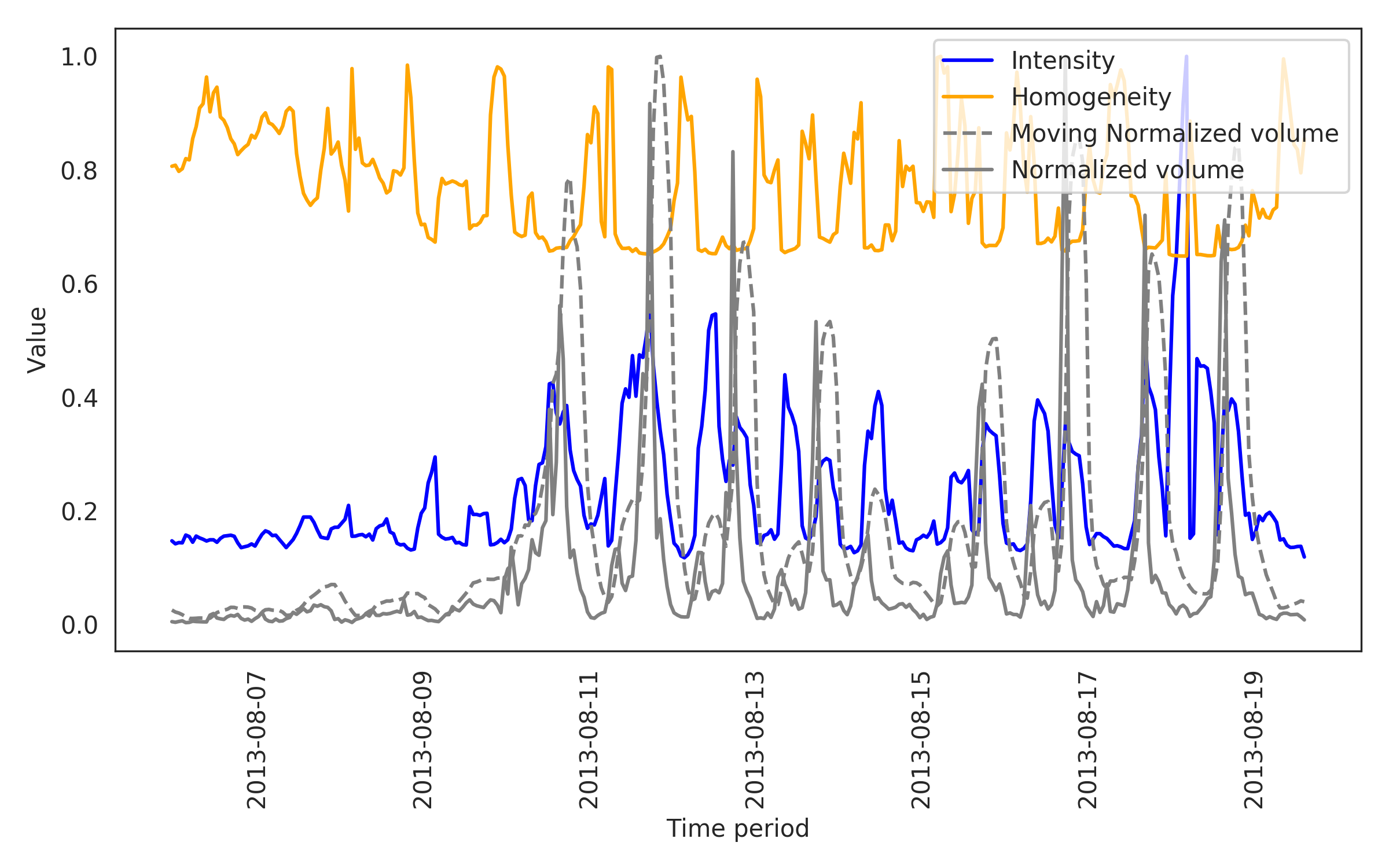}}
\end{tabular}
\caption{Visualization of temporal entanglement across \textit{MoscowAthletics2013}. In grey, volume over the period of time (dotted line for the aggregated volume over sliding window (c)). Intensity in blue and homogeneity in yellow.}
\label{fig:moscow}
\end{figure}

\begin{figure}[ht!]
\centering
\captionsetup{width=.90\linewidth}
\subcaptionbox{\textit{Cannes2013}, static window of 1h }{\includegraphics[width = 0.45\linewidth]{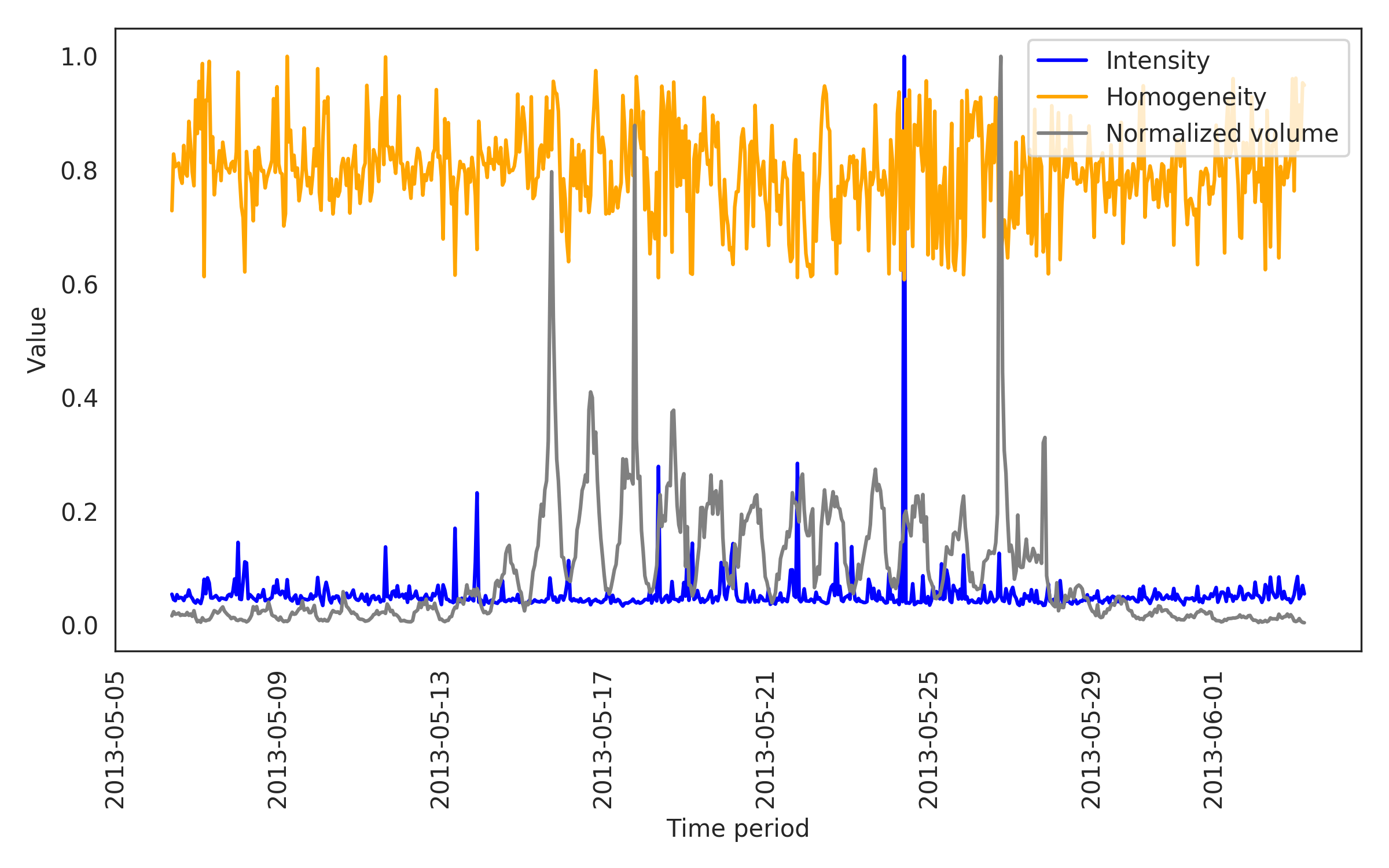}} \\
\begin{tabular}{cc}
\subcaptionbox{\textit{Cannes2013}, static window of 12h}{\includegraphics[width = 0.45\linewidth]{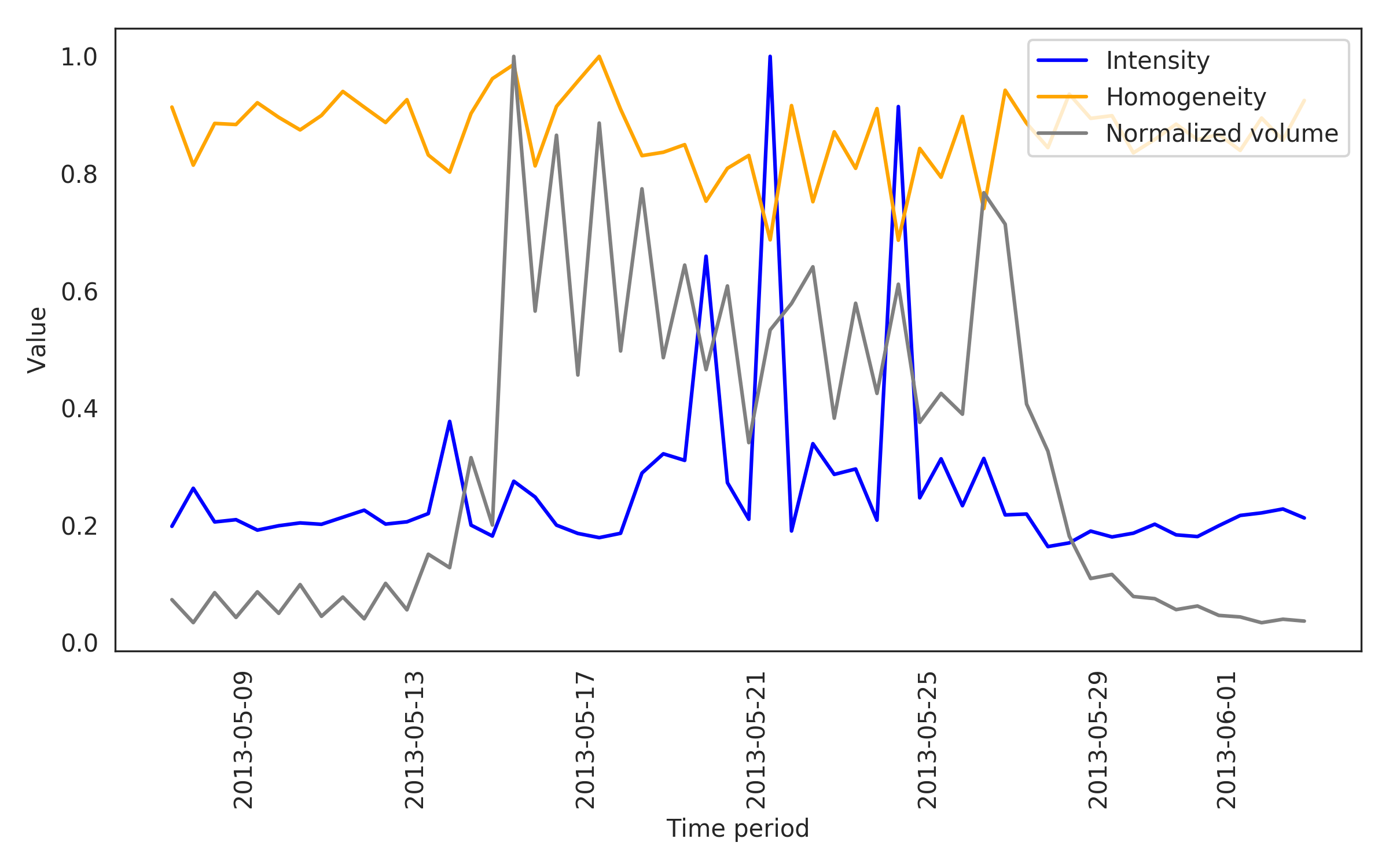}} &
\subcaptionbox{\textit{Cannes2013}, sliding window of 12h, \REVISE{with 1h steps}}{\includegraphics[width = 0.45\linewidth]{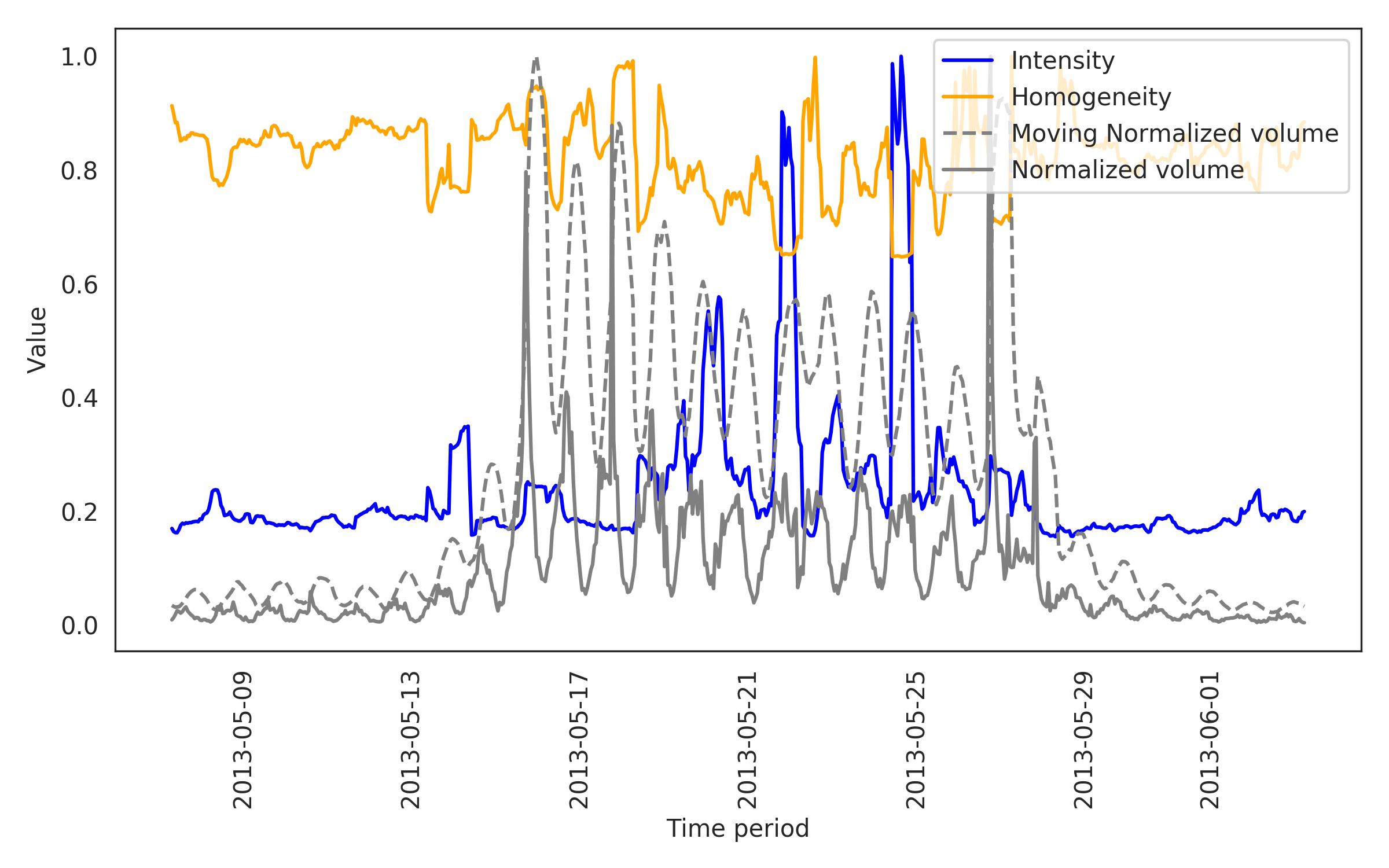}}
\end{tabular}
\caption{Visualization of temporal entanglement across \textit{Cannes2013}. In grey, volume over the period of time (dotted line for the aggregated volume over sliding window (c)). Intensity in blue and homogeneity in yellow.}
\label{fig:cannes}
\end{figure}

The first experiment considers fixed time windows of sizes 1h, 3h, 6h, and 12h. We compare with the activity volume in form of a total number of tweets -- as found in \cite{omodei2015characterizing}, Figure 1 for a 1h window size, here reported in Figures~\ref{fig:mlking}a,~\ref{fig:moscow}a, and~\ref{fig:cannes}a. We normalise here this volume so values are in $[0, 1]$.

We selected the coarse windows at their best readability for each dataset (3h for \textit{MLKing2013} in Figure~\ref{fig:mlking}b, 6h for \textit{MoscowAthletics2013} in Figure~\ref{fig:moscow}b, and 12h for \textit{Cannes2013} in Figure~\ref{fig:cannes}b) -- each coarsening is further illustrated in Appendix. 
A second experiment considers a moving window of the size corresponding to these best windows, sliding by the hours (Figures~\ref{fig:mlking}c,~\ref{fig:moscow}c, and~\ref{fig:cannes}c).

In the \textit{MLKing2013} data set (Figure~\ref{fig:mlking}), we can observe that spikes of intensity surround the main spike of volume activity. A smaller spike of intensity consistently coincides with a smaller spike of volume at the end of the main spike.

In the \textit{MoscowAthletics2013} data set (Figure~\ref{fig:moscow}), the 1h-time window does not show a consistent behaviour. However, we can see that spikes in coarser time windows coincide with the spikes in volume. A larger spike in intensity appears before the final spike in volume.

In the \textit{Cannes2013} data set (Figure~\ref{fig:cannes}), the 1h-time window shows some spikes in intensity, especially a major by the end of the period of activity in terms of volume. In coarser time windows, we can notice four main spikes: one before the beginning of volume of activity; the next two ones appear just before a slight increase in the daily volume; the last one appears the day before the last day of the volume activity. This last peak appears even more prominent from the sliding window example.

The volume captures Twitter activity, governed by the human activity following the day/night rhythm. Although entanglement intensity is also submitted to it, we see emerging patterns that seem proper to each type of event. The activity of entanglement shows definitely some relationship with volume while telling a different story. The sports event that is \textit{MoscowAthletics2013} may be much more subject to the day-by-day routine in which different disciplines are at play. On the other hand, the speech celebration in \textit{MLKing2013} has some very specific activity before (could it be anticipation?) and after (could it be ripples?) the event. The movie festival in \textit{Cannes2013} may be governed by sub-events of different importance in terms of networking activity. 

In accordance with the position of social networks in our evaluation of real-world networks in Section~\ref{sec:results-comparison}, we see a decrease in homogeneity whenever we see spiking of intensity. This may indicate that a lot of the network activity suddenly focuses on one specific modality of exchange (such as \textit{replies}). Entanglement study may help in targeting when this is driven by a particular modality.

Further studies on the nature of the events, and the specific topologies of the $LIN$ networks that gave rise to these entanglement values is necessary for a more in-depth analysis of each case. Since we see some spiking activity of entanglement before actual events took place, we may suspect that, beyond monitoring, there is a \textit{predictive power} of modelling time series from entanglement in past data (sliding windows).

\section{Discussion and conclusions}

In this work, we have revisited the notion of layer entanglement and extended it to coupled multilayer networks and temporal networks. To investigate entanglement, we have proposed a random generator for coupled multilayer networks, and generated a large set of synthetic ones. We have evaluated entanglement intensity and homogeneity in all cases, and compared to static and temporal real world networks.

Our analysis of the synthetic networks outlined that entanglement intensity is directly correlated with edge probability parameter -- the sparser the network, the lower the intensity. This result indicates the proposed generator indeed emits networks which adhere to this property. We have also observed that large parts of the generated networks are subject to high homogeneity with various degrees of entanglement intensity.

\REVISE{Entanglement in the synthetic networks appears very sensitive to the different probabilities characterising the model ($o$, $p$, and eventually $q$ for the coupled multilayer networks). The influence of each parameter should also be investigated theoretically in future work.}

The high homogeneity observed may be a byproduct of our computations. First, our random generation \REVISE{induces a lot of small connected components of the coupled multilayer networks, and small components tend to show higher homogeneity since there are not so many degrees of freedom for edges to overlap}. Because we are averaging the entanglement intensity and homogeneity over all components, this may go in favour of high homogeneity. Understanding this effect deserves more investigation. Second, entanglement homogeneity is a cosine measure, and the observed values may suffer from the skewness of cosine values when distributed in a linear space, amplifying the effect of having large values. Furthermore, it might also suffer from the curse of dimensionality in the case of a high number of layers. It would be worth considering normalizing this homogeneity with respect to the number of layers involved and the number of edges they cover. \REVISE{Instead of cosine, a Shannon's entropy measure may overcome some of these limitations.}

\REVISE{One of the aspects that was not extensively evaluated as a part of this work is the processing of the repeated links in a given time slice. The current implementation considers, for each time slice, the collection of \textit{unique} links, which are not weighted by their possible multiple occurrences. This way, the diversity of connections is emphasized, instead of link frequency. A more detailed study of how the links can be re-weighted will be considered in future work.}

We further demonstrated that the two measures offer interesting insights when computed across a wide array of real-world networks. The observed relationship between the intensity and homogeneity of layer entanglement with the family of dataset was previously reported for clusters of documents (in~\cite{renoust2013measuring}, Figure 5). In this previous experiments, clusters of documents were mostly located at the left frontier of high intensity for a varying homogeneity. Our current experiments showed that real networks cluster based on their type (\textit{e.g.} biological \textit{vs.} social), also close to this frontier. \REVISE{We have observed (from Figure~\ref{fig:real_overlay}) that the set of genetic networks tend to sit in areas with low entanglement intensity, which could correspond to lower edge probability $p$, but they also tend to show a wider span of entanglement homogeneity including our lowest values measured (from Figure~\ref{fig:real}), which could correspond to lower layer assignment probability $o$. \REVISE{Further work should be invested on finding the reason why genetic networks tend to show lower homogeneity.} This is opposed to social networks which tend to find their way in the higher probability area.} This should be further investigated, but this may be related to \textit{homophily}~\cite{mcpherson2001birds,borgatti2009network}. Homophily is the implied similarity of two entities in a social network, and the property of entities to agglomerate when \textit{being similar}. If the reason of \textit{`being similar'} could be modelled as a layer of interaction, the result of a group of entities in \textit{`being similar'} would lead to the formation of a clique in this layer, hence locating social networks in \REVISE{high probability areas}. 

The proposed work offers at least two prospects of multiplex network study which are in our belief worth exploring further. The difference between the genetic and social networks is possibly subject to very distinct topologies which emerge in individual layers. This claim may further be investigated via other measurements, such as graphlets, communities or other structures. Next, genetic networks are less homogeneous. Future work includes exploration of this fact, as it can be merely a property of the networks considered, empirical methodology used to obtain the networks or some other effect.

We believe that theoretical properties of the proposed network generator can also be further studied, offering potential insights into how multiplex networks behave and whether the human-made aspects are indeed representative of a given system's state. The model that we are currently exploring only takes into account a probability of linkage through (or within) layers without guarantee of connectivity. We made this choice to be able to compare between different fields, without prior assumption which could, for example, rule in favour of similarity to social network. Our future work will investigate other generation models including Erd\H{o}s-Rényi-based~\cite{caimo2020multilayer} or other with preferential attachment~\cite{nicosia2014nonlinear}.

The analysis of real-life temporal networks offers cues on evolution in layer entanglement which can happen prior to some other events. We have tested multiple time scales. 
Too small time windows mostly result in noisy time series carrying low amounts of useful information, while higher coarsening shows activity related to volume, but with a different light on the events that are captured. Future work will dive deeper into these events, and consider testing entanglement as a predictor using approaches such as of Prophet~\cite{taylor2018forecasting}.

When considering entanglement as a either a monitoring or a predictive variable, its utility largely depends on the time scale at which a given edge stream needs to be considered. We leave extensive, possibly automatic determination of a setting where entanglement would be of practical relevance for future work. 
To study the parameters driving the dynamics of entanglement in temporal networks, we will consider comparing entanglement measures with synthetic temporal networks in our future investigations.

\section*{Availability}
The code for reproduction of experiments is freely available at \url{https://gitlab.com/skblaz/entanglement-multiplex}.

\section*{Acknowledgements}
The work of the first author was funded by the Slovenian Research Agency through a young researcher grant.
The work of other authors was supported by the Slovenian Research Agency (ARRS) core research programme \textit{Knowledge Technologies} (P2-0103) and ARRS funded research project
\textit{Semantic Data Mining for Linked Open Data} (financed under the ERC Complementary Scheme, N2-0078). We also acknowledge Dagstuhl seminar 19061 where many ideas implemented in this paper emerged.

\section*{Competing interests}
The authors declare that they have no competing interests.

\section*{Author's contributions}
Both authors have contributed equally to the theoretical background, design of the experiments, and the writing of the manuscript. BR contributed to experiments, but the most of the experiments was handled by BS.

\bibliography{reference}
\bibliographystyle{bmc-mathphys}

\newpage
\section*{Appendix}

\subsection*{Dependency in synthetic networks over nodes and layers}

\begin{figure}[h!]
\centering
\captionsetup{width=.90\linewidth}
\begin{tabular}{cc}
\subcaptionbox{Transition $H \times n$}{\includegraphics[width = 0.45\linewidth]{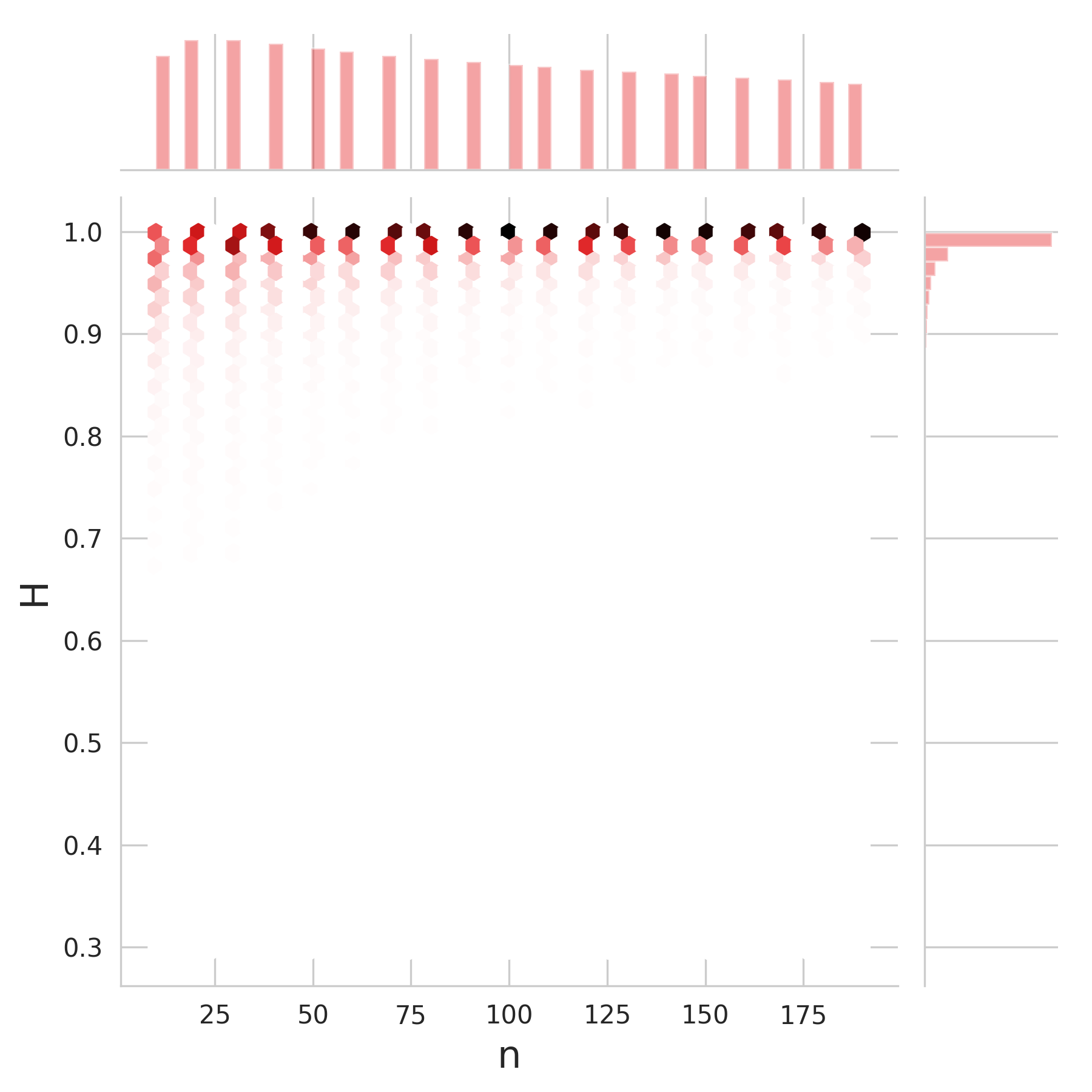}} &
\subcaptionbox{Transition $I \times n$}{\includegraphics[width = 0.45\linewidth]{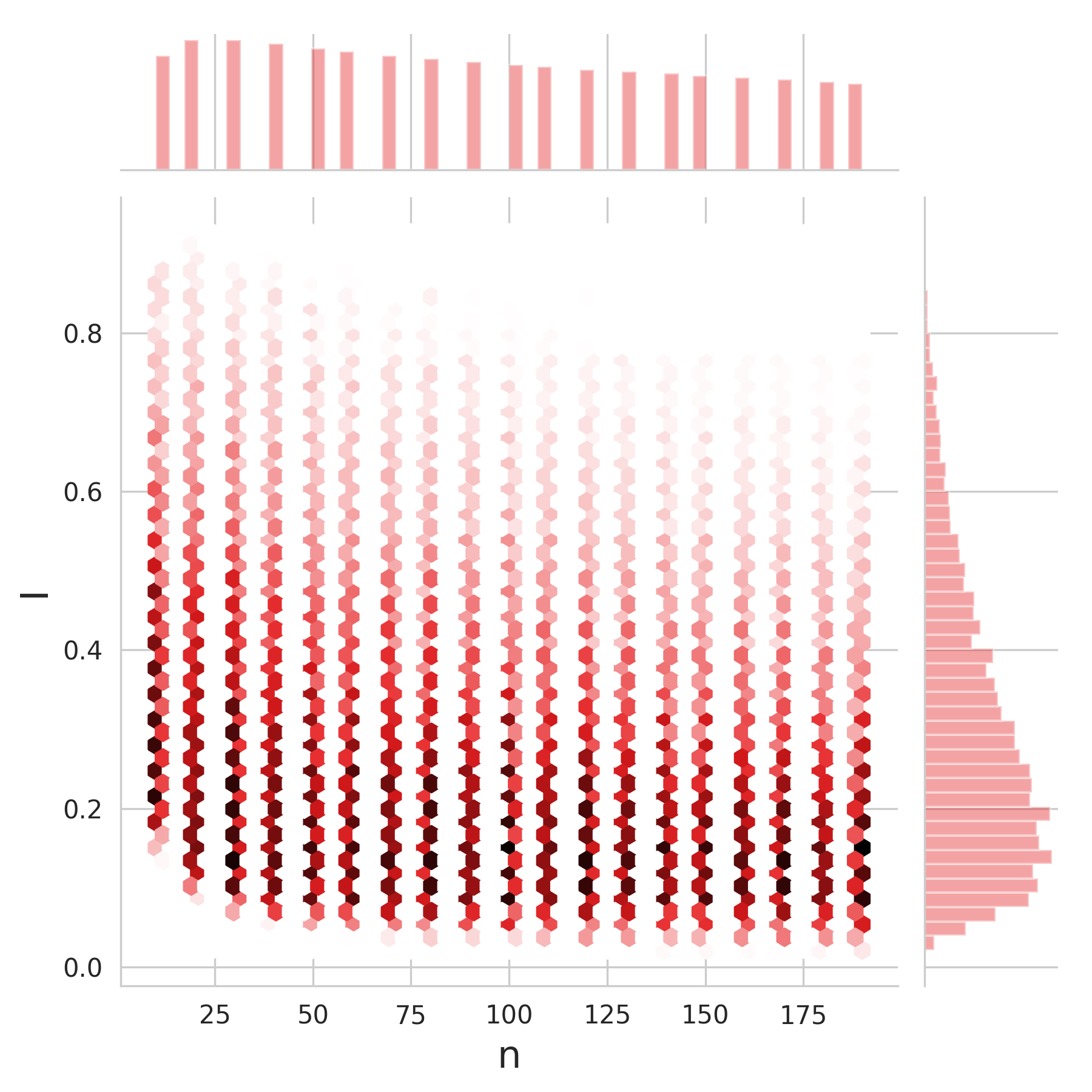}} \\
\subcaptionbox{Transition $H \times m$}{\includegraphics[width = 0.45\linewidth]{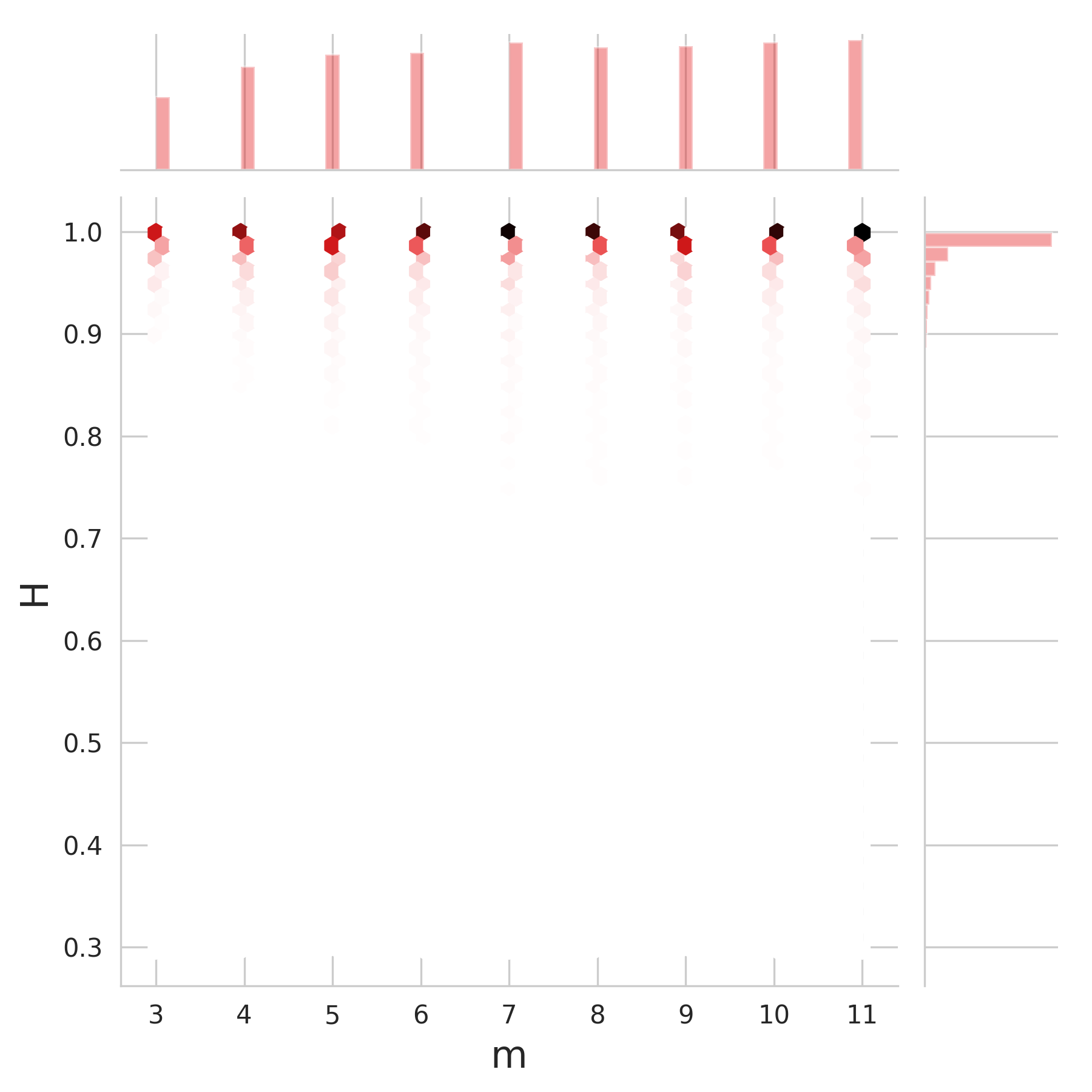}} &
\subcaptionbox{Transition $I \times m$}{\includegraphics[width = 0.45\linewidth]{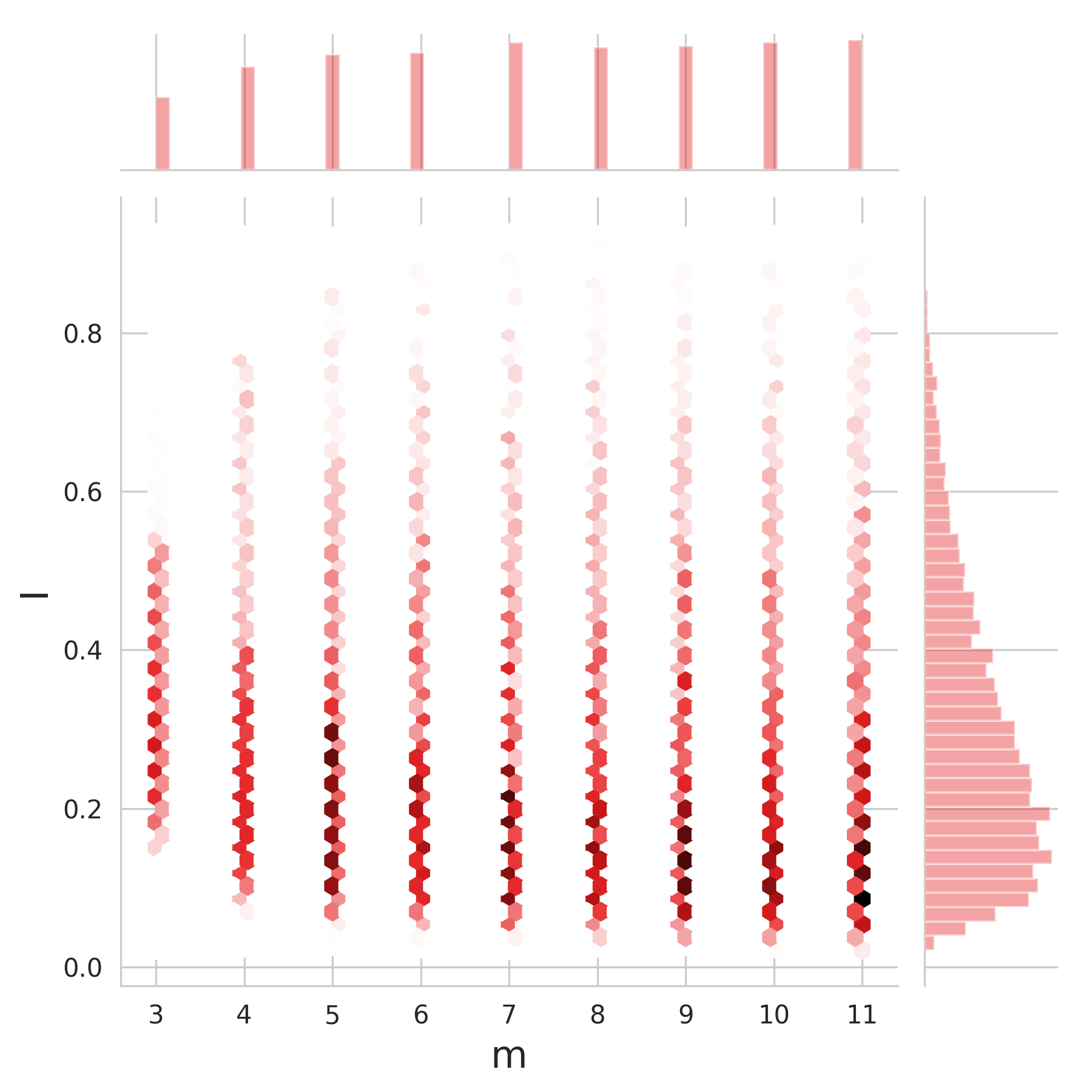}} \\
\end{tabular}
\caption{Dependency on the number of nodes and layers on the \REVISE{transition coupling} entanglement.}
\label{fig:apx_inter}
\end{figure}

One can predict of course a level of dependency over the number of nodes $n$ and layers $m$ for the \REVISE{transition coupling} case too. The dependency tends towards lower entanglement values since when increasing the number of nodes and layers, we increase the degree of freedom for layers to overlap. This trend, first illustrated in Figure~\ref{fig:inner_nm}, is confirmed in Figure~\ref{fig:apx_inter}.

\newpage
\subsection*{Independence of parameters}

\begin{figure}[h!]
\centering
\captionsetup{width=.90\linewidth}
\begin{tabular}{cc}

\subcaptionbox{Elementary $H \times q$}{\includegraphics[width = 0.45\linewidth]{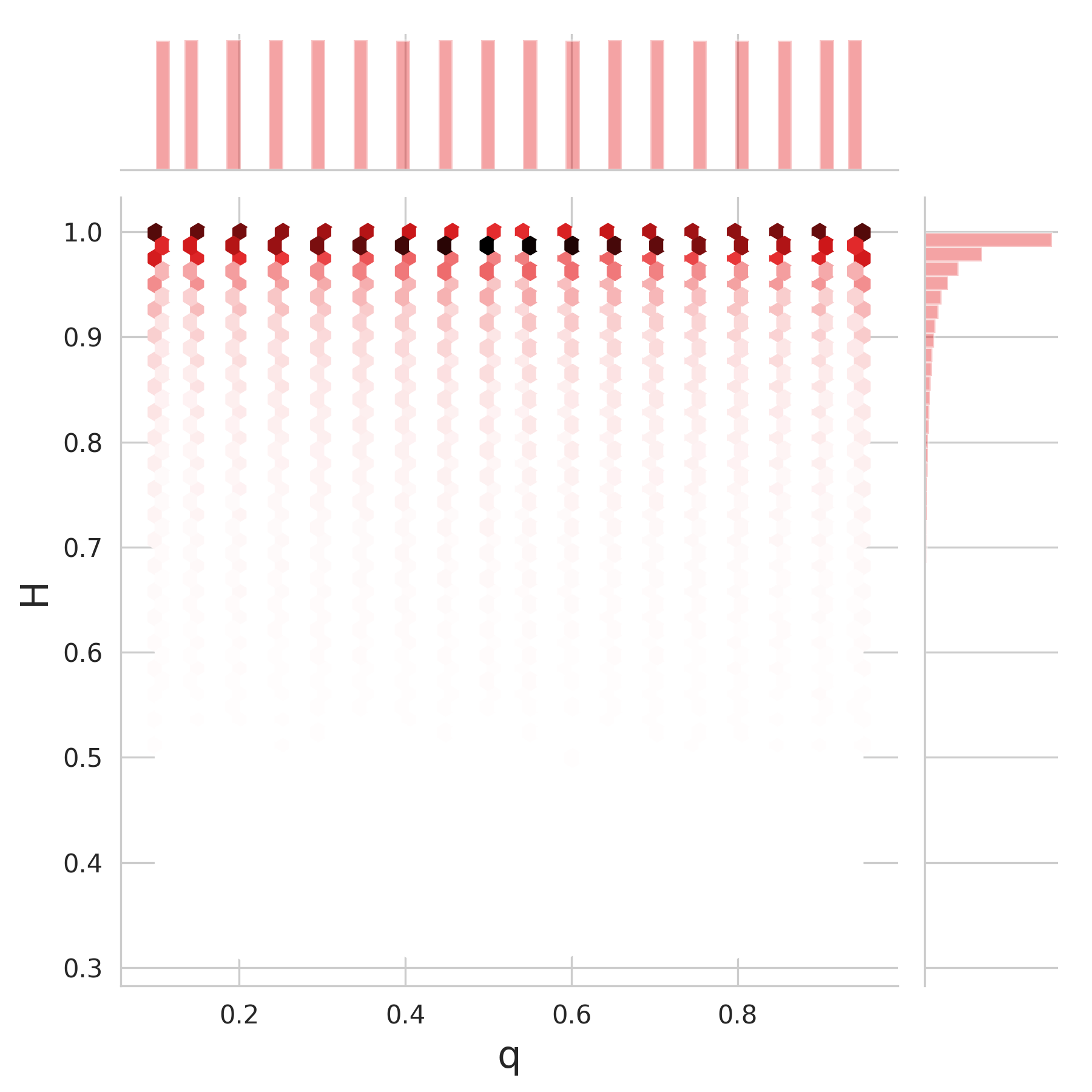}} &
\subcaptionbox{Elementary $I \times q$}{\includegraphics[width = 0.45\linewidth]{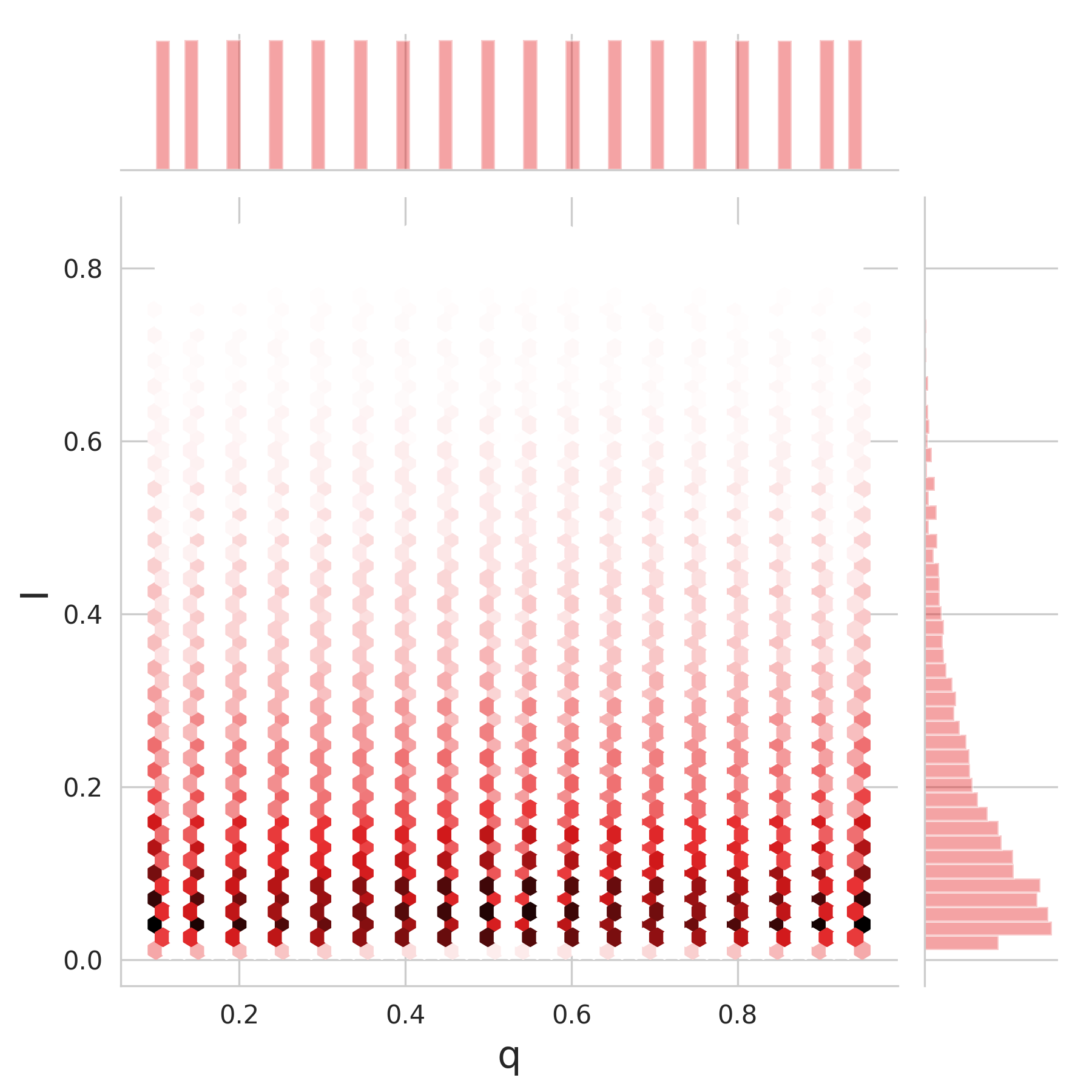}} \\
\subcaptionbox{Transition $H \times p$.}{\includegraphics[width = 0.45\linewidth]{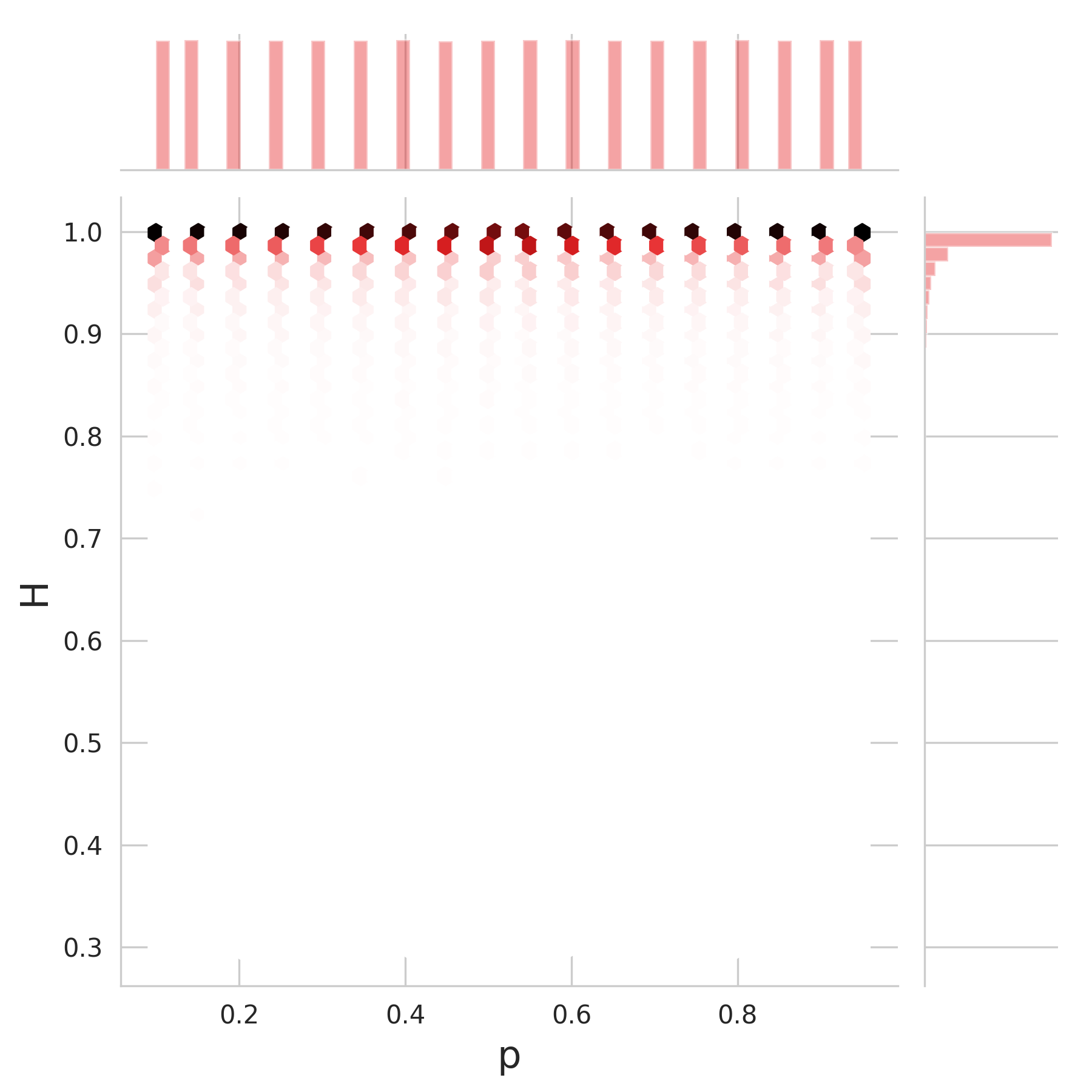}} &
\subcaptionbox{Transition $I \times p$.}{\includegraphics[width = 0.45\linewidth]{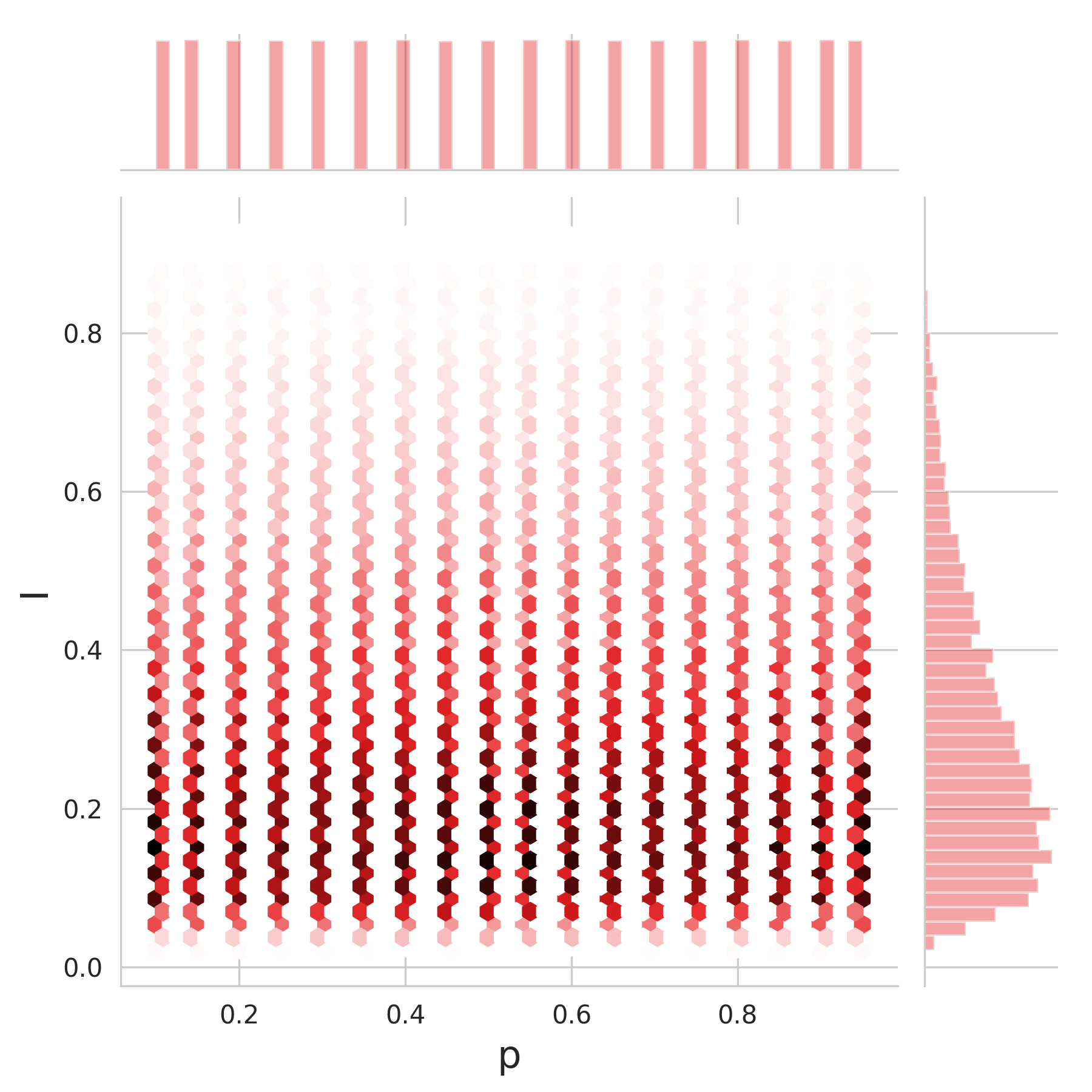}} \\

\end{tabular}
\caption{From the computation of entanglement over elementary layers (a, c, d) and \REVISE{transition coupling} (b, e, f), we see no dependency on parameters $p$ and $q$.}
\label{fig:syn_nodependency}
\end{figure}

The distribution of parameters of \REVISE{transition coupling} intensity and homogeneity over parameter $p$, and elementary layer intensity and homogeneity over parameter $q$, show no dependency as illustrated in Figure~\ref{fig:syn_nodependency}.

\newpage
\subsection*{Combining both elementary and \REVISE{transition coupling}}

\begin{figure}[h!]
\centering
\captionsetup{width=.90\linewidth}
\begin{tabular}{cc}
 \subcaptionbox{Lower combined $I$}{\includegraphics[width = 0.45\linewidth]{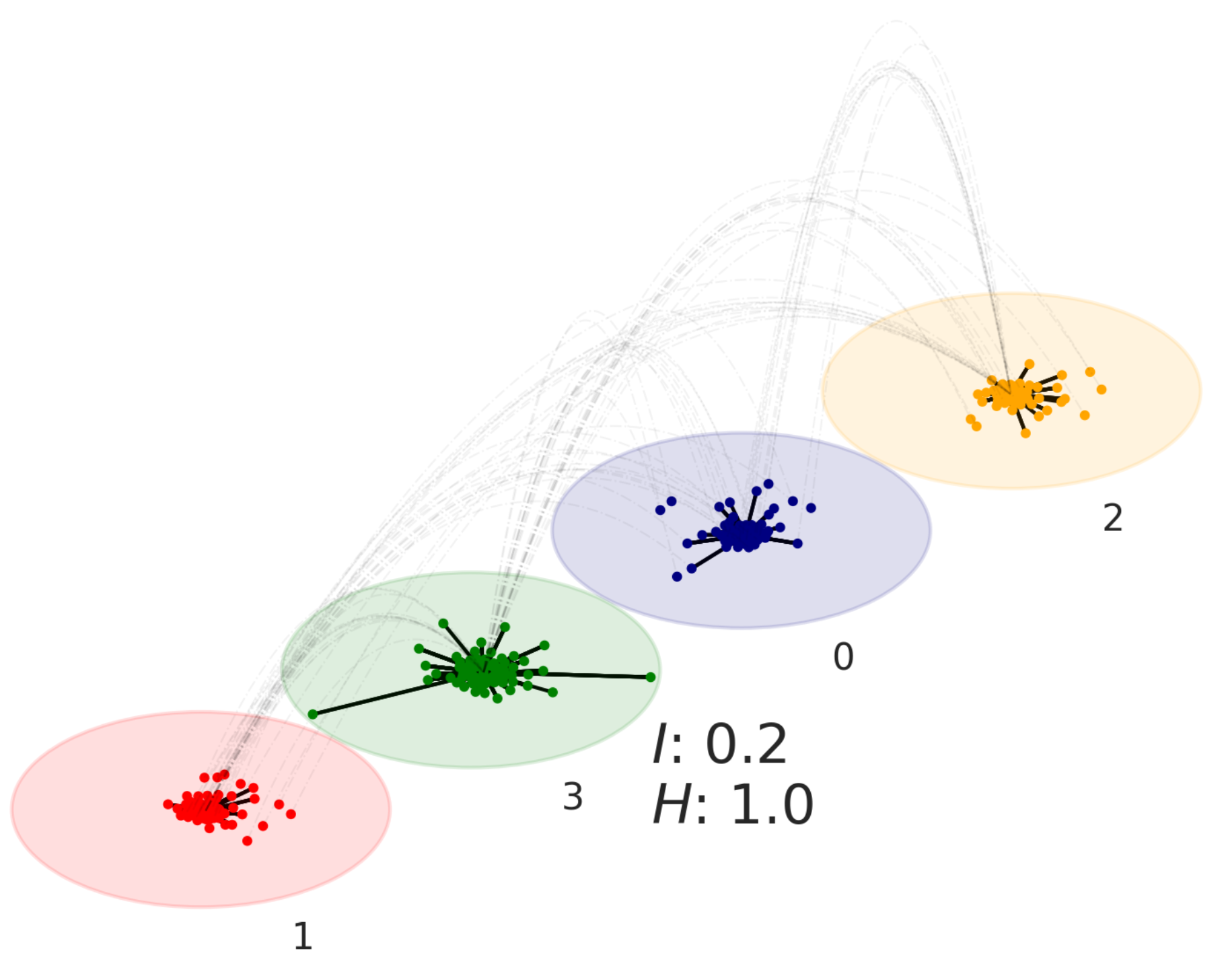}} &
 \subcaptionbox{Higher combined $I$}{\includegraphics[width = 0.45\linewidth]{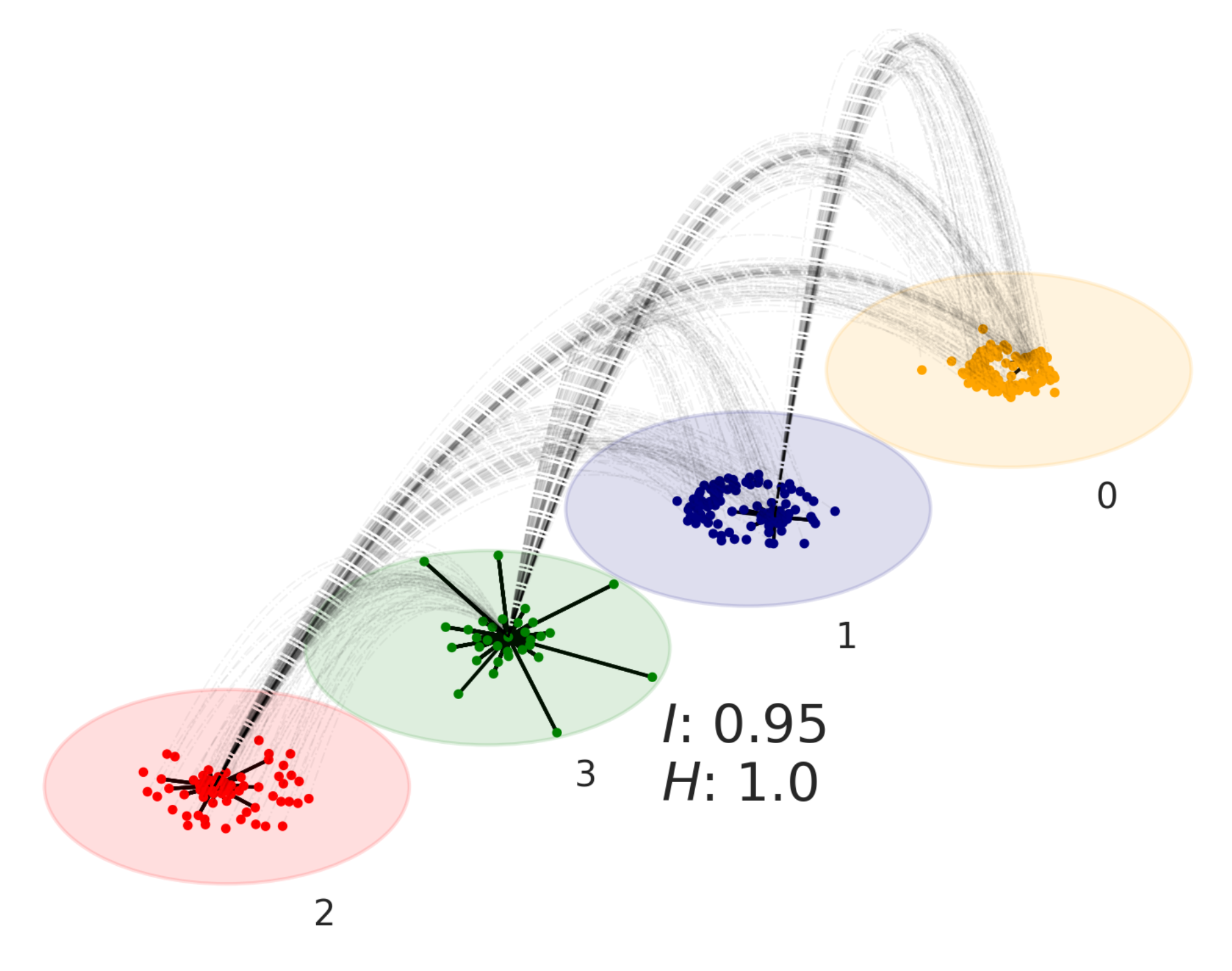}} \\
\end{tabular}

\caption{Visualization of both inner-layer and coupling edges in synthetic coupled multilayer networks.}
\label{fig:apx_illustration}
\end{figure}

\begin{figure}[h!]
\centering
\captionsetup{width=.90\linewidth}
\includegraphics[width = 0.75\linewidth]{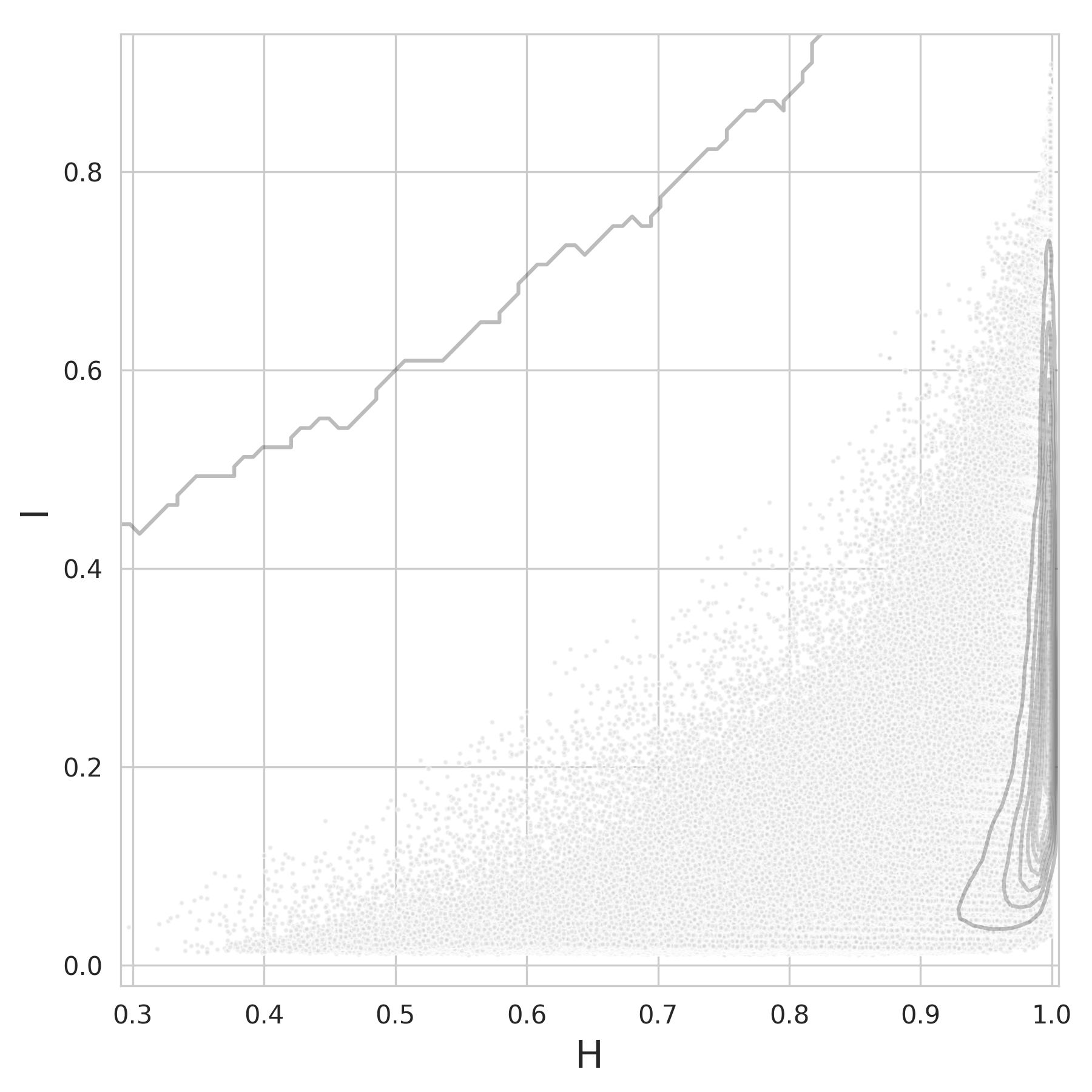}

\caption{Homogeneity and intensity $H \times I$ results on 1{,}329{,}696 synthetic multiplex networks considering their combined elementary layers and \REVISE{transition coupling} with density lines (Gaussian kernel density estimation).}
\label{fig:apx_t3plots}
\end{figure}

\begin{figure}[h!]
\centering
\captionsetup{width=.8\linewidth}
\begin{tabular}{cc}
\subcaptionbox{Combined $H \times n$}{\includegraphics[width = 0.45\linewidth]{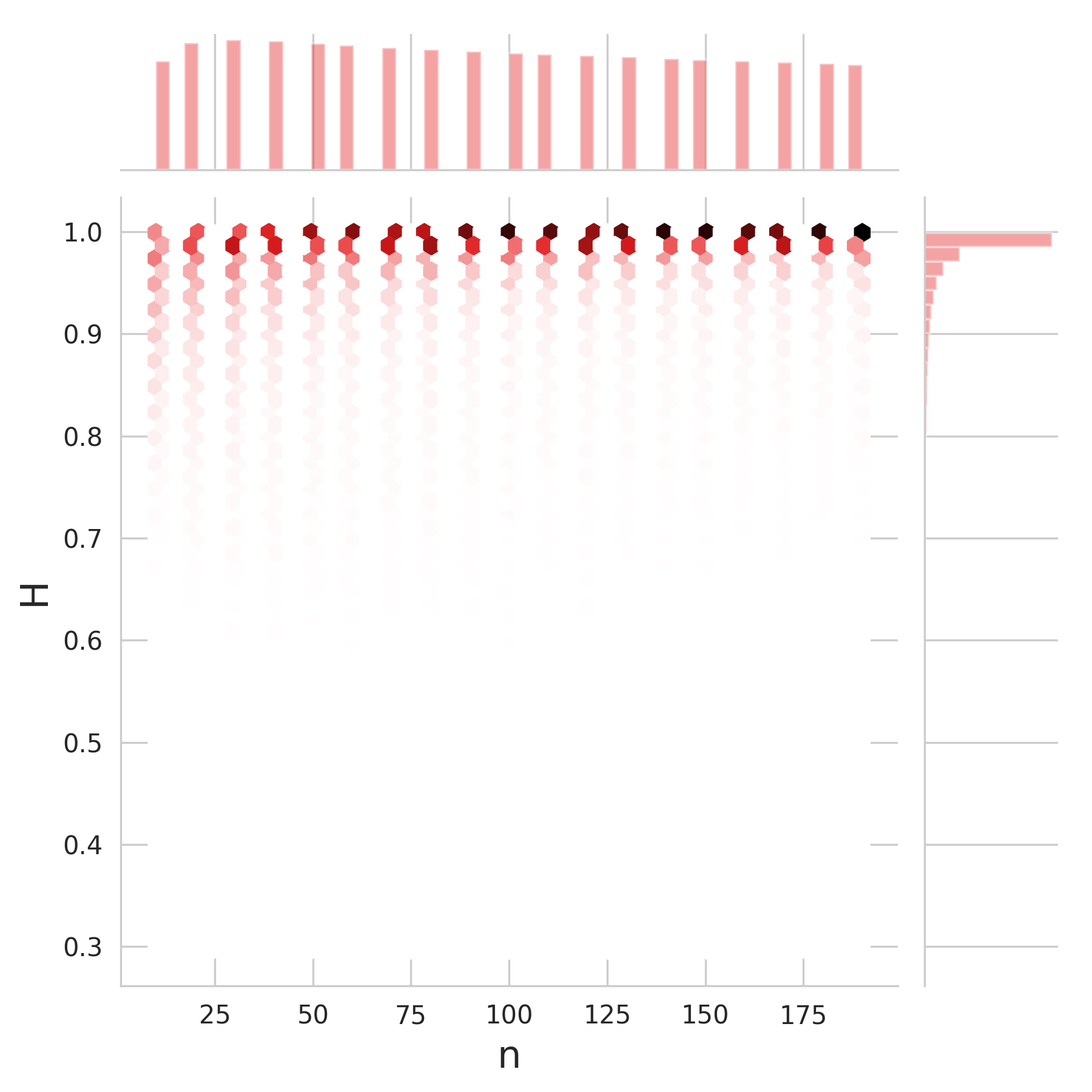}} &
\subcaptionbox{Combined $I \times n$}{\includegraphics[width = 0.45\linewidth]{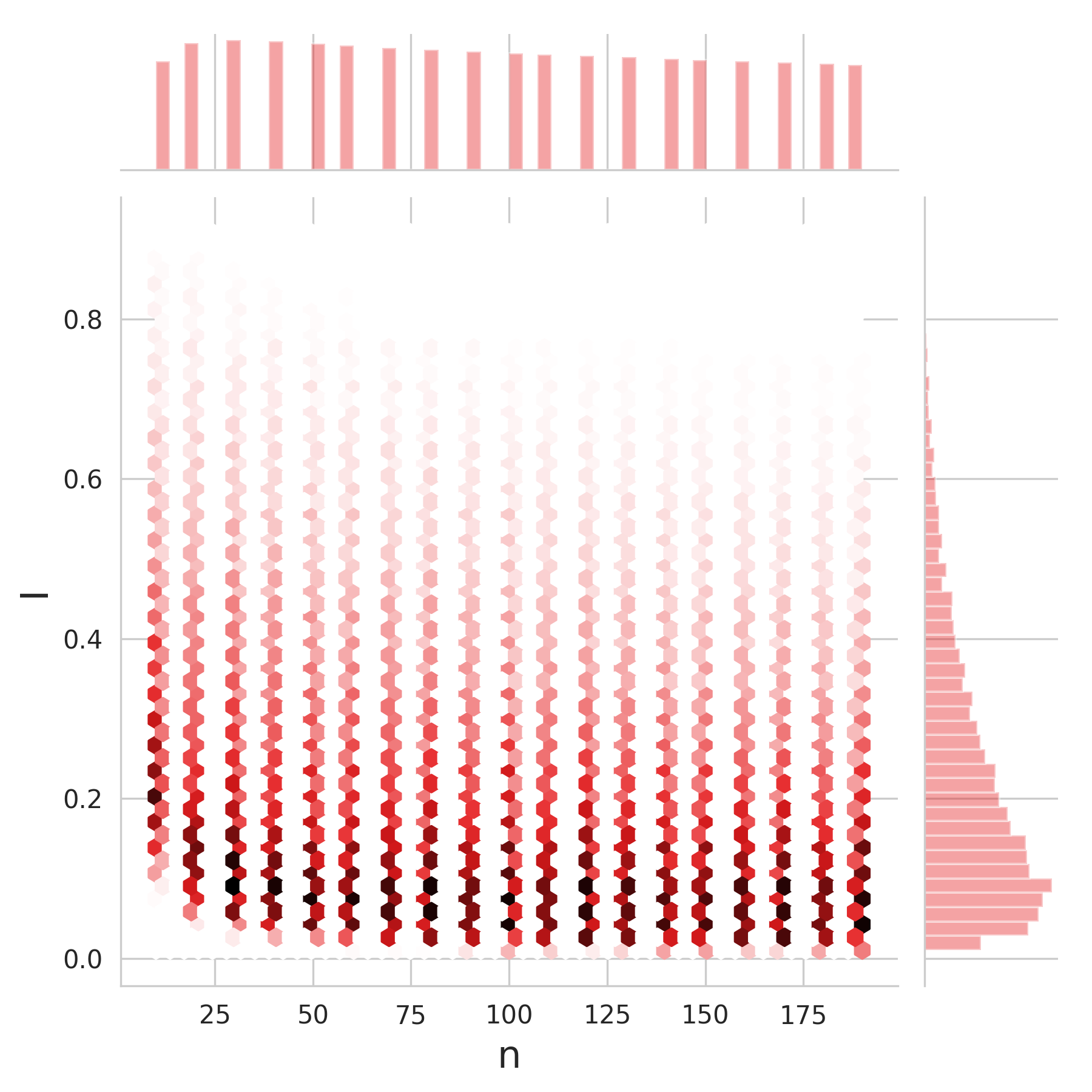}} \\
\subcaptionbox{Combined $H \times m$}{\includegraphics[width = 0.45\linewidth]{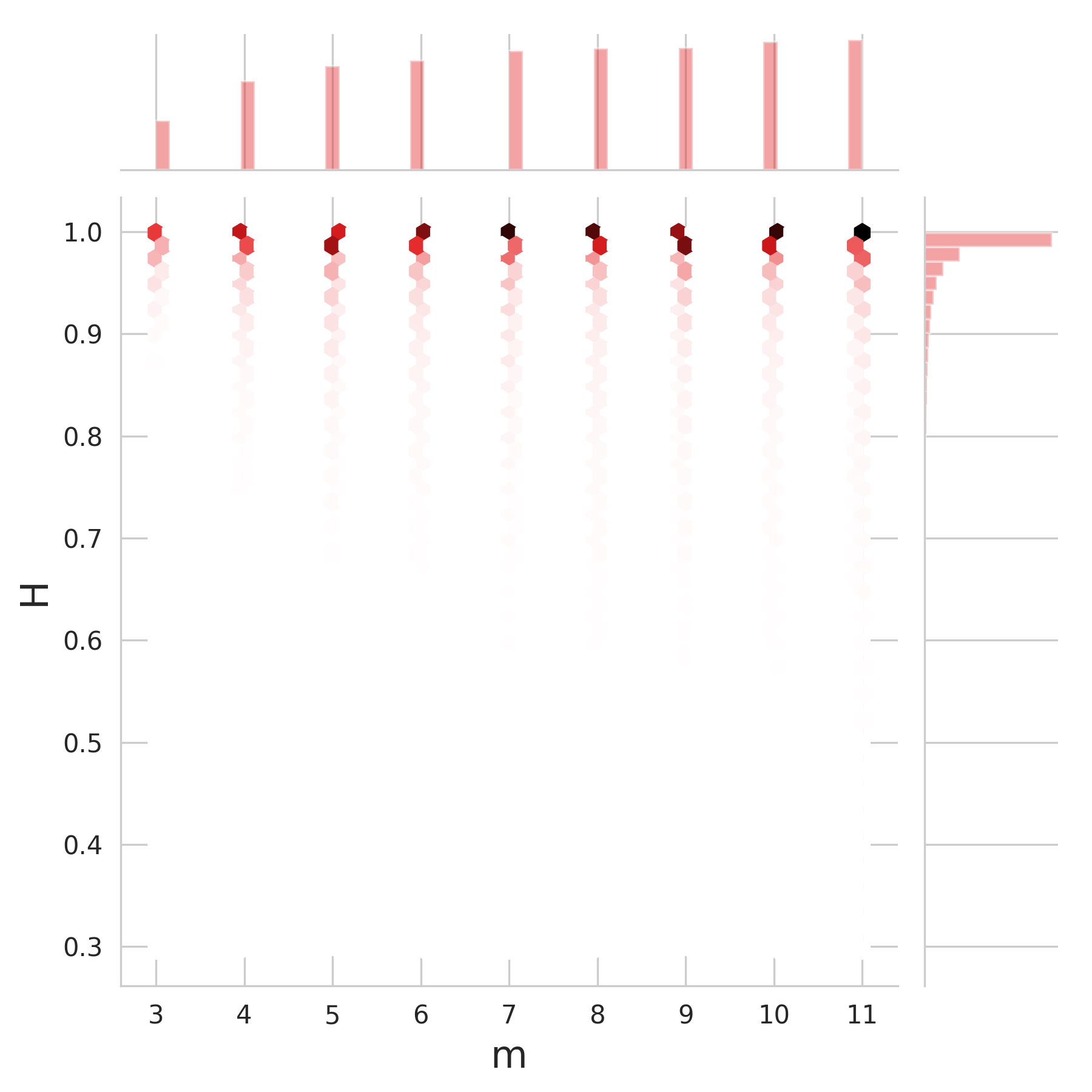}} &
\subcaptionbox{Combined $I \times m$}{\includegraphics[width = 0.45\linewidth]{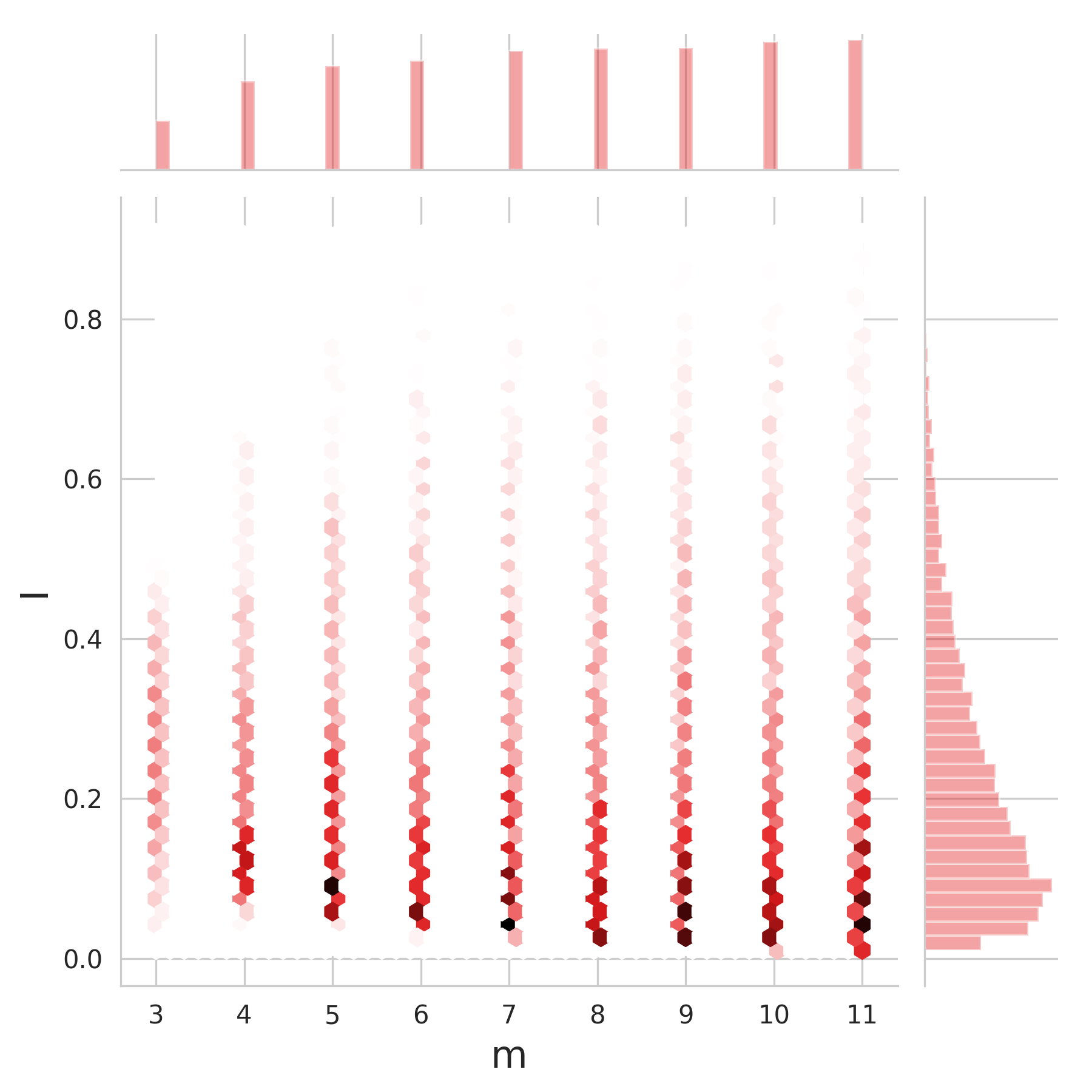}} \\
\end{tabular}
\caption{Dependency on the number of nodes and layers on the combined layers entanglement.}
\label{fig:apx_t3hives1}
\end{figure}

\begin{figure}[h!]
\centering
\captionsetup{width=.90\linewidth}
\begin{tabular}{cc}
\subcaptionbox{Combined $H \times o$}{\includegraphics[width = 0.45\linewidth]{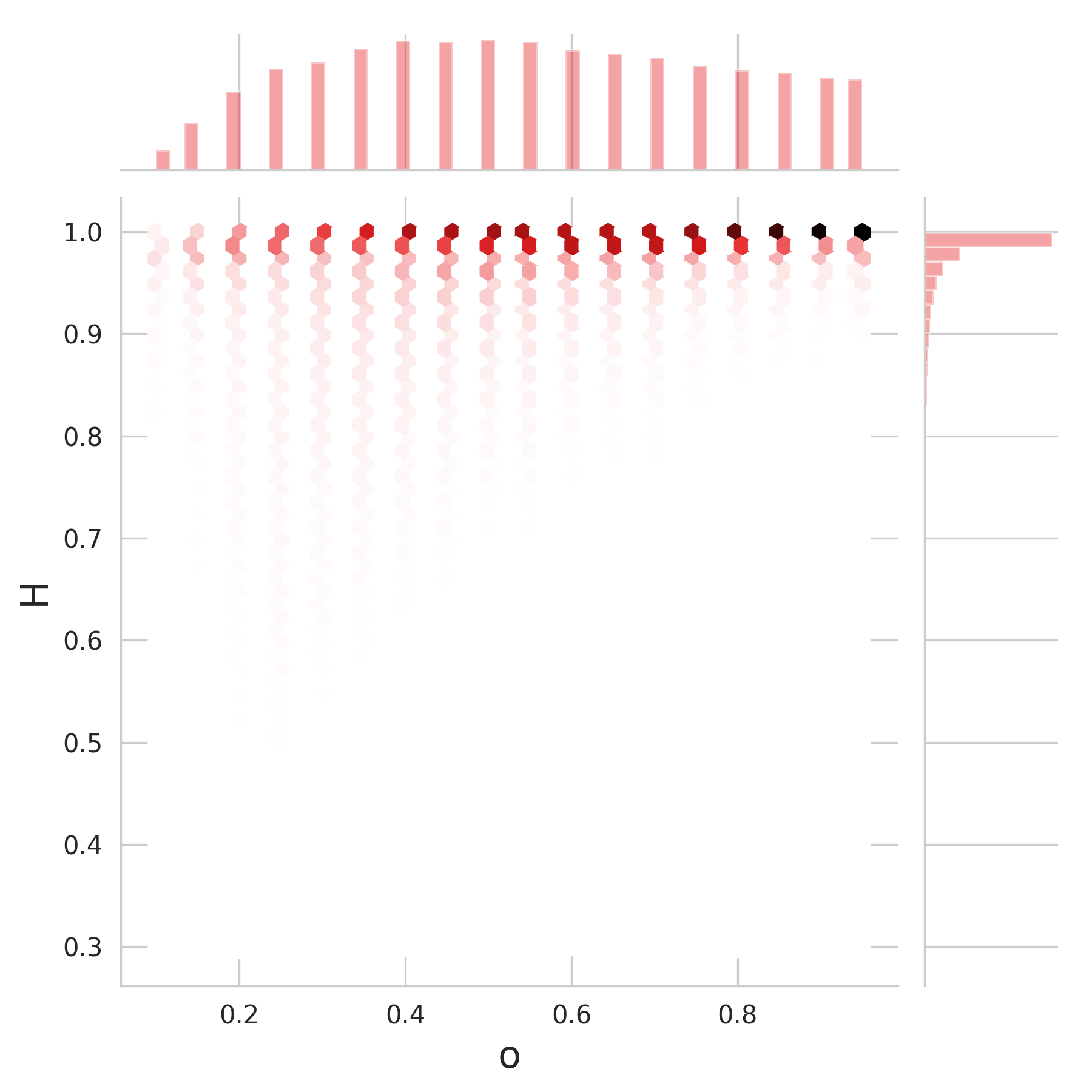}} &
\subcaptionbox{Combined $I \times o$}{\includegraphics[width = 0.45\linewidth]{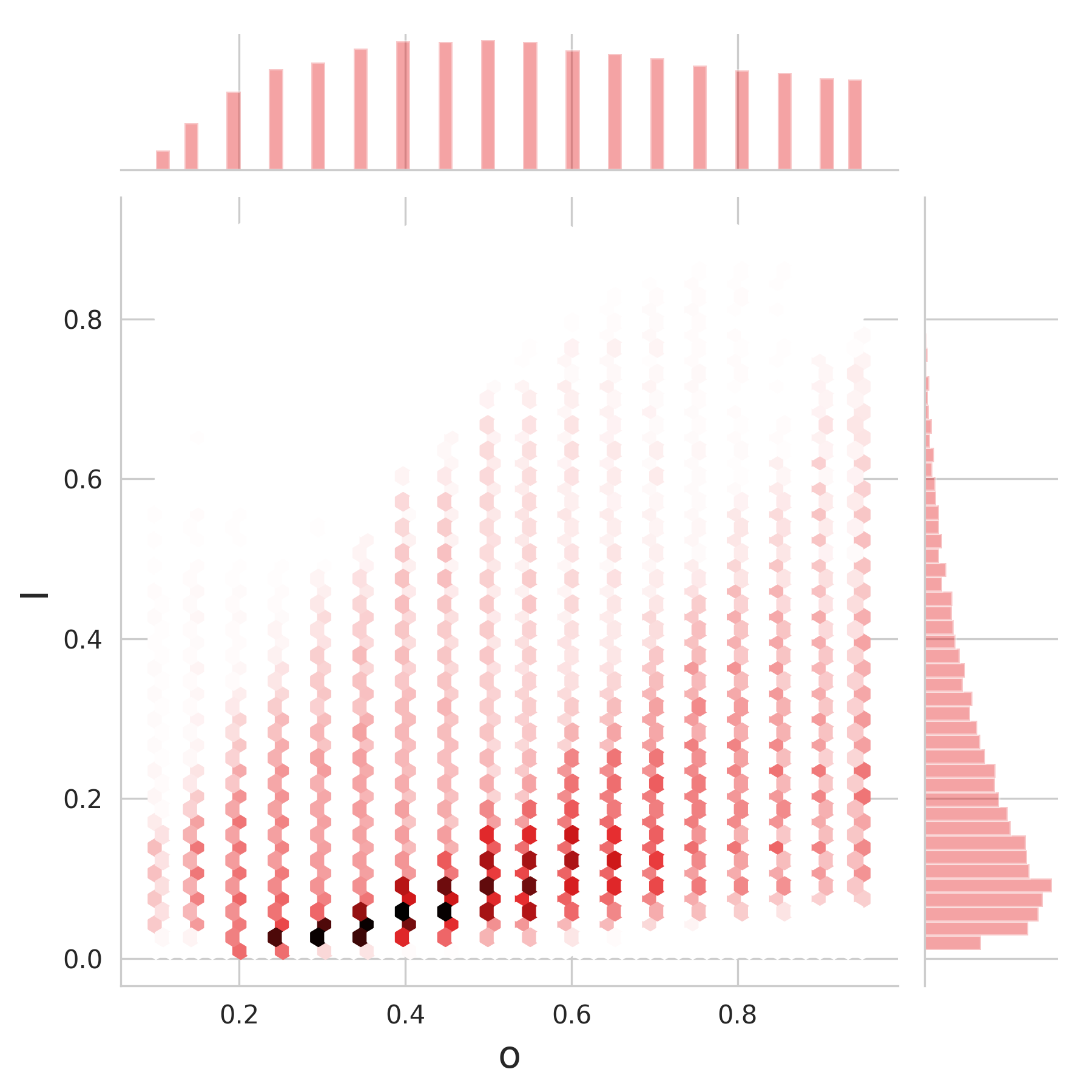}} \\
\subcaptionbox{Combined $H \times p$}{\includegraphics[width = 0.45\linewidth]{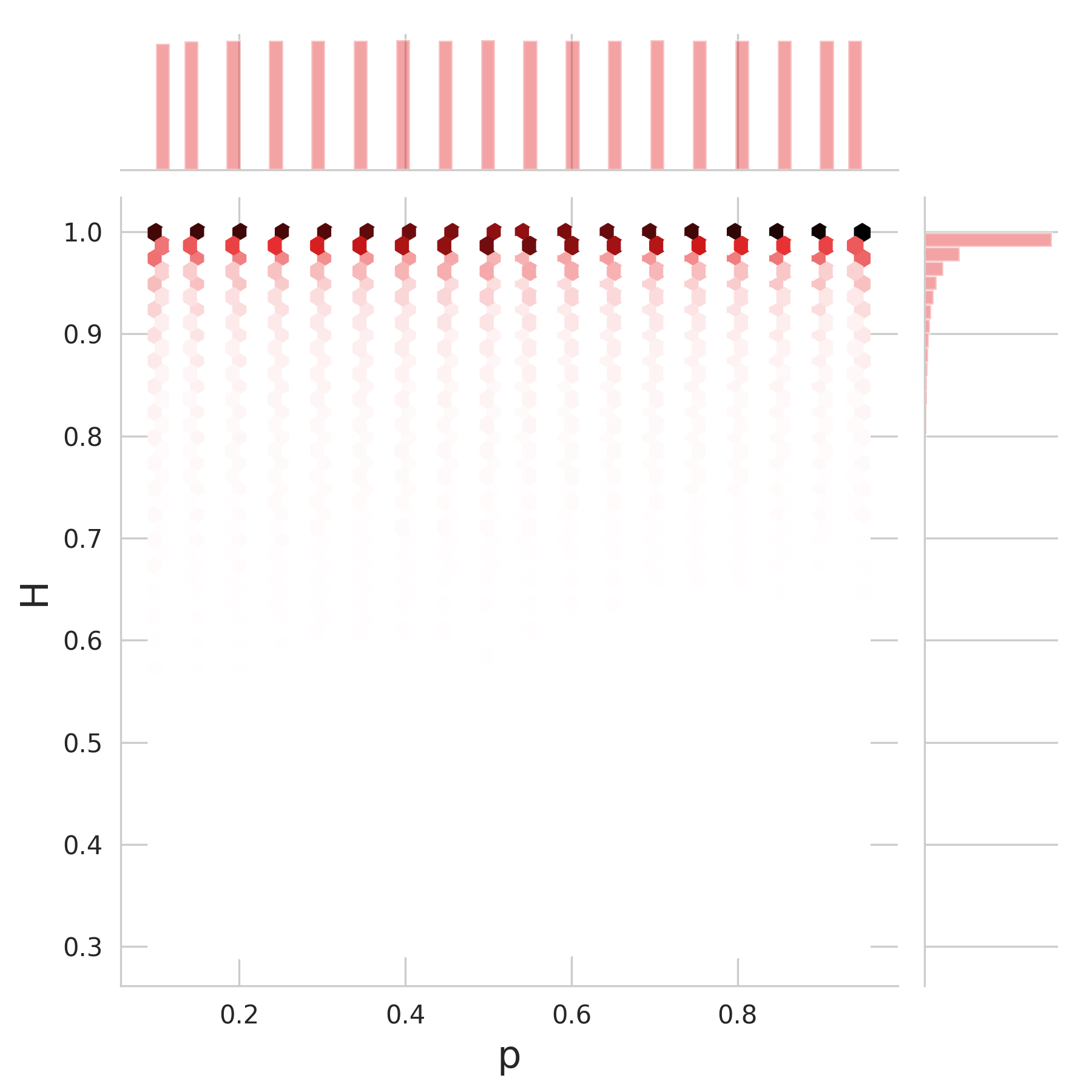}} &
\subcaptionbox{Combined $I \times p$}{\includegraphics[width = 0.45\linewidth]{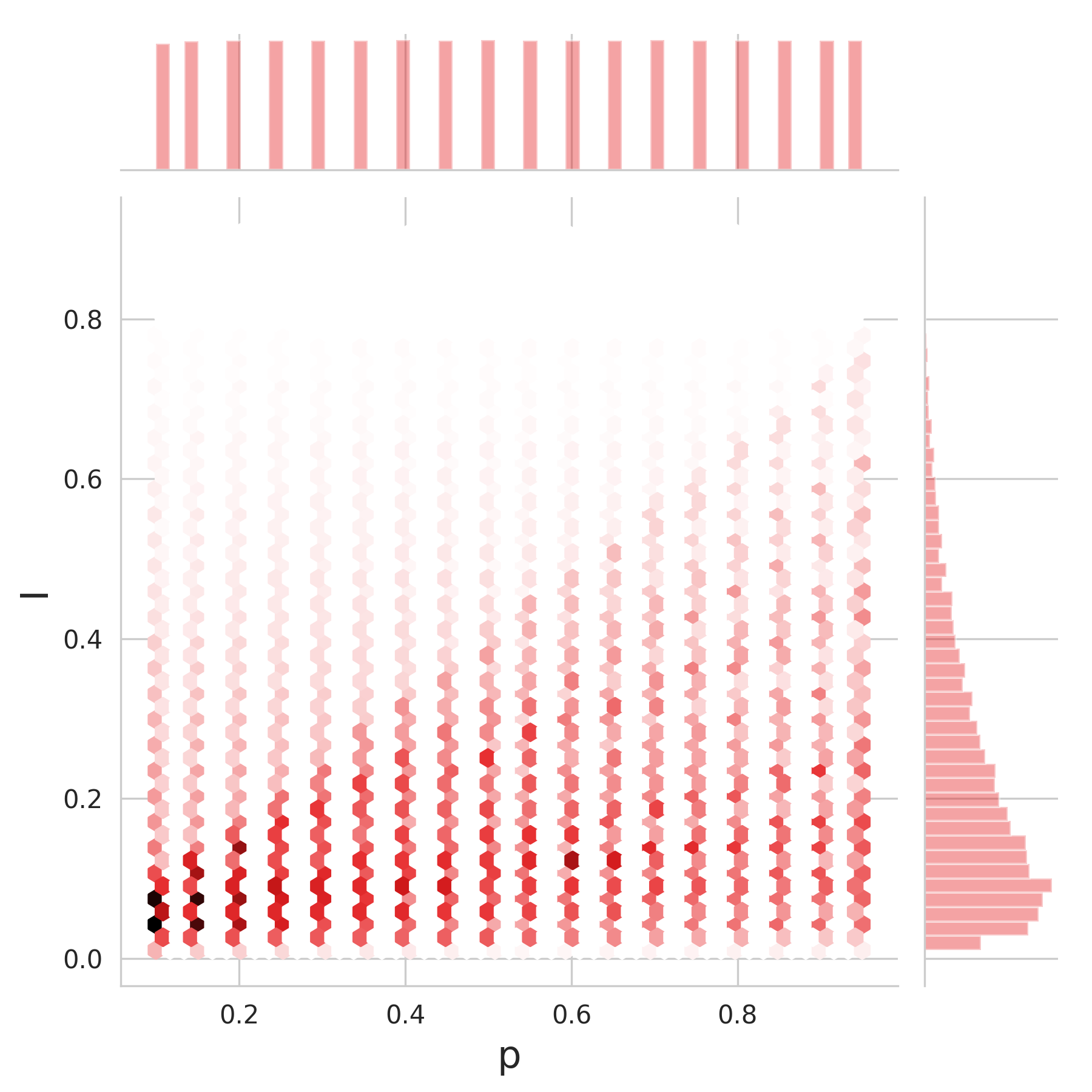}} \\
\subcaptionbox{Combined $H \times q$}{\includegraphics[width = 0.45\linewidth]{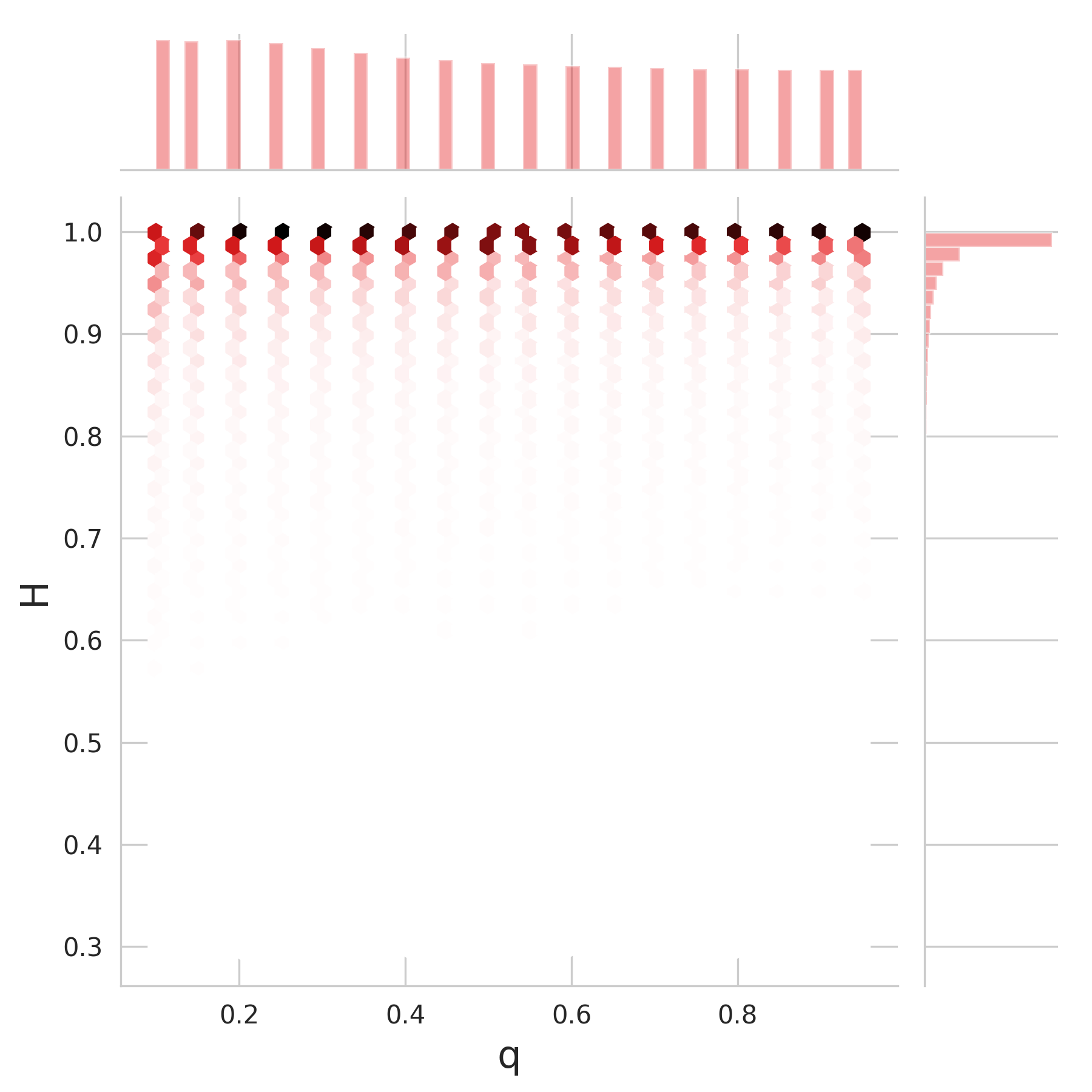}} &
\subcaptionbox{Combined $I \times q$}{\includegraphics[width = 0.45\linewidth]{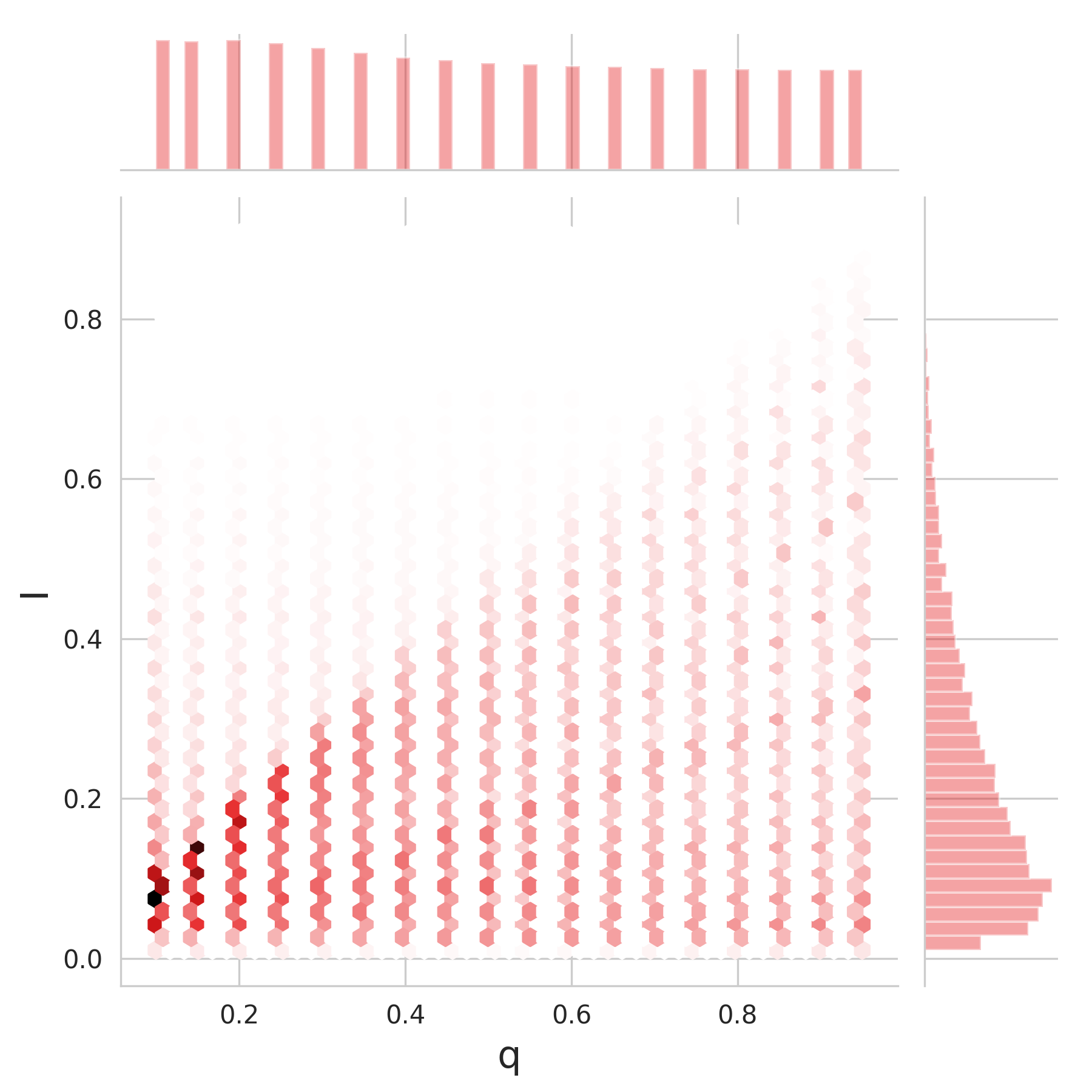}} \\
\end{tabular}
\caption{Dependency on the different probabilities $o$, $p$ and $q$ on the combined layers entanglement.}
\label{fig:apx_t3hives2}
\end{figure}

As we mentioned in Section~\ref{sec:interlayerdefinition}, one can compute entanglement over all the network, combining elementary layers and \REVISE{transition coupling} (as illustrated in Figure~\ref{fig:apx_illustration}). Although we have not identified practical use cases for this entanglement (often both categories of layers tell a different story), we report here the results over our synthetic networks in Figures~\ref{fig:apx_t3plots},~\ref{fig:apx_t3hives1}, and~\ref{fig:apx_t3hives2}. As expected we may observe a strong dependency over both $p$ and $q$ parameters combined (Figure~\ref{fig:apx_t3hives2}). Note that the current generator does not forbid the creation of loops enabling overlap between elementary layer and \REVISE{transition coupling}. A generation of \REVISE{transition coupling} edges that would connect \textit{different nodes} between layers would create even more overlap between elementary layers and \REVISE{transition coupling}. Such a parameter is actually available in the proposed code, but beyond the scope of this paper.

\newpage

\newpage

\newpage

\newpage
\subsection*{Choosing the right size of time window}

\begin{figure}[h!]
\centering
\captionsetup{width=.90\linewidth}
\begin{tabular}{cc}
\subcaptionbox{1h}{\includegraphics[width = 0.45\linewidth]{both_basic_slice_1h_1_mlking_1.png}} &
\subcaptionbox{3h}{\includegraphics[width = 0.45\linewidth]{both_basic_slice_3h_1_mlking_1.png}} \\
\subcaptionbox{6h}{\includegraphics[width = 0.45\linewidth]{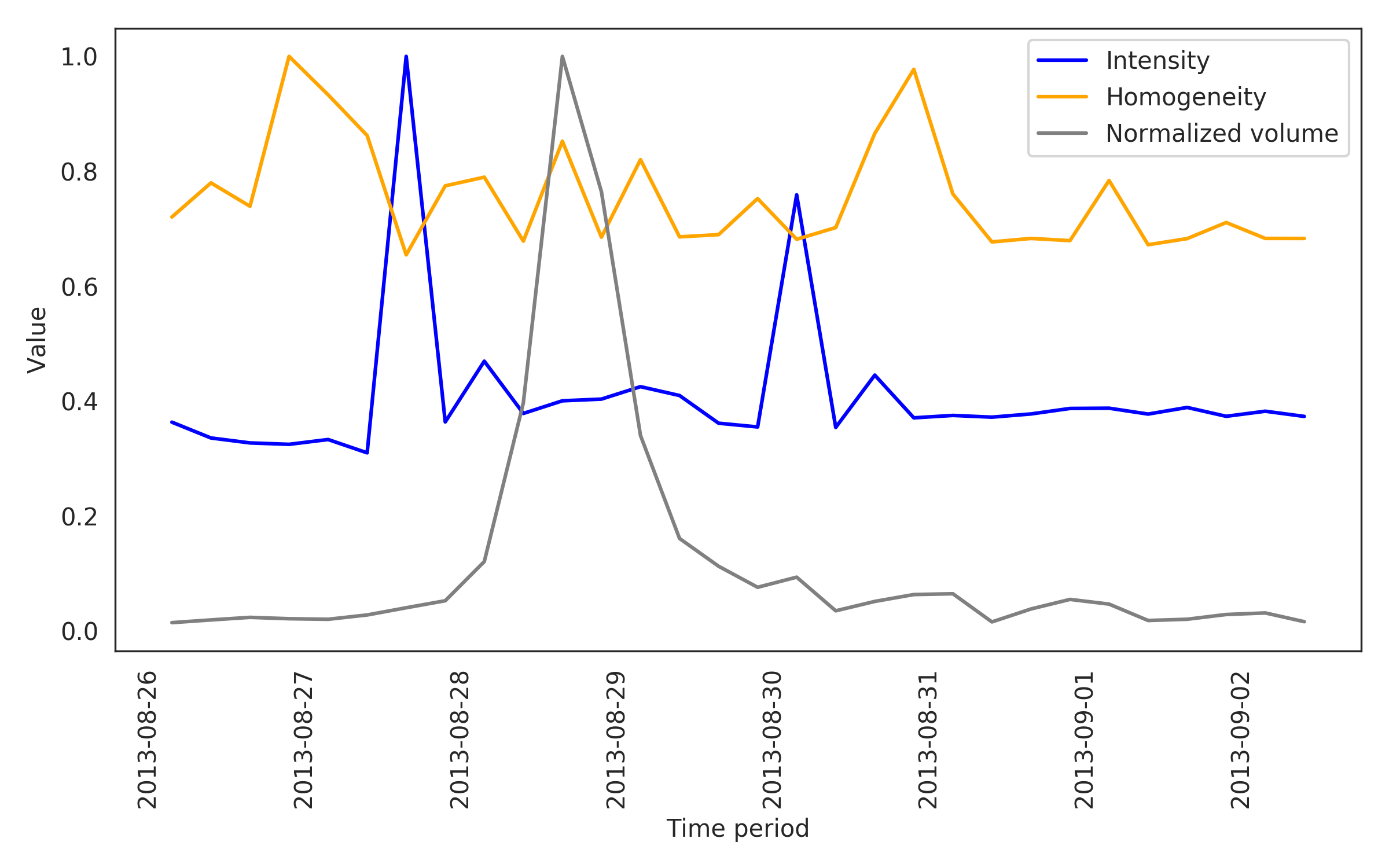}} &
\subcaptionbox{12\REVISE{h}}{\includegraphics[width = 0.45\linewidth]{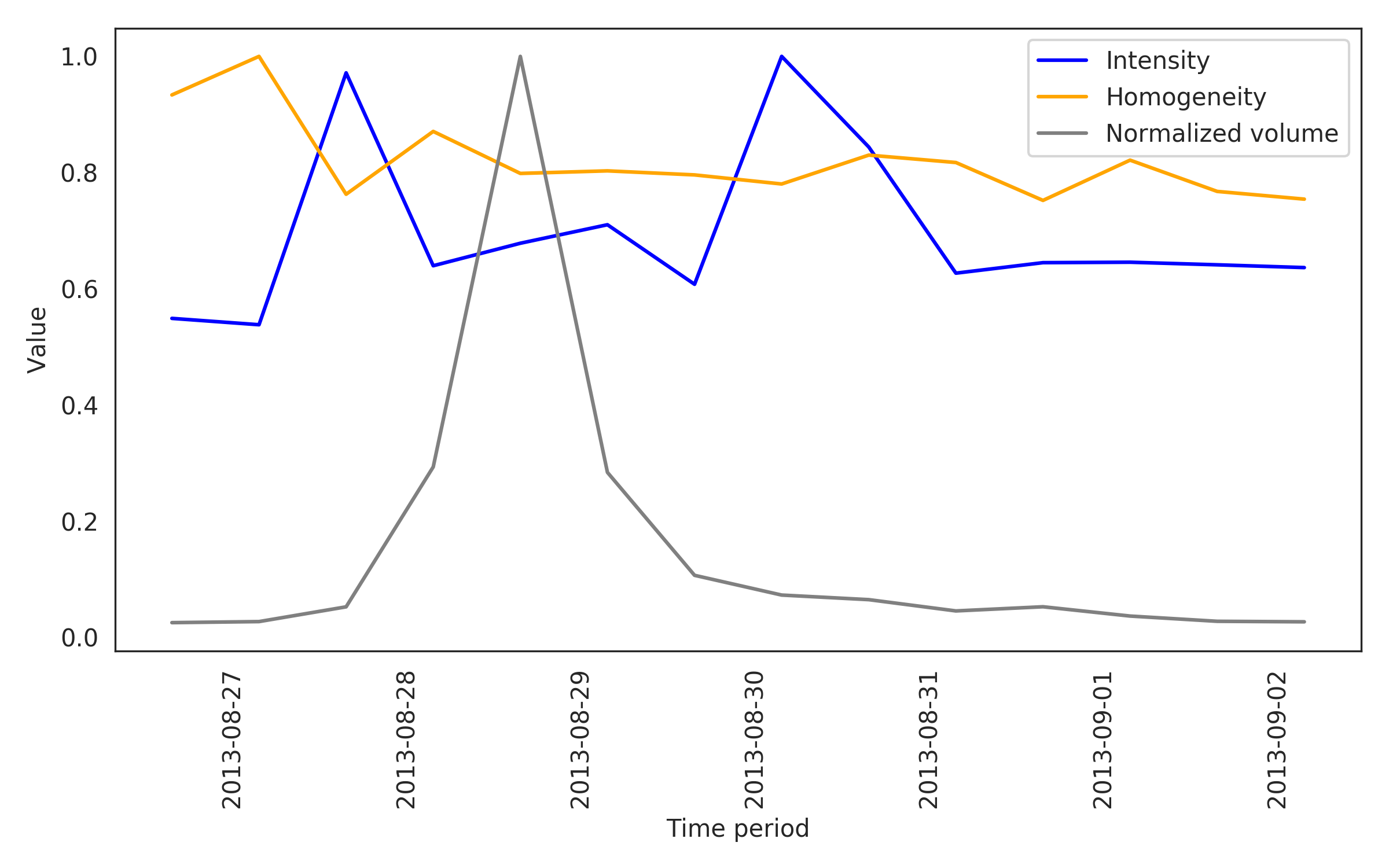}} \\
\end{tabular}
\caption{Different sizes of time windows for the \textit{MLKing2013} data set.}
\label{fig:apx_time_mlking}
\end{figure}

\begin{figure}[h!]
\centering
\captionsetup{width=.90\linewidth}
\begin{tabular}{cc}
\subcaptionbox{1h}{\includegraphics[width = 0.45\linewidth]{both_basic_slice_1h_1_moscow_1.png}} &
\subcaptionbox{3h}{\includegraphics[width = 0.45\linewidth]{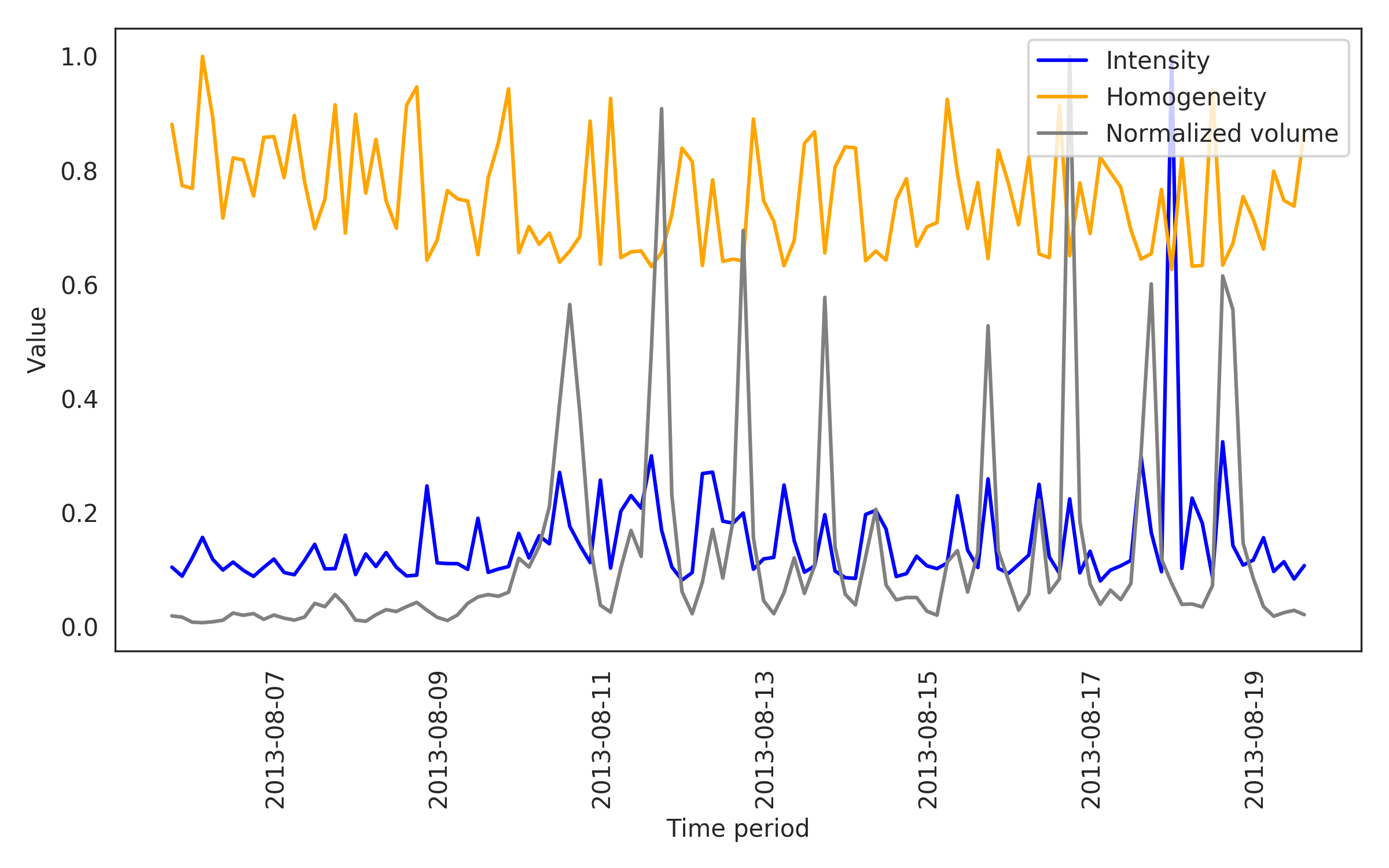}} \\
\subcaptionbox{6h}{\includegraphics[width = 0.45\linewidth]{both_basic_slice_6h_1_moscow_1.png}} &
\subcaptionbox{12\REVISE{h}}{\includegraphics[width = 0.45\linewidth]{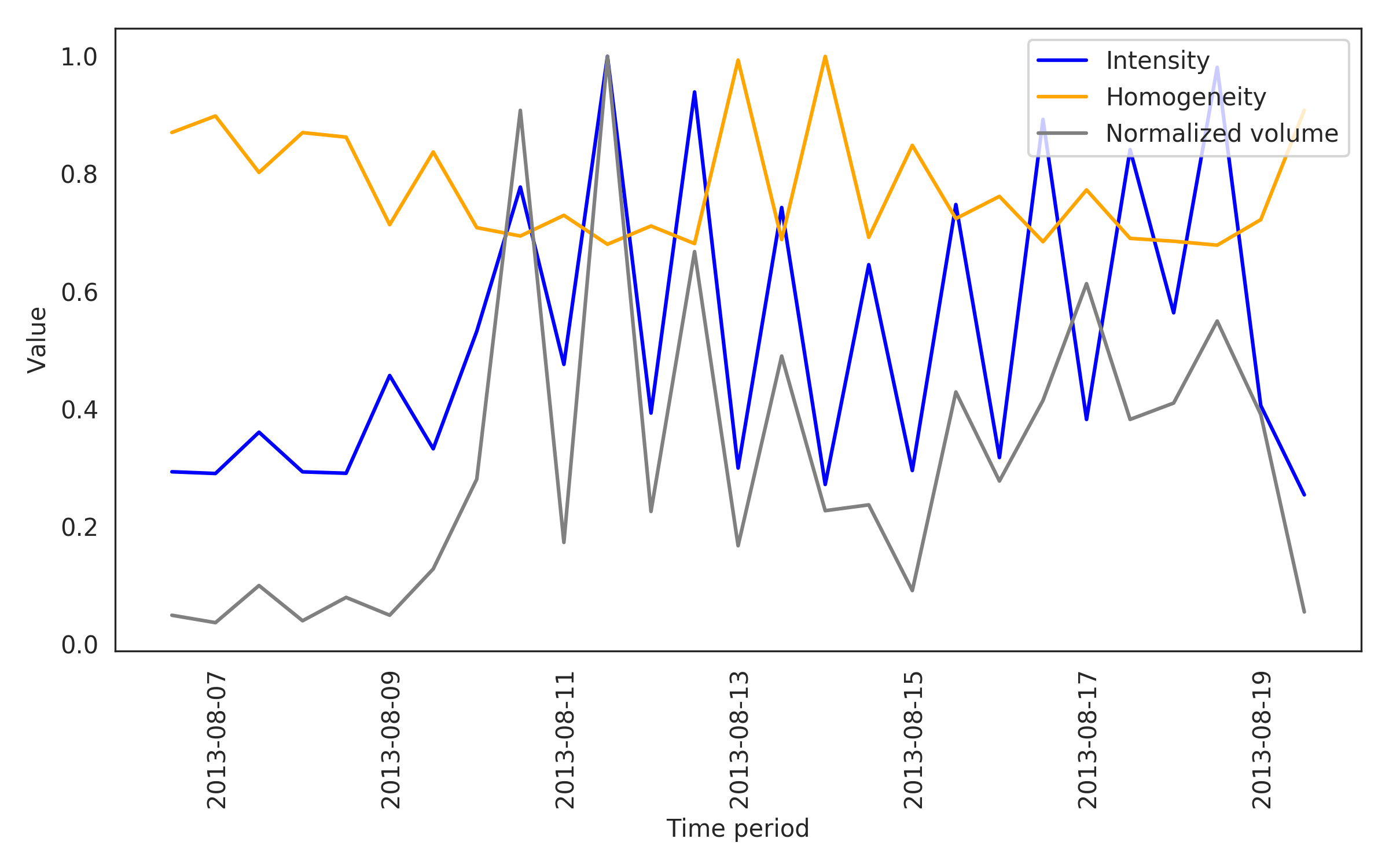}} \\
\end{tabular}
\caption{Different sizes of time windows for the \textit{MoscowAthletics2013} data set.}
\label{fig:apx_time_moscow}
\end{figure}

\begin{figure}[h!]
\centering
\captionsetup{width=.90\linewidth}
\begin{tabular}{cc}
\subcaptionbox{1h}{\includegraphics[width = 0.45\linewidth]{both_basic_slice_1h_1_cannes_1.png}} &
\subcaptionbox{3h}{\includegraphics[width = 0.45\linewidth]{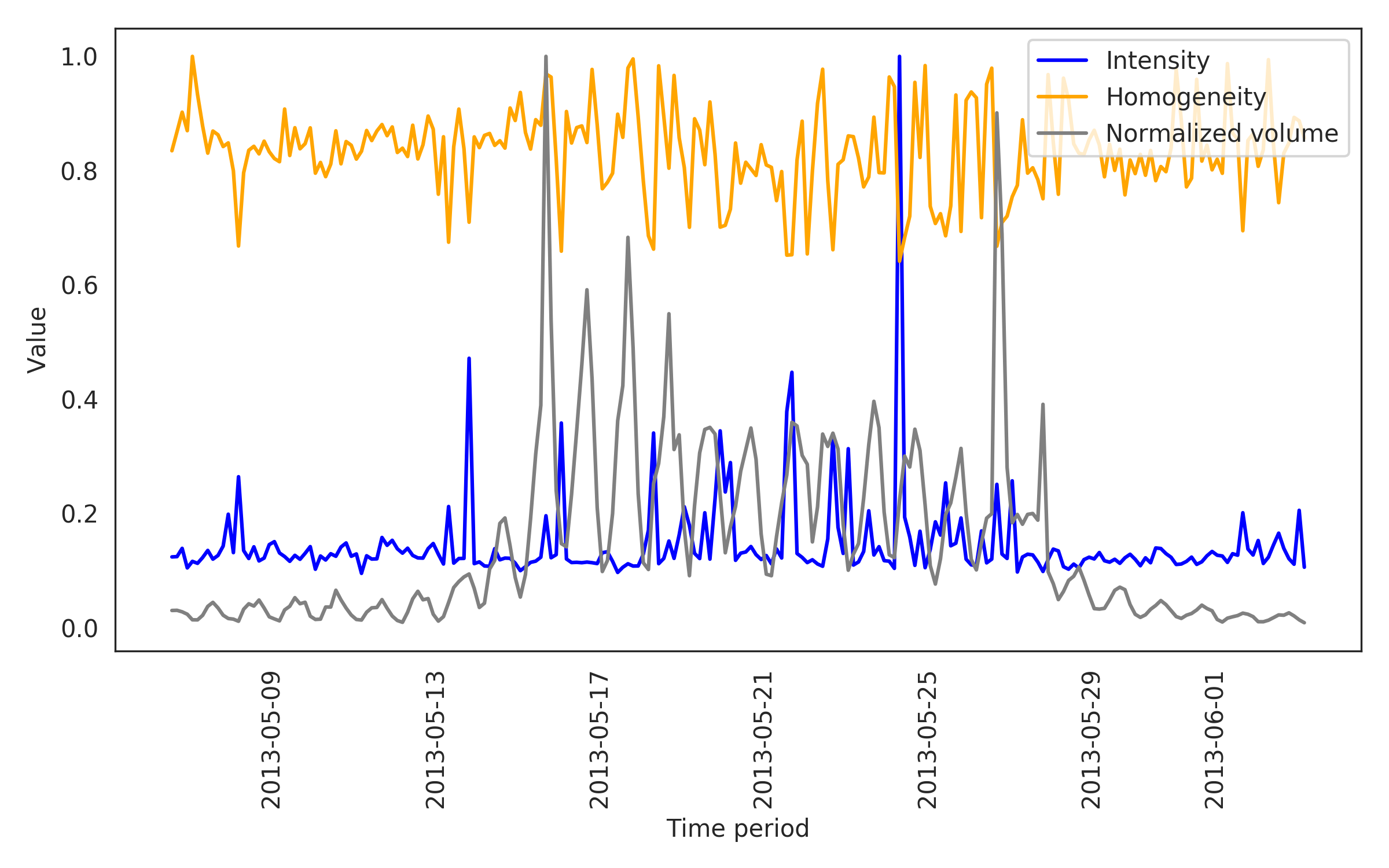}} \\
\subcaptionbox{6h}{\includegraphics[width = 0.45\linewidth]{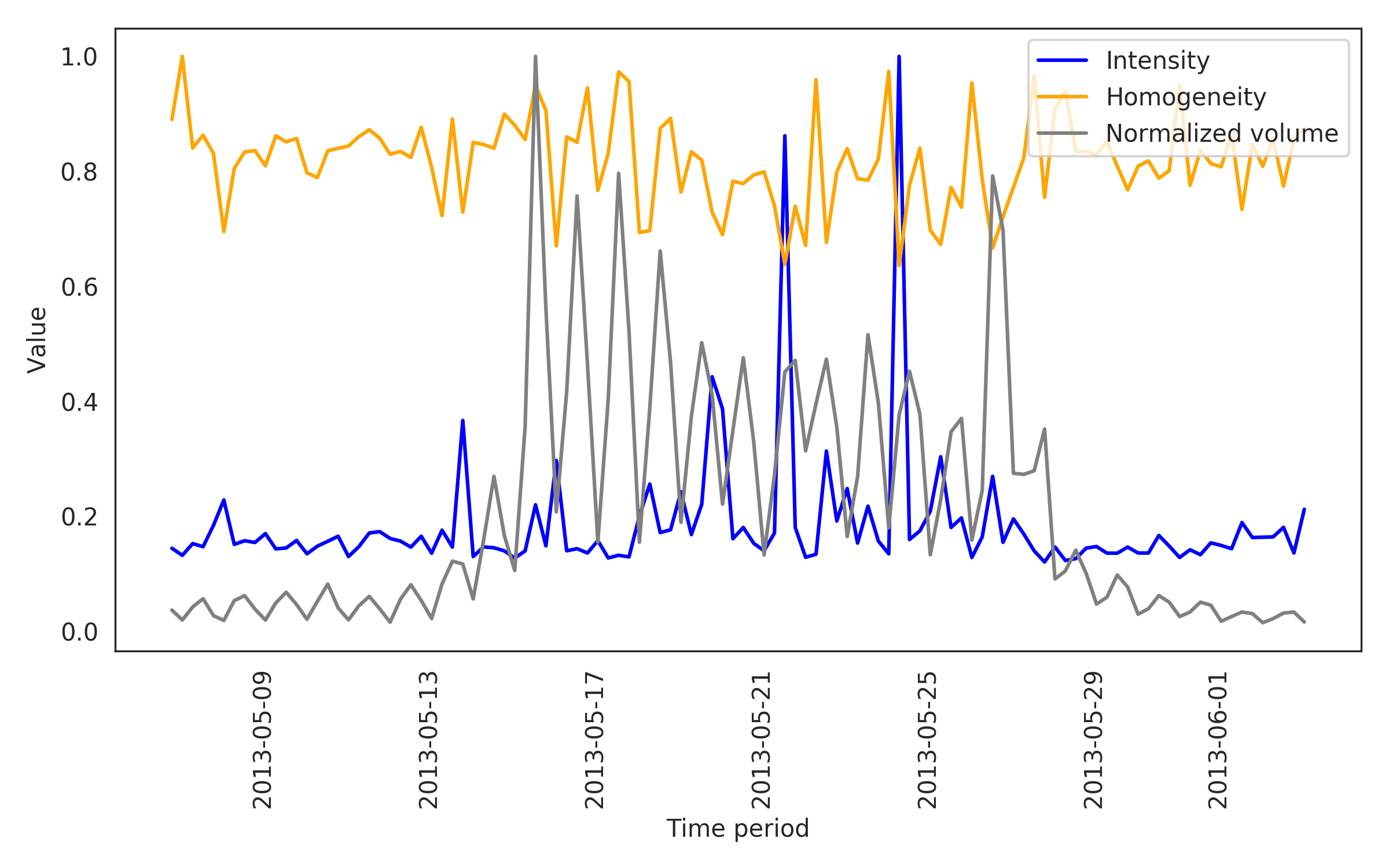}} &
\subcaptionbox{12\REVISE{h}}{\includegraphics[width = 0.45\linewidth]{both_basic_slice_12h_1_cannes_1.png}} \\
\end{tabular}
\caption{Different sizes of time windows for the \textit{Cannes2013} data set.}
\label{fig:apx_time_cannes}
\end{figure}

Choosing the right size of time-window fundamentally depends on the dataset we observe. We report all variations of fixed time window coarsening we have explored, among 1h, 3h, 6h, and 12h-long windows for each of the \textit{MLKing2013}  (Figures~\ref{fig:apx_time_mlking}), \textit{MoscowAthletics2013} (Figure~\ref{fig:apx_time_moscow}), and \textit{Cannes2013} (Figure~\ref{fig:apx_time_cannes}) events. Too fine selection displays a lot of noise, too coarse eludes most of the content.

\newpage

\newpage

\subsection*{\REVISE{Correlation analysis}}
\label{sec:correlation-analysis}

\begin{figure}[h!]
\centering
\captionsetup{width=.90\linewidth}
\begin{tabular}{cc}
\subcaptionbox{Std}{\includegraphics[width = 0.45\linewidth]{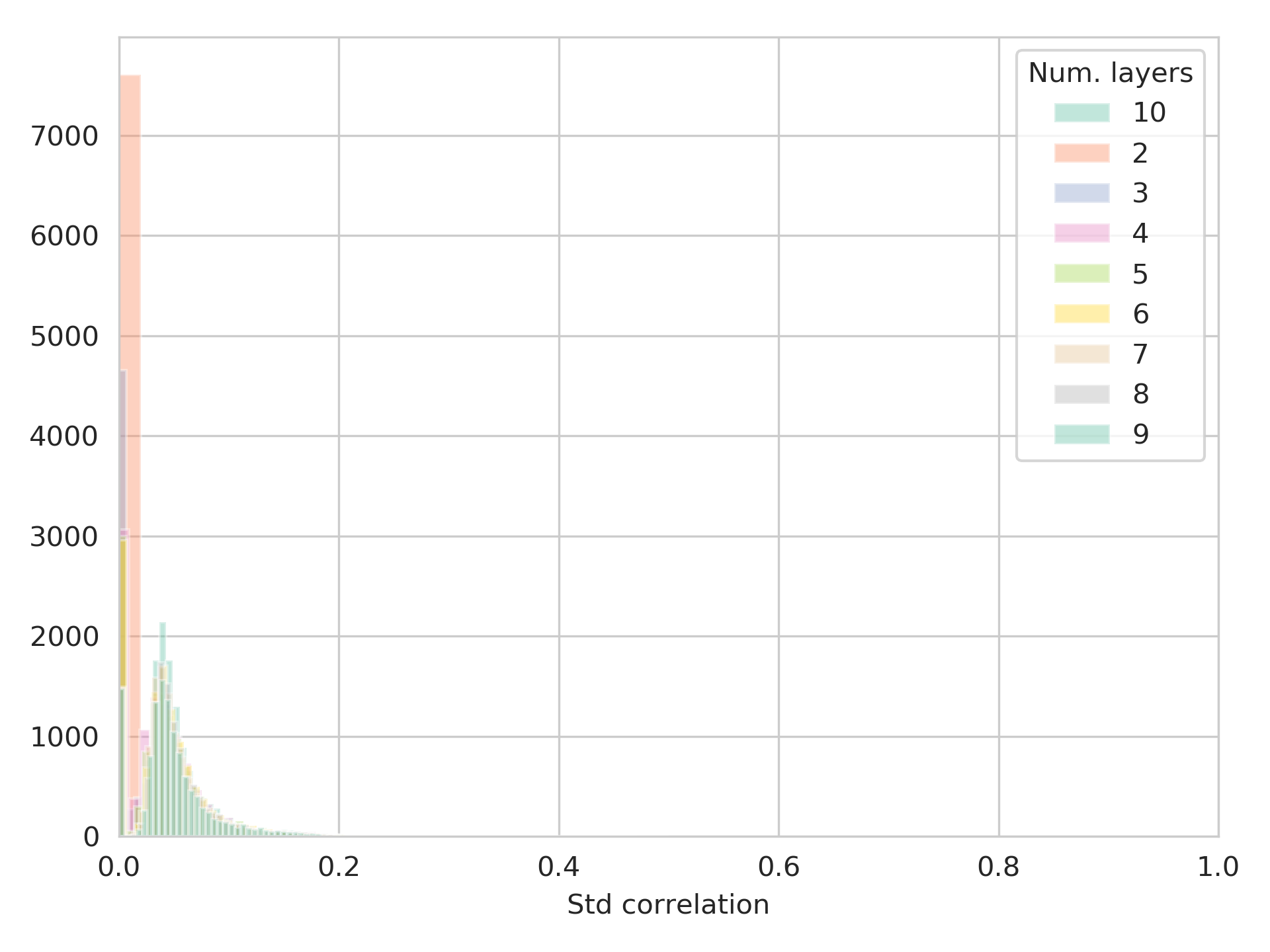}} &
\subcaptionbox{Max}{\includegraphics[width = 0.45\linewidth]{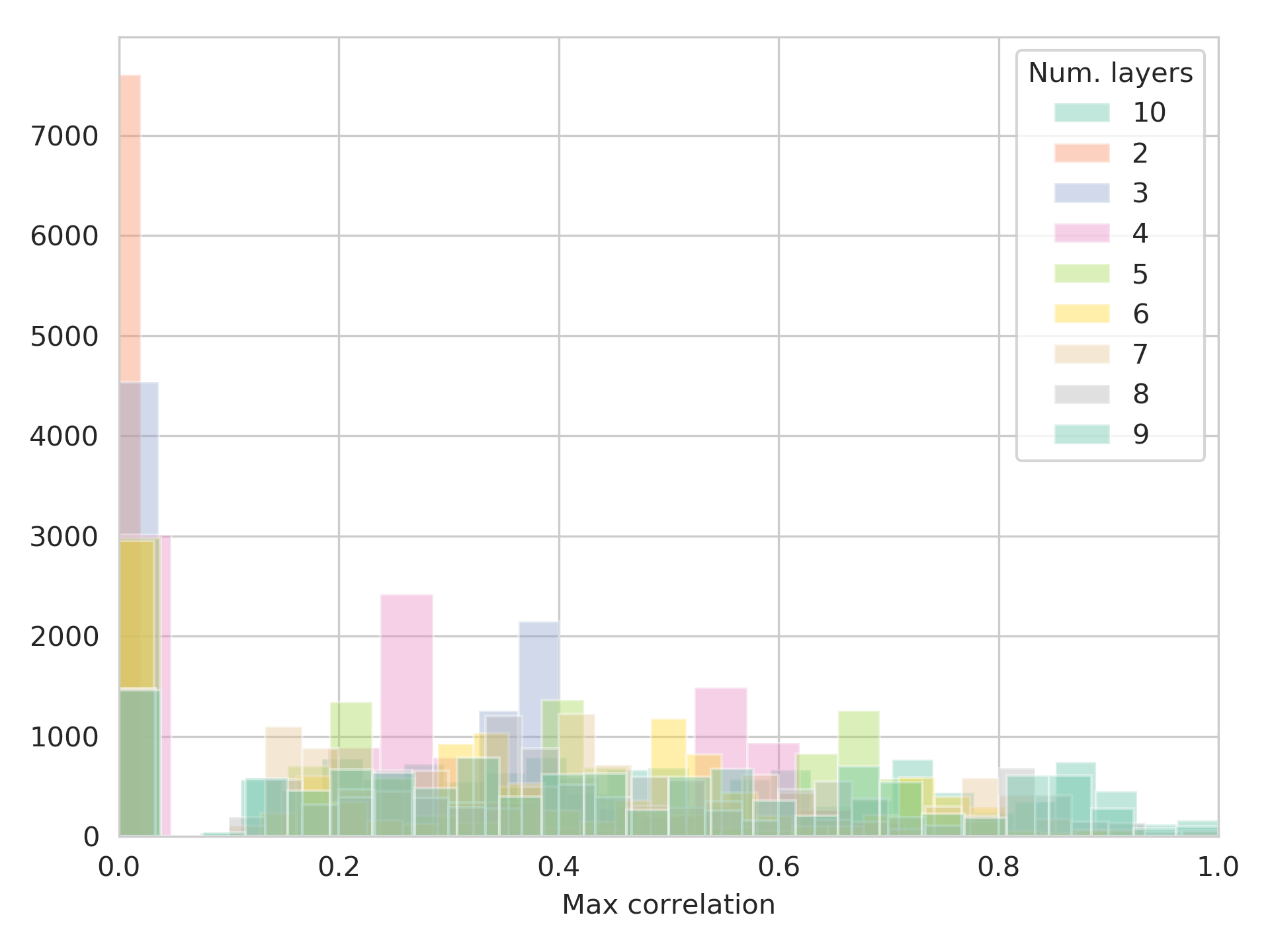}} \\
\subcaptionbox{Min}{\includegraphics[width = 0.45\linewidth]{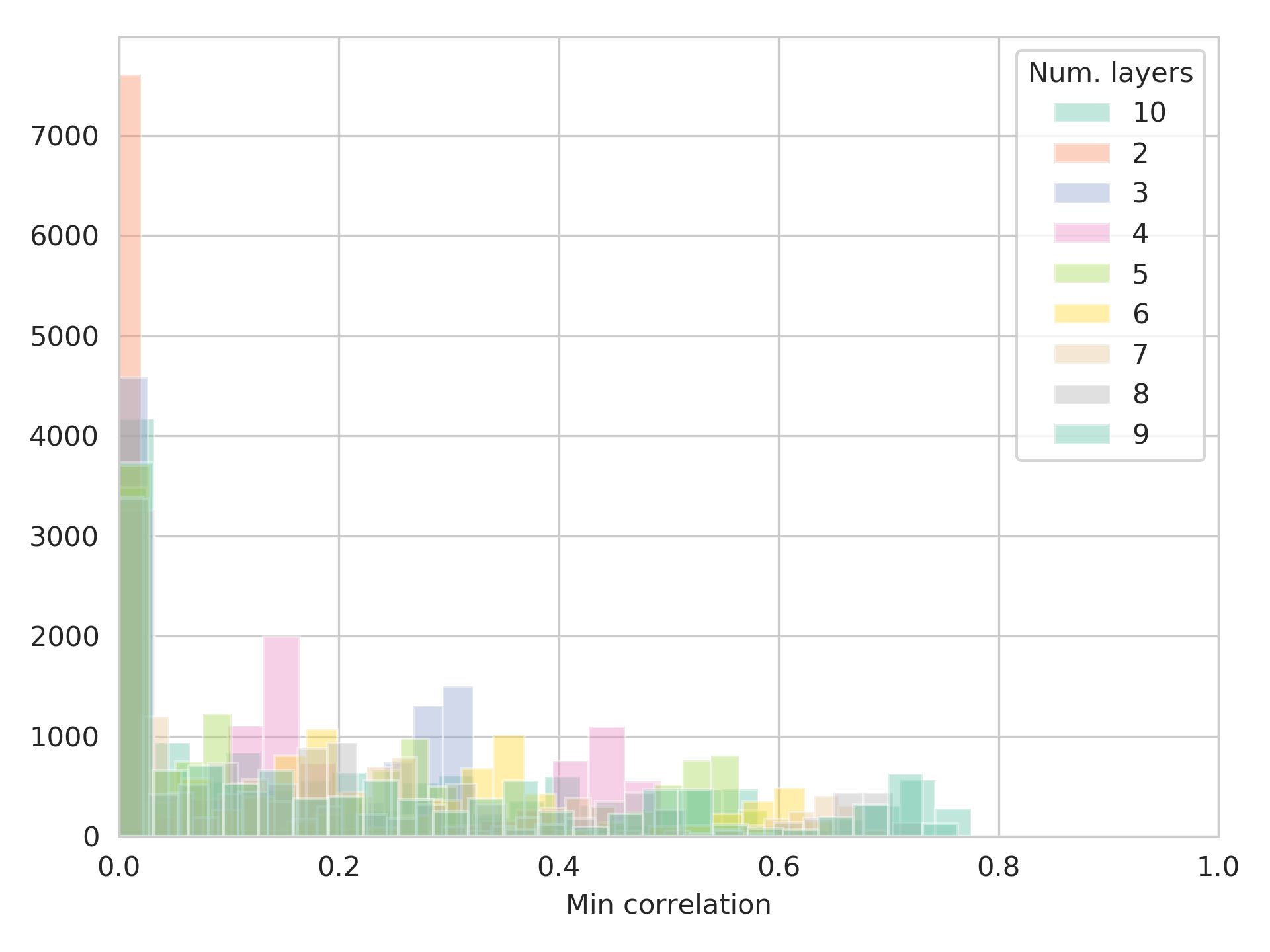}} &
\subcaptionbox{Mean}{\includegraphics[width = 0.45\linewidth]{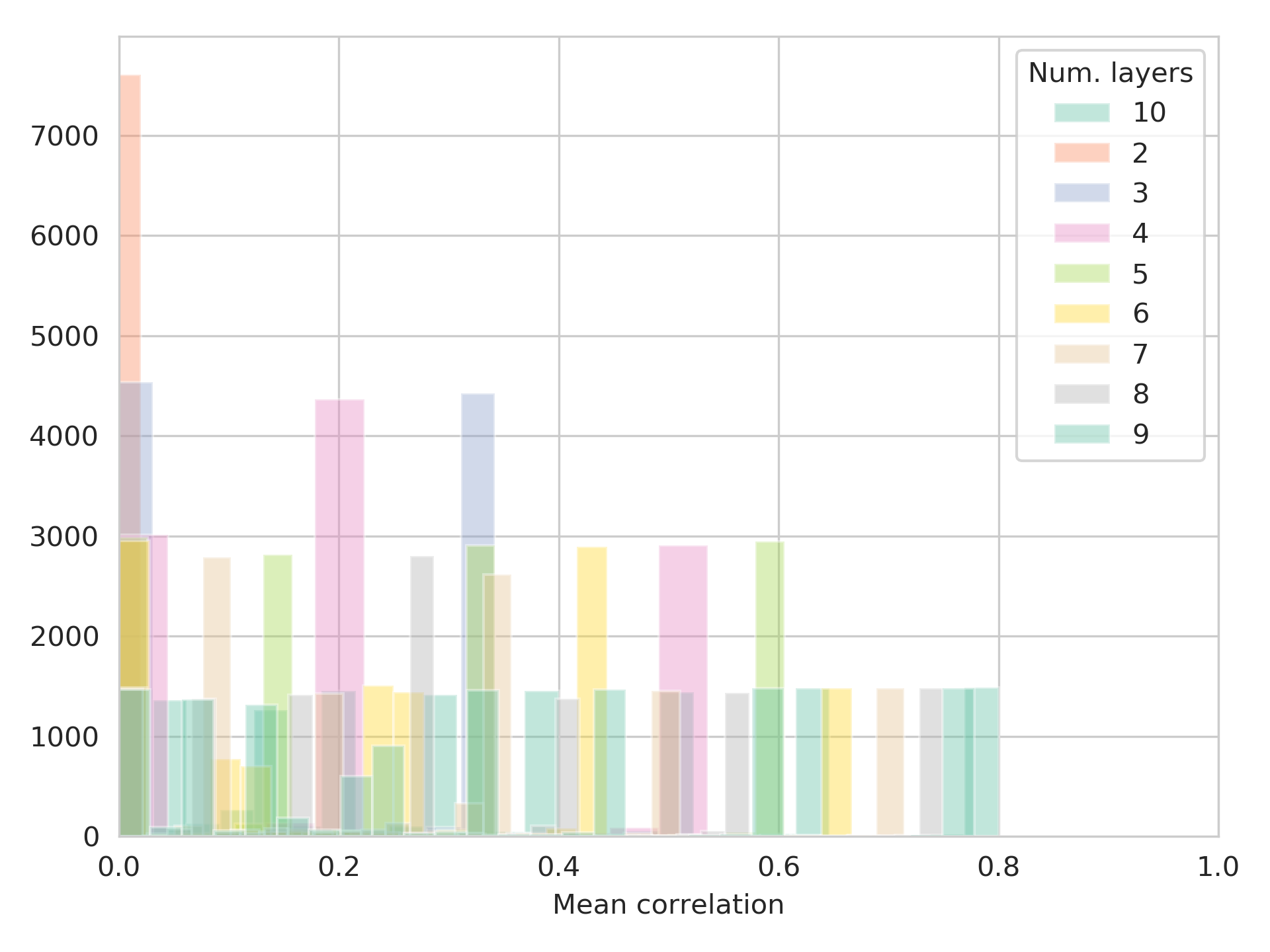}} \\
\end{tabular}

\caption{\REVISE{Results of correlation analysis. It can be observed that the larger layers are subject to the most dispersed correlation distributions (Std peak in (a)). Further, when considering only pairs of layers, very little correlation was observed in a large body of generated networks, indicating that by increasing the number of layers, correlation also increases.}}
\label{fig:correlations}
\end{figure}

\REVISE{
For completeness, we also computed the correlation between the occurrence of a given pair of nodes, as discussed in \cite{nicosia2015measuring}. As a measure of correlations of layer activity, we have computed the \textit{pairwise multiplexity}. 
Following the original notation, we computed, for each synthetic network:}
\begin{equation*}
    Q_{\alpha,\beta} = \frac{1}{|N|}\sum_i b_i^{[\alpha]} b_i^{[\beta]},
\end{equation*}
\REVISE{
where $\alpha \in L$ and $\beta \in L$ are two distinct layers. Here, $b_i \in [0,1]$ represents the presence of a given node $i$, hence, the product equals zero if a given node is not co-present on both considered layers. Given the large space of synthetic multiplex networks, it is not sensible to analyse individual ones, as performed in \cite{nicosia2014nonlinear}, hence we further computed:
}
\begin{equation*}
    Q_d = \{Q_{\alpha,\beta}\}_{\alpha, \beta \in L^2 \wedge (\alpha \neq \beta)},
\end{equation*}
\REVISE{
i.e., the distribution of all possible pairwise correlations. In Figure~\ref{fig:correlations}, we present the statistical properties of this distribution further segmented according to the number of layers. 
}

\begin{figure}[h!]
\centering
\captionsetup{width=.90\linewidth}
\begin{tabular}{cc}
\subcaptionbox{$m$}{\includegraphics[width = 0.45\linewidth]{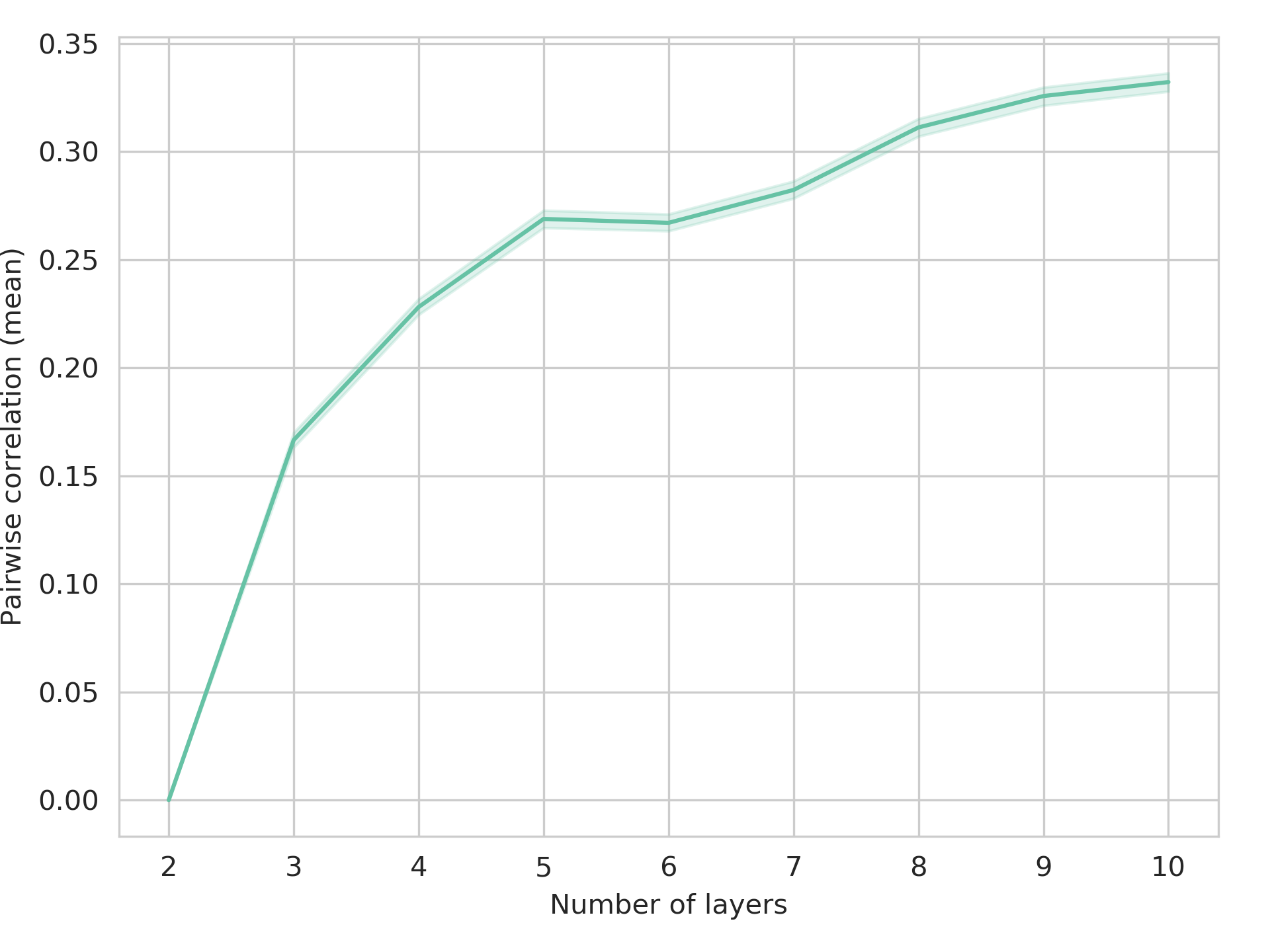}} &
\subcaptionbox{$n$}{\includegraphics[width = 0.45\linewidth]{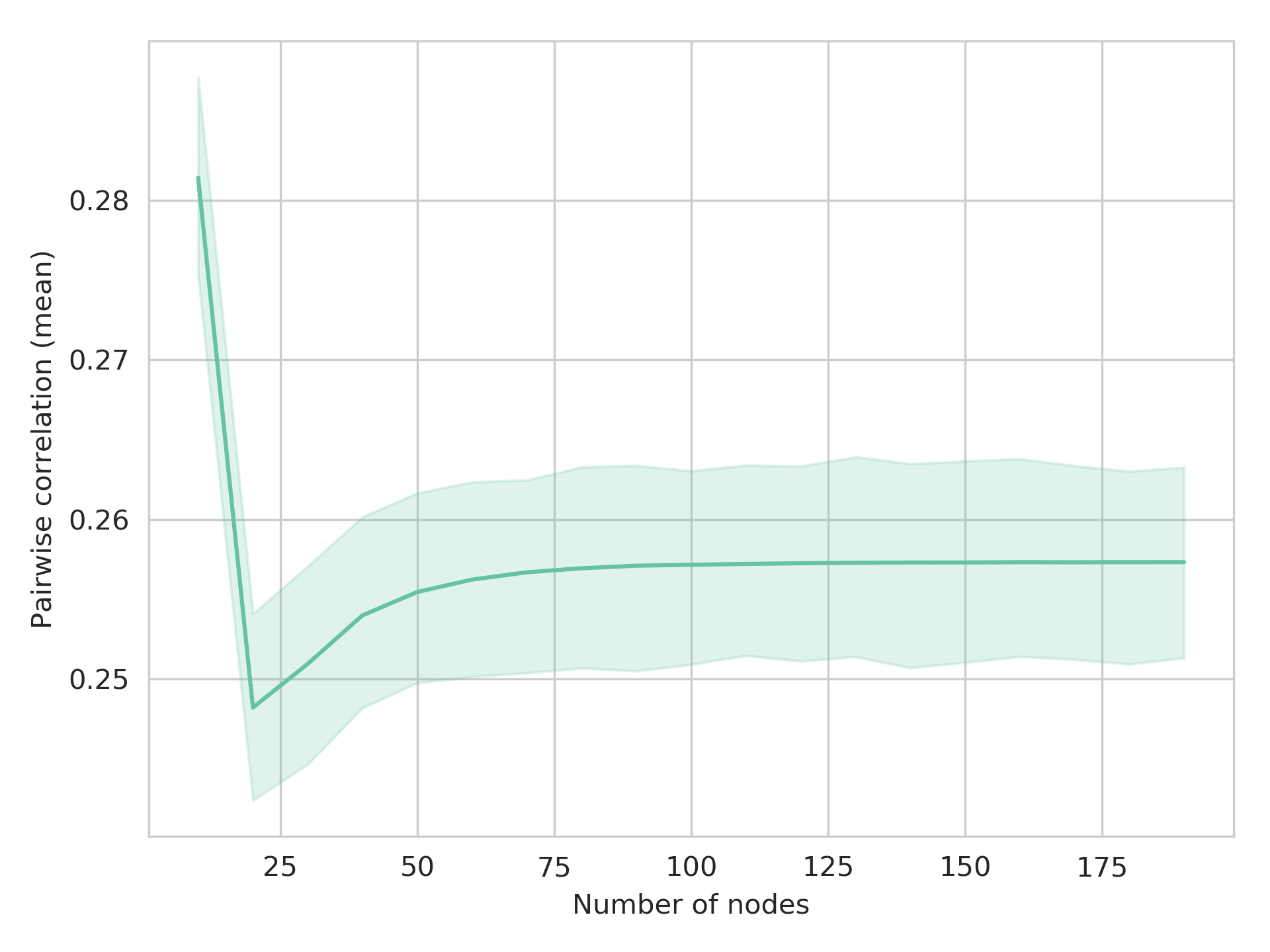}} \\
\subcaptionbox{$o$}{\includegraphics[width = 0.45\linewidth]{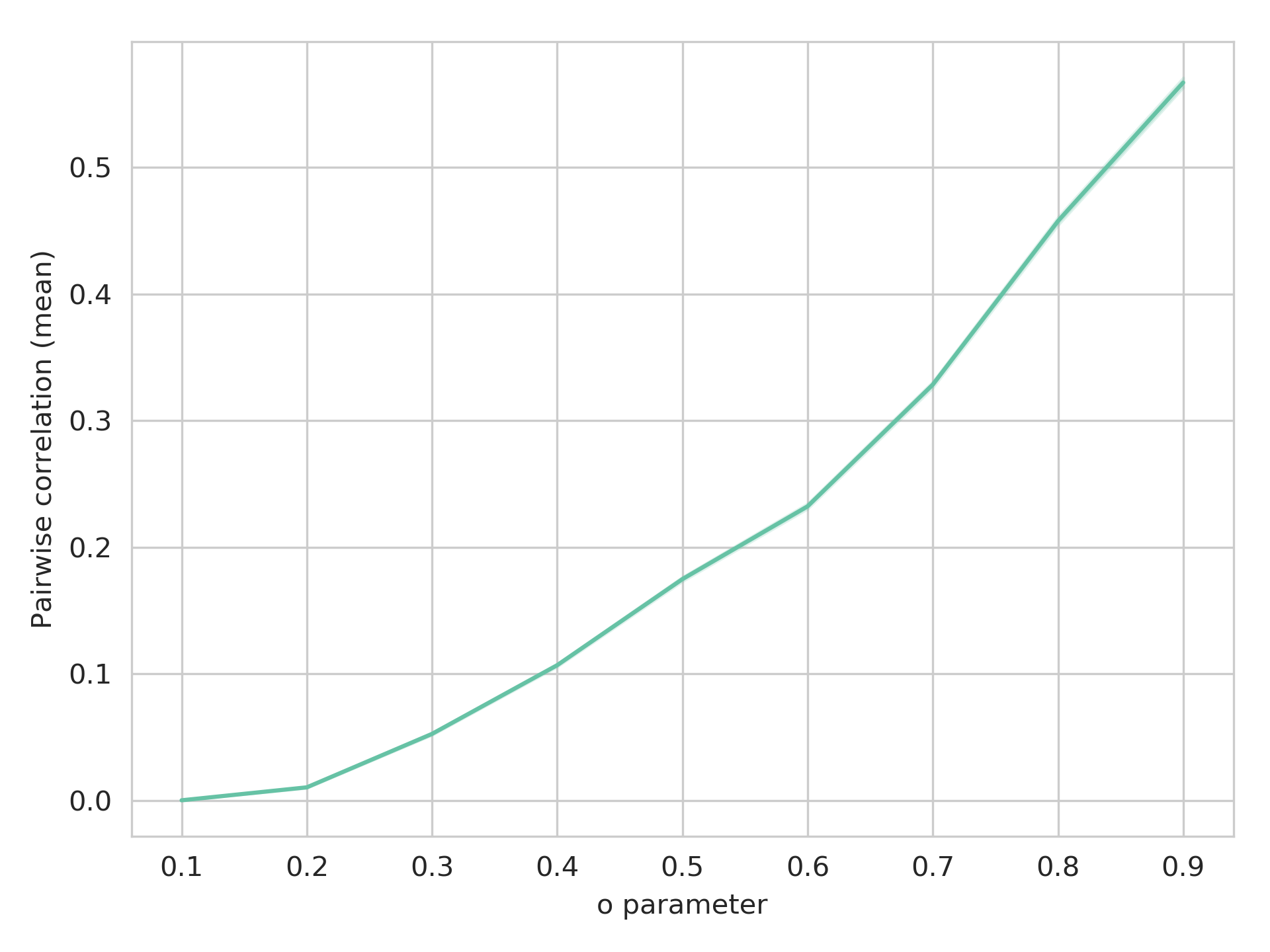}} &
\subcaptionbox{$p$}{\includegraphics[width = 0.45\linewidth]{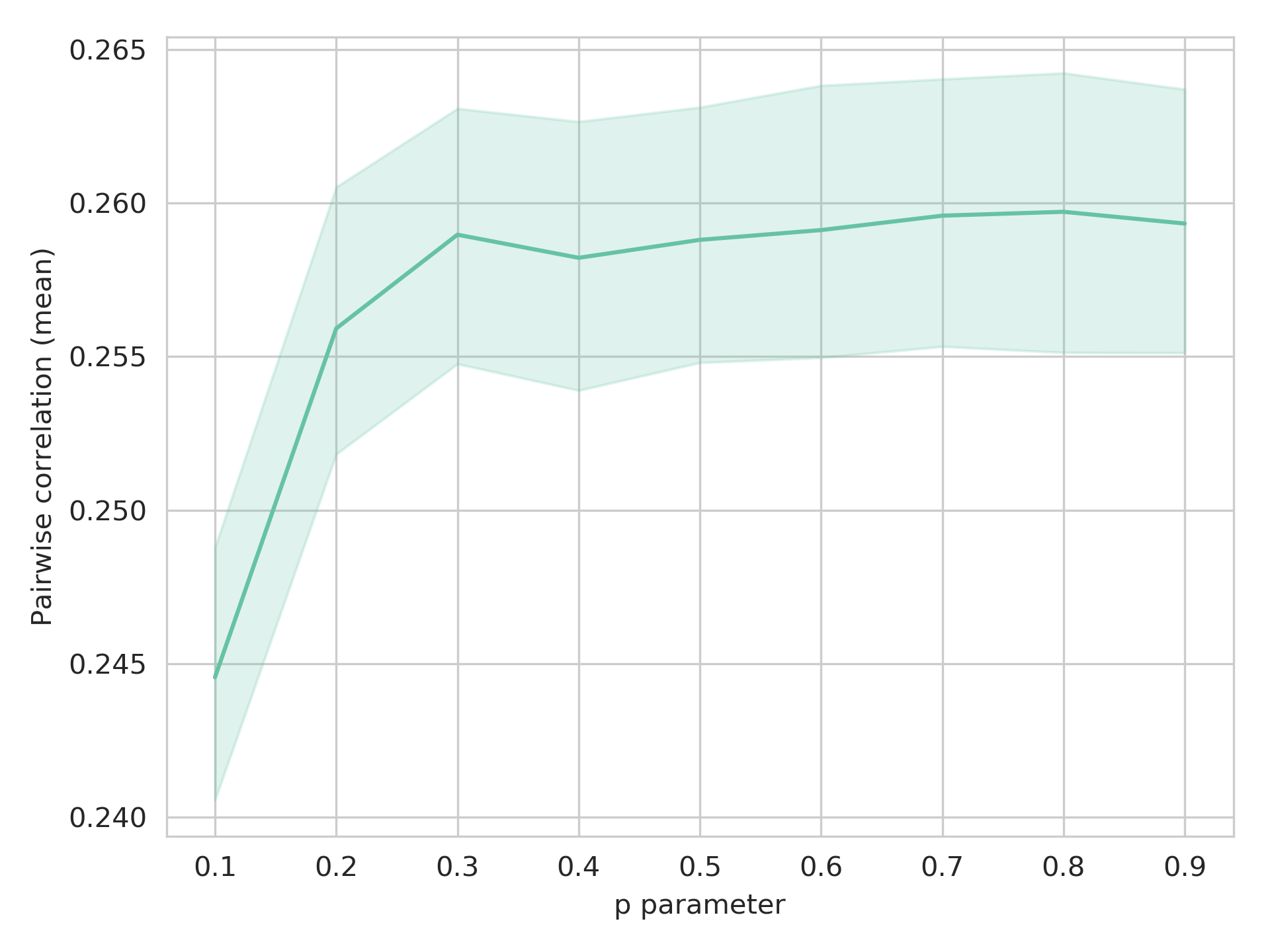}} \\
\end{tabular}
\subcaptionbox{$q$}{\includegraphics[width = 0.45\linewidth]{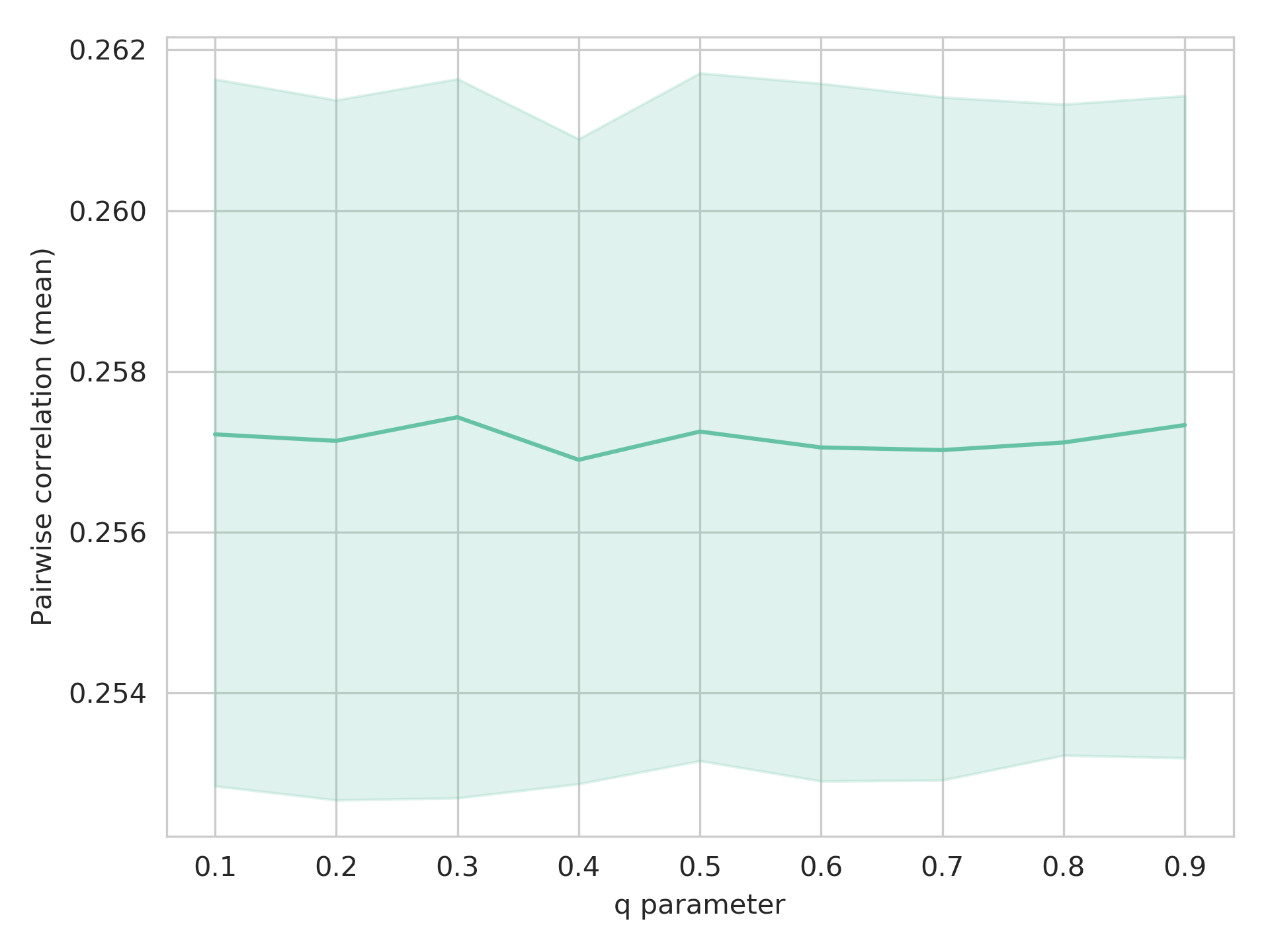}}

\caption{\REVISE{Comparison of the distribution of (averaged, with min/max) correlations as a function of: (a) number of layer $m$, (b) number of nodes $n$, (c) layer assignment probability $o$, (d) inner-layer edge probability $p$, and (e) transition coupling edge probability $q$.}}
\label{fig:correlations_parameters}
\end{figure}

\REVISE{
We further compare this distribution with each parameter of our generator in Figure~\ref{fig:correlations_parameters}. We first could expect some level of relationship with the number of layers and the node layer probability. This is confirmed, and we can observe a direct dependency with the number of layers $m$, and the node-layer assignment probability $o$. The relationship with the number of layers slowly decreases with a minimum of 5 layers, since with more layers, there are more degrees of freedom for nodes to be assigned on layers. The relationship with $o$ is almost linear. There is almost no dependency with the number of nodes (once enough nodes are assigned). The inner-layer edge probability $p$ does not influence this measure, although we may observe a sudden increase first, it is an artefact of our algorithm because nodes with degree zero in a layer are discarded from the layer to speed up computations. The transition coupling edge probability $q$ shows no correlation because this algorithmic artefact does not apply to it.}

\end{document}